\documentclass[a4paper]{article} 
\usepackage{geometry}
\usepackage{tgpagella}
\usepackage{fontawesome5}
\usepackage[usenames,dvipsnames,svgnames,table]{xcolor}
\definecolor{mybg}{HTML}{282A36}
\definecolor{mycl}{HTML}{44475A}
\definecolor{myfg}{HTML}{F8F8F2}
\definecolor{mycomment}{HTML}{6272A4}
\definecolor{mycyan}{HTML}{8BE9FD}
\definecolor{mygreen}{HTML}{50FA7B}
\definecolor{myorange}{HTML}{FFB86C}
\definecolor{mypink}{HTML}{FF79C6}
\definecolor{mypurple}{HTML}{BD93F9}
\definecolor{myred}{HTML}{FF5555}
\definecolor{myyellow}{HTML}{F1FA8C}
\definecolor{VividPurple}{HTML}{3E0097}
\definecolor{SlateGrey}{HTML}{2E2E2E}
\definecolor{LightGrey}{HTML}{666666}
\definecolor{DarkPastelRed}{HTML}{450808}
\definecolor{PastelRed}{HTML}{8F0D0D}
\definecolor{GoldenEarth}{HTML}{E7D192}
\definecolor{awesome-emerald}{HTML}{00A388}
\definecolor{awesome-emerald-dark}{HTML}{00806A} 
\definecolor{awesome-skyblue}{HTML}{0395DE}
\definecolor{awesome-skyblue-dark}{HTML}{0376B0}
\definecolor{awesome-red}{HTML}{DC3522}
\definecolor{awesome-red-dark}{HTML}{B02A1C}
\definecolor{awesome-pink}{HTML}{EF4089}
\definecolor{awesome-pink-dark}{HTML}{EC136D}
\definecolor{awesome-orange}{HTML}{FF6138}
\definecolor{awesome-orange-dark}{HTML}{FF3300}
\definecolor{awesome-nephritis}{HTML}{27AE60}
\definecolor{awesome-nephritis-dark}{HTML}{219150}
\definecolor{awesome-concrete}{HTML}{95A5A6}
\definecolor{awesome-concrete-dark}{HTML}{74898B}
\definecolor{awesome-darknight}{HTML}{131A28}
\definecolor{awesome-darknight-dark}{HTML}{101623}
\definecolor{awesome-snowwhite}{HTML}{F9FBFD}
\definecolor{awesome-snowwhite-dark}{HTML}{F3F6FB}
\definecolor{awesome-blue-dark}{HTML}{0000FF}
\definecolor{awesome-golden}{HTML}{E1AD21}
\definecolor{awesome-silver}{HTML}{AAA9AD}
\definecolor{darktext}{HTML}{414141}
\definecolor{darktext-dark}{HTML}{262626}
\definecolor{text}{HTML}{333333}
\definecolor{graytext}{HTML}{5D5D5D}
\definecolor{lighttext}{HTML}{999999}
\definecolor{VividRed}{HTML}{7e2635}
\definecolor{DarkRed}{HTML}{a5402d}
\definecolor{SlateGrey}{HTML}{2E2E2E}
\definecolor{LightGrey}{HTML}{666666}
\UseRawInputEncoding
\usepackage[english]{babel}
\usepackage{siunitx}
\usepackage{amsmath, nccmath} 
\usepackage{gensymb}
\usepackage{amssymb}
\usepackage{graphics}
\usepackage{graphicx}
\usepackage{epsfig}
\usepackage{fancyhdr}
\usepackage{wasysym}
\usepackage{latexsym}
\usepackage{amsfonts}
\usepackage{color,colortbl}
\usepackage{cite}
\usepackage{notoccite} 
\usepackage{soul}
\usepackage{multirow}
\usepackage{etoolbox}
\usepackage{authblk}
\usepackage{blindtext}
\usepackage{dcolumn}
\usepackage{bm}
\usepackage{qcircuit}
\usepackage{braket}
\usepackage{physics}
\usepackage{braket}	
\usepackage{varwidth}
\usepackage{empheq}
\usepackage[]{algorithm2e}
						   
\usepackage{algorithmic}
\usepackage{framed}
\usepackage{enumitem}
\usepackage[utf8x]{inputenc} 
\usepackage[english]{babel}	
\usepackage{comment}					   
\usepackage{url}
\usepackage{hyperref}
\usepackage{caption}
\usepackage[font={color=PastelRed,bf,footnotesize}]{caption}

\usepackage{enumitem}
\usepackage{multirow}
\usepackage{textcomp}
\usepackage[T1]{fontenc}
\usepackage{listings}
\usepackage{xparse}
\NewDocumentCommand{\codeword}{v}{%
\texttt{\textcolor{blue}{#1}}%
}
\usepackage{subcaption}
\usepackage{textcomp}
\usepackage{mathrsfs}
\usepackage{hyperref}

\usepackage{cleveref}
\usepackage{xr}
\graphicspath{{./figures}}
\usepackage{amsthm}
\usepackage{enumerate}
\usepackage{array}
\usepackage{nicematrix}
\usepackage{threeparttablex}
\usepackage{floatrow}
\usepackage{xltabular}
\usepackage{pgfplots}
\usepackage{pgfplotstable}
\usepackage[parfill]{parskip}
\usepackage{amscd}
\usepackage{multicol}
\usepackage{tcolorbox}
\usepackage{media9}
\usepackage{pifont}
\usepackage{pdflscape}
\usepackage{afterpage}
\usepackage{capt-of}
\usepackage{rotating}
\usepackage{booktabs}
\usepackage[table]{xcolor}
\usepackage{longtable}
\usepackage{tabularx}
\usepackage{colortbl}
\usepackage{tablestyles}
\usepackage[acronym]{glossaries}
\usepackage{alphalph}
\usepackage{mathtools}
\usepackage{blindtext}
\usepackage{subfiles}
\usepackage{fullpage}
\usepackage{makeidx}
\usepackage{fancyhdr}
\usepackage[calcwidth]{titlesec}
\usepackage{setspace}%
\providecommand{\U}[1]{\protect\rule{.1in}{.1in}}
\usepackage[symbol]{footmisc}
\usepackage{bookmark}
\usepackage{pdfpages}	

\usepackage[compat=1.1.0]{tikz-feynman}
\usepackage{tikz}

\usetikzlibrary{spy,calc}
\usetikzlibrary{matrix,backgrounds}
\usetikzlibrary{shapes.geometric, arrows}
\usetikzlibrary{decorations.pathmorphing}
\tikzstyle{startstop} = [rectangle, rounded corners, minimum width=3cm, minimum height=1cm,text centered, draw=black, fill=red!30]
\tikzstyle{io} = [trapezium, trapezium left angle=70, trapezium right angle=110, minimum width=3cm, minimum height=1cm, text centered, draw=black, fill=blue!30]
\tikzstyle{process} = [rectangle, minimum width=3cm, minimum height=1cm, text centered, text width=3cm, draw=black, fill=orange!30]
\tikzstyle{decision} = [diamond, minimum width=3cm, minimum height=1cm, text centered, draw=black, fill=green!30]
\tikzstyle{arrow} = [thick,->,>=stealth]

\tikzstyle{decision} = [diamond, draw, fill=blue!20, 
text width=4.5em, text badly centered, node distance=3cm, inner sep=0pt]
\tikzstyle{block} = [rectangle, draw, fill=blue!20, 
text width=5em, text centered, rounded corners, minimum height=4em]
\tikzstyle{line} = [draw, -latex']
\tikzstyle{cloud} = [draw, ellipse,fill=red!20, node distance=3cm,
minimum height=2em]		

\usetikzlibrary{shadows, patterns}
\tikzset{button/.style={
preaction={fill=blue,path fading=circle with fuzzy edge 20 percent,
opacity=.7,transform canvas={xshift=1mm,yshift=-1mm}},
preaction={pattern=#1,
path fading=circle with fuzzy edge 20 percent},
preaction={top color=white,
bottom color=red!50,
shading angle=180,
path fading=circle with fuzzy edge 20 percent,
opacity=0.4},
preaction={path fading=fuzzy ring 15 percent,
top color=black!5,
bottom color=black!80,
shading angle=180},
inner sep=2ex
},
button/.default=horizontal lines light blue,
circle}	

\newif\ifblackandwhitecycle
\gdef\patternnumber{0}

\pgfkeys{/tikz/.cd,
zoombox paths/.style={
draw=orange,
very thick
},
black and white/.is choice,
black and white/.default=static,
black and white/static/.style={ 
draw=white,   
zoombox paths/.append style={
draw=white,
postaction={
draw=black,
loosely dashed
}
}
},
black and white/static/.code={
\gdef\patternnumber{1}
},
black and white/cycle/.code={
\blackandwhitecycletrue
\gdef\patternnumber{1}
},
black and white pattern/.is choice,
black and white pattern/0/.style={},
black and white pattern/1/.style={    
draw=white,
postaction={
draw=black,
dash pattern=on 2pt off 2pt
}
},
black and white pattern/2/.style={    
draw=white,
postaction={
draw=black,
dash pattern=on 4pt off 4pt
}
},
black and white pattern/3/.style={    
draw=white,
postaction={
draw=black,
dash pattern=on 4pt off 4pt on 1pt off 4pt
}
},
black and white pattern/4/.style={    
draw=white,
postaction={
draw=black,
dash pattern=on 4pt off 2pt on 2 pt off 2pt on 2 pt off 2pt
}
},
zoomboxarray inner gap/.initial=5pt,
zoomboxarray columns/.initial=2,
zoomboxarray rows/.initial=2,
subfigurename/.initial={},
figurename/.initial={zoombox},
zoomboxarray/.style={
execute at begin picture={
\begin{scope}[
spy using outlines={%
zoombox paths,
width=\imagewidth / \pgfkeysvalueof{/tikz/zoomboxarray columns} - (\pgfkeysvalueof{/tikz/zoomboxarray columns} - 1) / \pgfkeysvalueof{/tikz/zoomboxarray columns} * \pgfkeysvalueof{/tikz/zoomboxarray inner gap} -\pgflinewidth,
height=\imageheight / \pgfkeysvalueof{/tikz/zoomboxarray rows} - (\pgfkeysvalueof{/tikz/zoomboxarray rows} - 1) / \pgfkeysvalueof{/tikz/zoomboxarray rows} * \pgfkeysvalueof{/tikz/zoomboxarray inner gap}-\pgflinewidth,
magnification=3,
every spy on node/.style={
zoombox paths
},
every spy in node/.style={
zoombox paths
}
}
]
},
execute at end picture={
\end{scope}
\node at (image.north) [anchor=north,inner sep=0pt] {\subcaptionbox{\label{\pgfkeysvalueof{/tikz/figurename}-image}}{\phantomimage}};
\node at (zoomboxes container.north) [anchor=north,inner sep=0pt] {\subcaptionbox{\label{\pgfkeysvalueof{/tikz/figurename}-zoom}}{\phantomimage}};
\gdef\patternnumber{0}
},
spymargin/.initial=0.5em,
zoomboxes xshift/.initial=1,
zoomboxes right/.code=\pgfkeys{/tikz/zoomboxes xshift=1},
zoomboxes left/.code=\pgfkeys{/tikz/zoomboxes xshift=-1},
zoomboxes yshift/.initial=0,
zoomboxes above/.code={
\pgfkeys{/tikz/zoomboxes yshift=1},
\pgfkeys{/tikz/zoomboxes xshift=0}
},
zoomboxes below/.code={
\pgfkeys{/tikz/zoomboxes yshift=-1},
\pgfkeys{/tikz/zoomboxes xshift=0}
},
caption margin/.initial=4ex,
},
adjust caption spacing/.code={},
image container/.style={
inner sep=0pt,
at=(image.north),
anchor=north,
adjust caption spacing
},
zoomboxes container/.style={
inner sep=0pt,
at=(image.north),
anchor=north,
name=zoomboxes container,
xshift=\pgfkeysvalueof{/tikz/zoomboxes xshift}*(\imagewidth+\pgfkeysvalueof{/tikz/spymargin}),
yshift=\pgfkeysvalueof{/tikz/zoomboxes yshift}*(\imageheight+\pgfkeysvalueof{/tikz/spymargin}+\pgfkeysvalueof{/tikz/caption margin}),
adjust caption spacing
},
calculate dimensions/.code={
\pgfpointdiff{\pgfpointanchor{image}{south west} }{\pgfpointanchor{image}{north east} }
\pgfgetlastxy{\imagewidth}{\imageheight}
\global\let\imagewidth=\imagewidth
\global\let\imageheight=\imageheight
\gdef\columncount{1}
\gdef\rowcount{1}

},
image node/.style={
inner sep=0pt,
name=image,
anchor=south west,
append after command={
[calculate dimensions]
node [image container,subfigurename=\pgfkeysvalueof{/tikz/figurename}-image] {\phantomimage}
node [zoomboxes container,subfigurename=\pgfkeysvalueof{/tikz/figurename}-zoom] {\phantomimage}
}
},
color code/.style={
zoombox paths/.append style={draw=#1}
},
connect zoomboxes/.style={
spy connection path={\draw[draw=none,zoombox paths] (tikzspyonnode) -- (tikzspyinnode);}
},
help grid code/.code={
\begin{scope}[
x={(image.south east)},
y={(image.north west)},
font=\footnotesize,
help lines,
overlay
]
\foreach \x in {0,1,...,9} { 
\draw(\x/10,0) -- (\x/10,1);
\node [anchor=north] at (\x/10,0) {0.\x};
}
\foreach \y in {0,1,...,9} {
\draw(0,\y/10) -- (1,\y/10);                        \node [anchor=east] at (0,\y/10) {0.\y};
}
\end{scope}    
},
help grid/.style={
append after command={
[help grid code]
}
},
}

\newcommand\phantomimage{%
\phantom{%
\rule{\imagewidth}{\imageheight}%
}%
}
\newcommand\zoombox[2][]{
\begin{scope}[zoombox paths]
\pgfmathsetmacro\xpos{
(\columncount-1)*(\imagewidth / \pgfkeysvalueof{/tikz/zoomboxarray columns} + \pgfkeysvalueof{/tikz/zoomboxarray inner gap} / \pgfkeysvalueof{/tikz/zoomboxarray columns} ) + \pgflinewidth
}
\pgfmathsetmacro\ypos{
(\rowcount-1)*( \imageheight / \pgfkeysvalueof{/tikz/zoomboxarray rows} + \pgfkeysvalueof{/tikz/zoomboxarray inner gap} / \pgfkeysvalueof{/tikz/zoomboxarray rows} ) + 0.5*\pgflinewidth
}
\edef\dospy{\noexpand\spy [
#1,
zoombox paths/.append style={
black and white pattern=\patternnumber
},
every spy on node/.append style={#1},
x=\imagewidth,
y=\imageheight
] on (#2) in node [anchor=north west] at ($(zoomboxes container.north west)+(\xpos pt,-\ypos pt)$);}
\dospy
\pgfmathtruncatemacro\pgfmathresult{ifthenelse(\columncount==\pgfkeysvalueof{/tikz/zoomboxarray columns},\rowcount+1,\rowcount)}
\global\let\rowcount=\pgfmathresult
\pgfmathtruncatemacro\pgfmathresult{ifthenelse(\columncount==\pgfkeysvalueof{/tikz/zoomboxarray columns},1,\columncount+1)}
\global\let\columncount=\pgfmathresult
\ifblackandwhitecycle
\pgfmathtruncatemacro{\newpatternnumber}{\patternnumber+1}
\global\edef\patternnumber{\newpatternnumber}
\fi
\end{scope}
}				

\usepackage{fancyvrb,newverbs}
\definecolor{cverbbg}{gray}{0.93}

\newverbcommand{\cverb}
{\setbox\verbbox\hbox\bgroup}
{\egroup\colorbox{cverbbg}{\box\verbbox}}

\definecolor{anti-flashwhite}{rgb}{0.95, 0.95, 0.96}
\definecolor{codegreen}{rgb}{0,0.6,0}
\definecolor{codepurple}{rgb}{0.58,0,0.82}
\lstdefinelanguage{NeMO}{
keywords={},
ndkeywords={solver},
keywordstyle=\color{blue},
ndkeywordstyle=\color{codepurple},
commentstyle=\color{codegreen},
stringstyle=\color{cyan},
sensitive=true
}
\hypersetup{colorlinks=true,
linktocpage=true,
colorlinks=true,
linkcolor=awesome-blue-dark,
filecolor=awesome-red,
citecolor=awesome-pink-dark,      
urlcolor=awesome-orange-dark,
pdftitle={Materials~Informatics:~An~Algorithmic~Design~Rule},
bookmarks=true,
pdfauthor={Bhupesh~Bishnoi},
bookmarks=true,	
bookmarksopen=true,
pdfpagemode=UseOutlines,
pdfstartview={FitH},	
unicode   =true,
pdftoolbar=true,
pdfmenubar=true,
}
\makeatletter
\def\@fnsymbol#1{\ensuremath{\ifcase#1\or \pmb\ddagger \else\@ctrerr\fi}}
\makeatother
\title {\Huge {Materials Informatics: An Algorithmic Design Rule}}
\author{\bf{Bhupesh~Bishnoi\thanks{\href{mailto:bishnoi@computer.org}{\textcolor{PastelRed}{\texttt{\bf{bishnoi[At]computer[Dot]org}}}}~~~\href{mailto:bishnoi@ieee.org}{\textcolor{PastelRed}{\texttt{\bf{bishnoi[At]ieee[Dot]org}}}} }}}
\date{}
\begin{document}
\maketitle
\begin{abstract} 
Materials informatics, data-enabled investigation, is a ``fourth paradigm" in materials science research after the conventional empirical approach, theoretical science, and computational research. Materials informatics has two essential ingredients: fingerprinting materials proprieties and the theory of statistical inference and learning. We have researched the organic semiconductor's enigmas through the materials informatics approach. By applying diverse neural network topologies, logical axiom, and inferencing information science, we have developed data-driven procedures for novel organic semiconductor discovery for the semiconductor industry and knowledge extraction for the materials science community. We have reviewed and corresponded with various algorithms for the neural network design topology for the materials informatics dataset.
\end{abstract}	

\section*{Introduction}
\begin{figure}[H]
\centering
\includegraphics[scale=0.9]{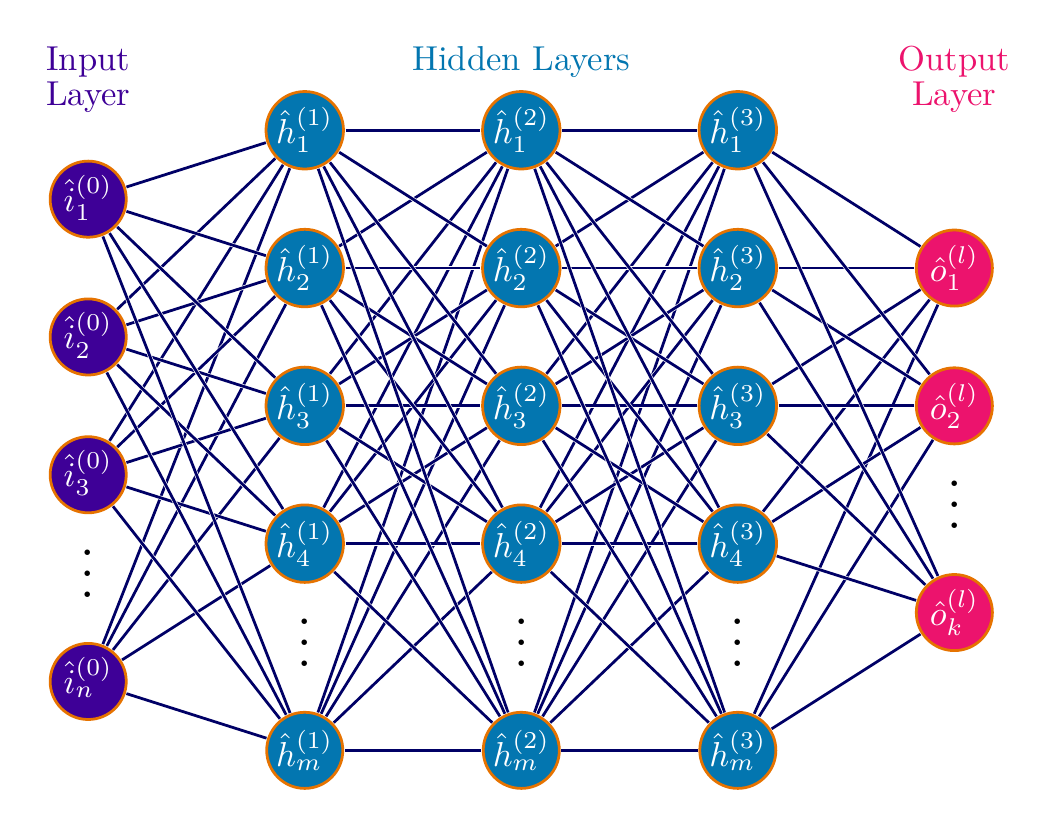}
\caption{\textcolor{VividPurple}{A Generalized Neural Network Topology}}
\label{fig-A4}
\end{figure}

We have researched the organic semiconductor's enigmas through the material informatics approach. By applying diverse neural network topologies, logical axiom, and inferencing information science, we have developed data-driven procedures for novel organic semiconductor discovery for the semiconductor industry and knowledge extraction for the material science community. We have reviewed and corresponded with various algorithms for the neural network design topology for the material informatics dataset, as shown in \cref{fig-A4}, a generalized neural network topology. We have used four chemical compound space databases for model training and validation in this research notebook. The first one is the general quantum chemistry structures and properties of 134-kilo molecules (\texttt{QM9}) of computed geometric, energetic, electronic, and thermodynamic properties for 134-kilo stable small organic molecules made up of C, H, O, N, F for the novel design of new drugs and materials. \cite{ramakrishnan_dral_rupp_anatolevonlilienfeld_2014} The second dataset is for the compounds of molecular organic light-emitting diodes (OLED) materials for high-throughput virtual screening and efficient design. \cite{gomez-bombarelli_design_2016,Aspuru-Guzik2017May} The third dataset is related to sustainable energy storage materials for the quantum chemistry compounds of Redox flow battery materials for accelerated design and discovery. \cite{doan2020quantum} The final fourth dataset is a statistical study of 51,000 organic photovoltaic solar cell molecules designed with the non-fullerene acceptor. \cite{Lopez2017Dec}

We have used various encoding and descriptors algorithms in this work. In the one-hot encoding scheme, we convert simplified molecular-input line-entry system (SMILES), \cite{weininger1988smiles,lin2019bigsmiles} and Self-referencing embedded strings (SELFIES), \cite{Krenn2020Oct} strings to 2-D pixel images to use convolutional neural network, (CNN)\cite{Fukushima1980,LeCun1989Dec} recurrent neural network (RNN),\cite{Elman1990Apr} and variational autoencoders (VAE), \cite{kingma2013auto, kusner2017grammar, jin2018junction} networks taking advantage of image-based learning. In the organic semiconductor molecular design through variational autoencoders (VAE) combining convolutional neural networks (CNN)  as encoder and recurrent neural network (RNN)  as decoder section. We have also used the RDKit 2-D and 3-D descriptors, \cite{Morgan1965May,greg_landrum_2023_7828379} cheminformatics molecular similarity 166-bit MACCS (Molecular ACCess System) keys,\cite{Durant2002Nov}  Morgan Extended-connectivity fingerprints (ECFP6),\cite{Rogers2010May} extended reduced graph approach pharmacophore-2D type node descriptions, \cite{Stiefl2006Jan} and breaking of retrosynthetically interesting chemical substructures (BRICS) algorithm, \cite{Degen2008Oct,Liu2017Apr} to describe the information to the network. We have used sklearn, min-max, and standard-scaler preprocessor for the data preparation to train the various network topologies. We have used several data-splitting techniques to train, validate, and test the models. We have extensively used the no-split, no-select, repeated 5-fold cross-validation, leave one group out, and leave out percentage techniques. For extrication and feature engineering tasks, we have used the diverse strategy of the learning curve, ensemble model feature selector, sklearn feature selector, and standard scaler algorithms.

\begin{figure}[H]
\centering
\includegraphics[scale=0.9]{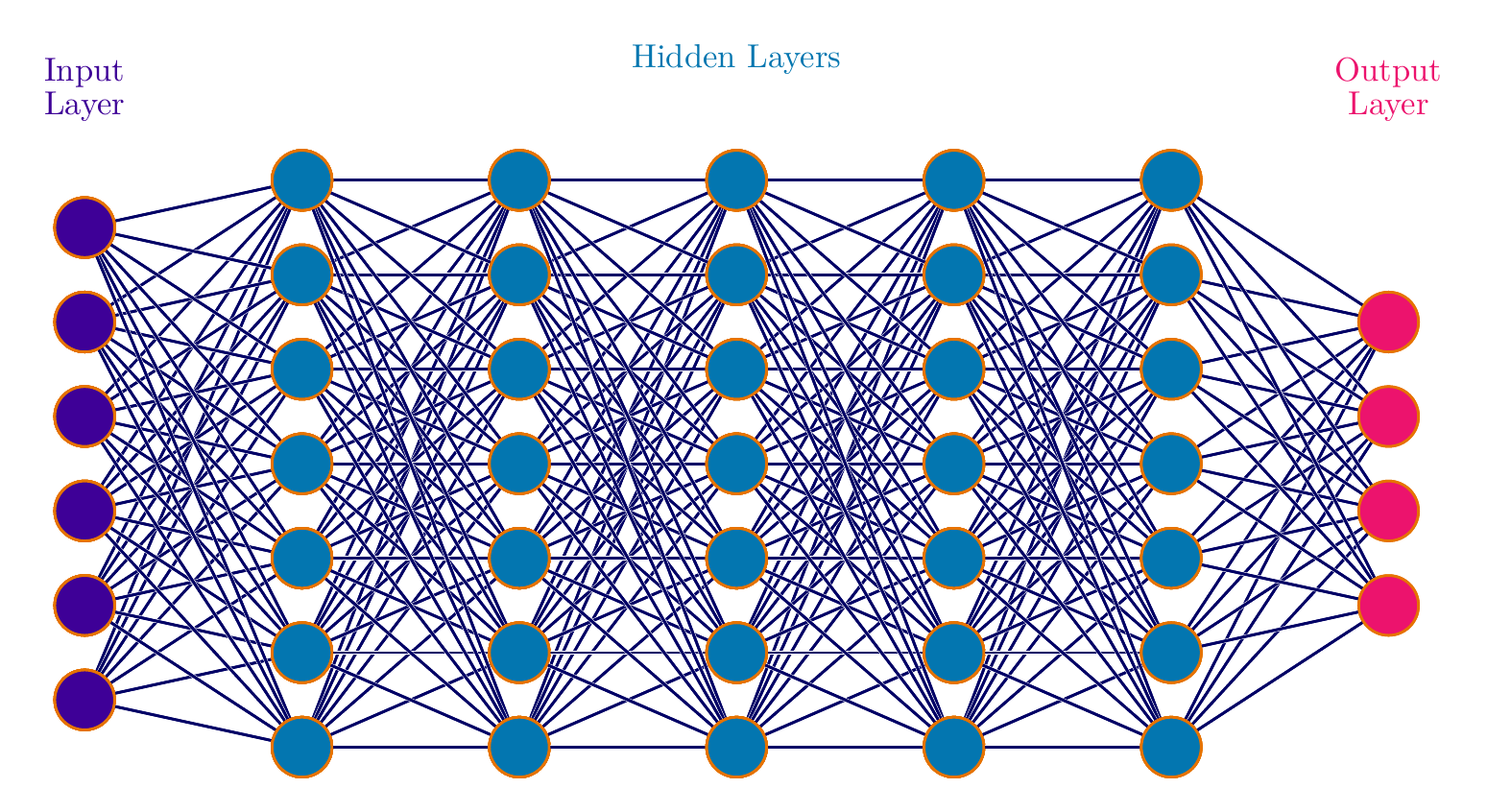}
\caption{\textcolor{VividPurple}{A Deep Learning  Design}}
\label{fig-A6}
\end{figure}

We have used a variety of regression analysis techniques. We have trained models with linear regressor,\cite{Fabian_Scikit_2011,varoquaux2015scikit} Kernel ridge regressor,\cite{pml2Book} Keras regressor, \cite{chollet2015keras} Gaussian process regressor,\cite{rasmussen2003gaussian, Rasmussen06gaussianprocesses} Random forest regressor,\cite{breiman2001random} Multi-layer perceptron regressor,\cite{Rosenblatt1961Mar} Bagging regressor, Extreme gradient boosting regressor, \cite{friedman2001greedy} Extreme gradient boosting multi-layer perceptron regressor,\cite{sawada_boosting_2018} Extreme gradient boosting Keras regressor, Extreme gradient boosting kernel ridge regressor for the material informatics dataset. \cite{DBLP:journals/corr/ChenG16} For the dimensionality reduction, projection, and classification task in the material informatics dataset, we have employed Principal component analysis (PCA),\cite{Jolliffe2016Apr} t-stochastic neighborhood embedding (t-SNE),\cite{JMLR:v9:vandermaaten08a, JMLR:v15:vandermaaten14a, linderman2017clustering}, and Uniform manifold approximation and projection for dimension reduction (UMAP) algorithms. \cite{mcinnes2018umap,mcinnes2020umap} Further, We have trained models on the convolutional neural network (CNN), recurrent neural network (RNN), radial basis function network, variational autoencoders (VAE), graph neural network (GNN), message-passing neural network (MPNN),\cite{gilmer2017neural,chemprop_code,mcgill2021predicting} directed message-passing neural network, materials graph network-based variational autoencoder, attention network (AN),\cite{vaswani2017attention,Veli2017graph} geometric learning network,\cite{monti2016geometric} active learning network,\cite{Bassman2018Dec, farache2021active} and Bayesian optimization network,\cite{snoek2012practical,shahriari2016bayes, ikebata_bayesian_2017, Bayesian, couperthwaite2020materials, mcdannald2021onthefly,H_se_2021} Evolutionary algorithm based neural network,\cite{DEAP_JMLR2012,dey_design_2017, Kern2021Dec} genetic algorithm network,\cite{Huan2016Feb, Mannodi-Kanakkithodi2016Mar, Kim2018Aug, Kuenneth2021Jul} multi-fidelity batch reification,\cite{Thomison2017May, chen2021learning, greenman2022multi} model correlation estimation, and fusion optimization,\cite{ghoreishi2018multi,khatamsaz2021efficiently} and optimal design of experiments network.\cite{iwasaki_predicting_2020} We have investigated a deep learning design, as shown in \cref{fig-A6}, to predict quantitatively accurate and desirable material properties by constructing a relationship between the molecular structure and its property through a material graph-based neural network. \cite{Gori2005Jul,scarselli2005graph,scarselli2008computational} 

For the error analysis and uncertainty quantification of the statistical prediction of the network, we have used nested cross-validation and optimized the models. We have also used a random forest regressor with repeated 5-fold cross-validation and a Gaussian process regressor with repeated 5-fold cross-validation for uncertainty quantification in the predicted norm.

\section*{Variational Autoencoder Network}

\begin{figure}[H]
\centering
\includegraphics[scale=0.9]{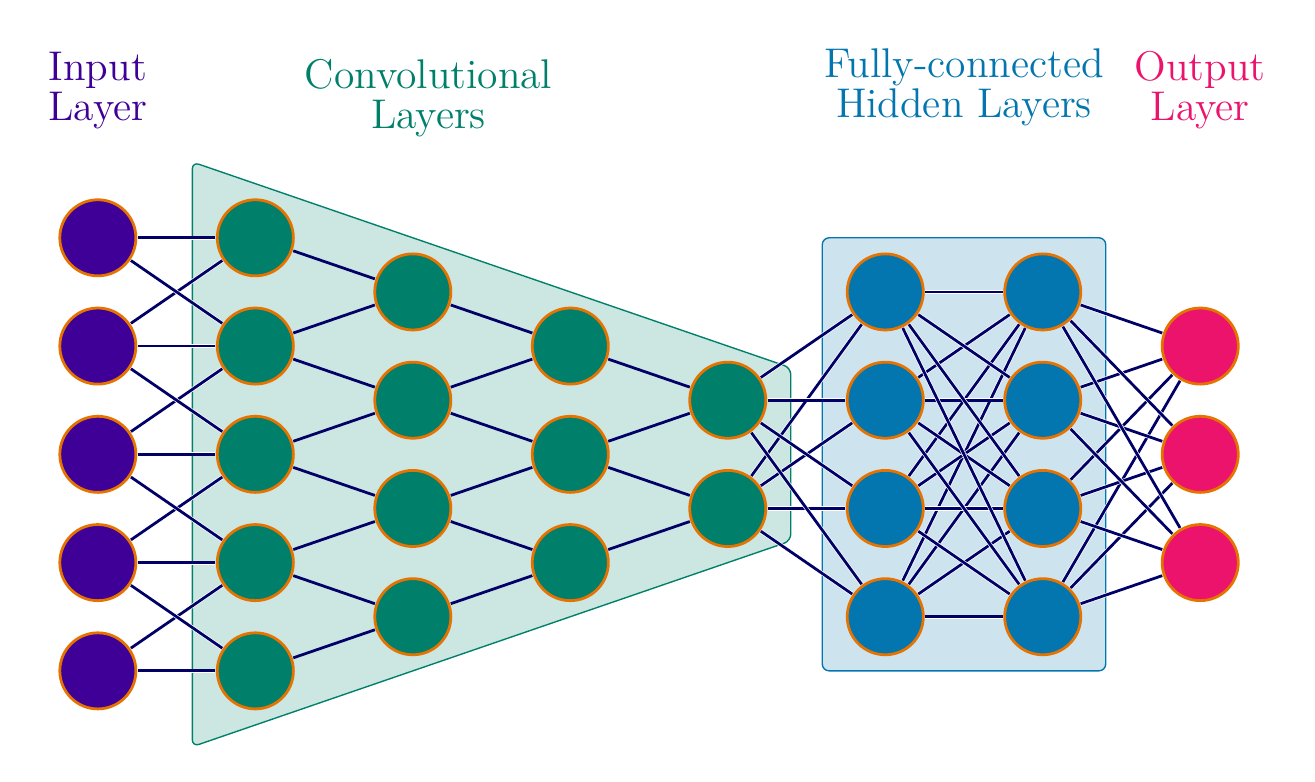}
\caption{\textcolor{VividPurple}{A Convolutional Neural Network}}
\label{fig-A7}
\end{figure}

\begin{figure}[H]
\centering
\includegraphics[scale=0.9]{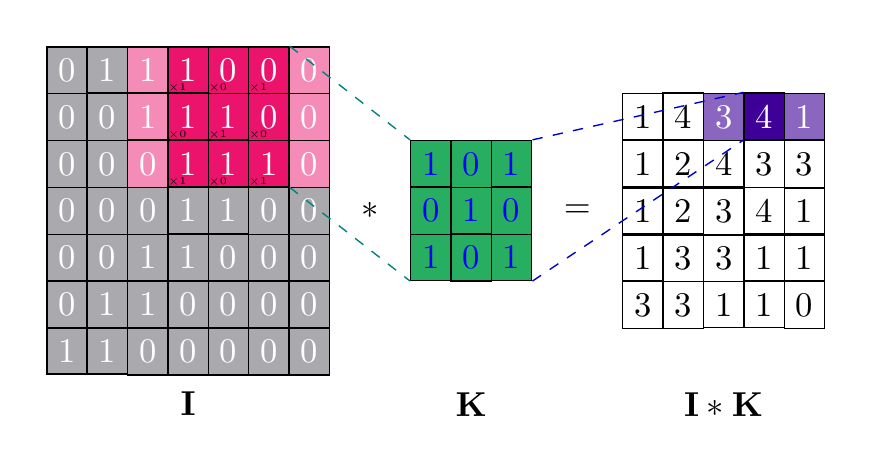}
\caption{\textcolor{VividPurple}{A Convolutional Operation}}
\label{fig-A10}
\end{figure}

\begin{figure}[H]
\centering
\includegraphics[scale=0.9]{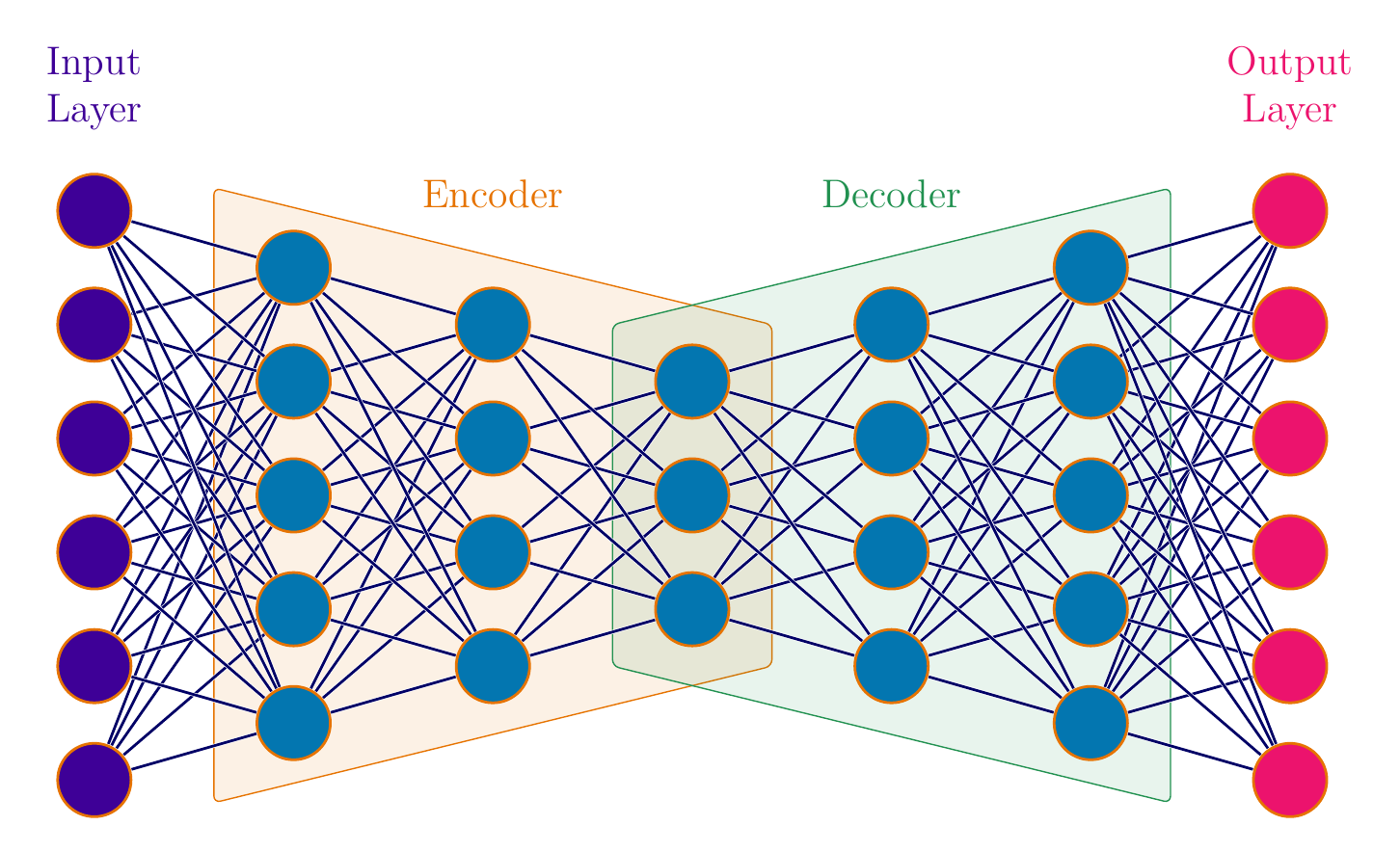}
\caption{\textcolor{VividPurple}{An Autoencoder Network Topology}}
\label{fig-A8}
\end{figure}

\begin{figure}[H]
\centering
\includegraphics[scale=0.9]{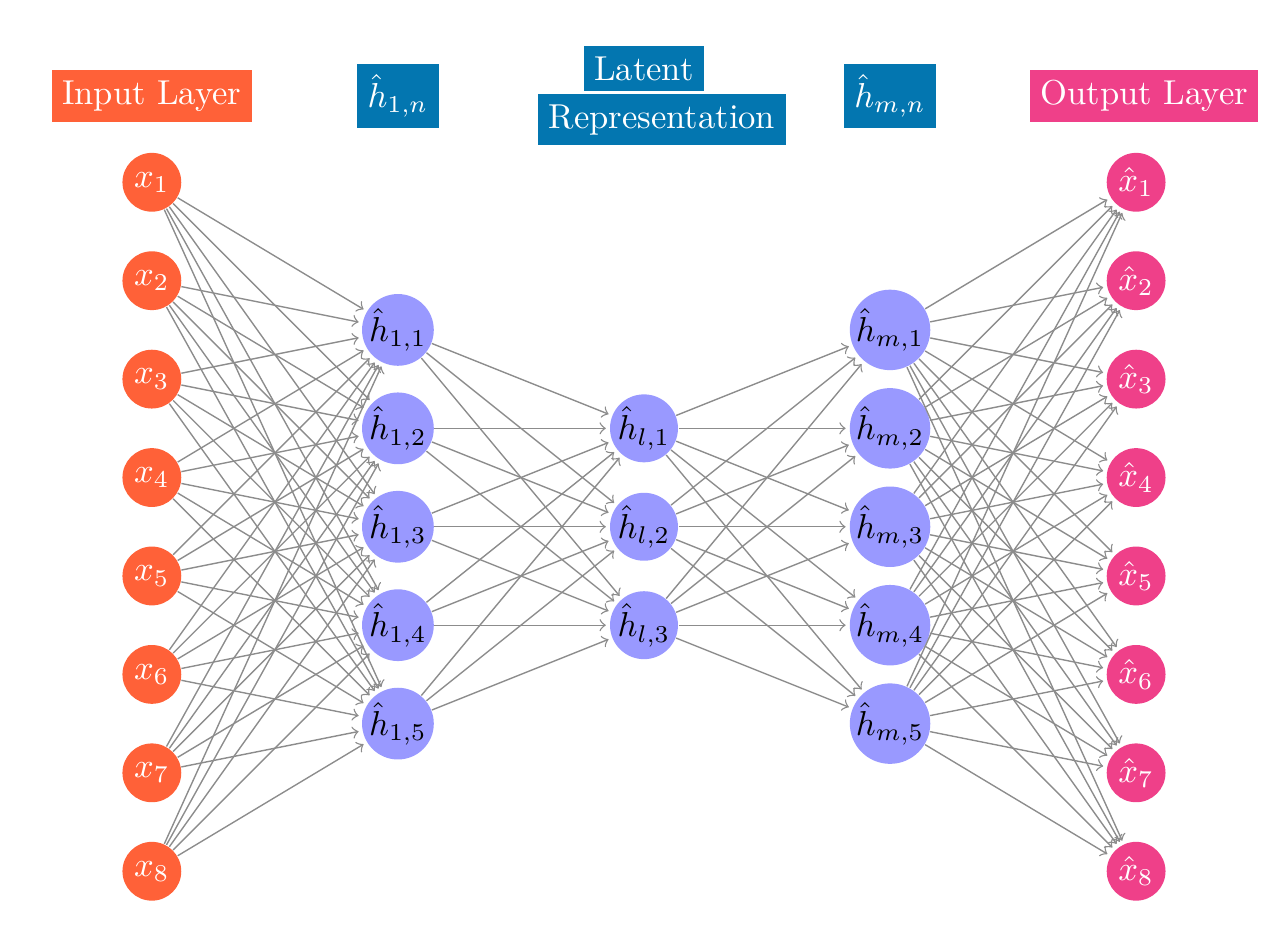}
\caption{\textcolor{VividPurple}{A Generalized Variational Autoencoder Network}}
\label{fig-1}
\end{figure}

First, we will briefly discuss the intuitive high-level concept of the deep learning approach, e.g., convolutional neural networks (CNN) and its operation, as shown in \cref{fig-A7}, and \cref{fig-A10} and variational autoencoders (VAE), as shown in \cref{fig-A8}, and \cref{fig-1} to demystify it. 

In a traditional feed-forward neural network, neurons arrange in layers. Each layer passes its output to the following layer, which receives it as an input and performs Activision function operations. 

\begin{figure}[H]
\centering
\includegraphics[scale=0.55]{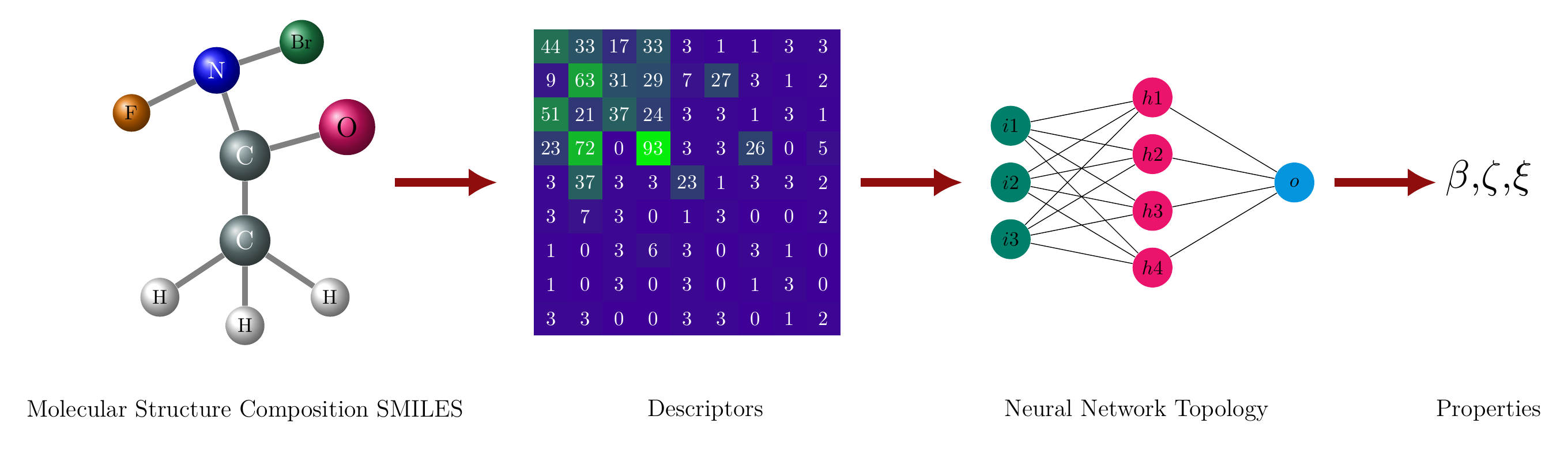}
\caption{\textcolor{VividPurple}{A Molecular Structure Composition SMILES}}
\label{fig-A17}
\end{figure}

Afterward, we will explore python libraries packages, e.g.,  Karas, to use in material informatics to design novel materials. In the Deep neural network, the number of layers grows in architecture to process a large amount of data with advances statistical algorithms. In traditional machine learning approaches, we have to do feature engineering from inputs that go into the network in the deep learning model representations; the data itself do this. We are building a convolutional neural network for predicting the electronic properties of organic molecules. \cite{10.1145/3448250} A convolutional neural network has at least one layer that performs convolution operation, an element by element matrix multiplication, and transforming the input data in a specific way by filter or kernel to give a feature map for the network. The molecules measuring electronic properties will feature as license molecules that involve some crystal structure information, fractions of elements, or SMILES string to represent the material to queries from the database to predict the new organic molecules as shown in the \cref{fig-A17}. SMILES is an ASCII string standard for describing the structure of chemical species to use in molecule editors.\cite{doi:10.1021/ci00057a005} 

CNN is heavily exploited in the image and video recognition task by using the pixel data of frame for featuring the nearby pixel in the image to get a sense of the entire image. We will use CNN with SMILES string to predict molecular properties using the same concept where input is a 2-D images string of crystal graph.\cite{PhysRevLett.120.145301} An autoencoder in a neural network architecture aims for the representation learning task to reconstruct the input data as a self-supervised/unsupervised learning technique as shown in the \cref{fig-A8}.

\begin{figure}[H]
\centering
\includegraphics[scale=0.9]{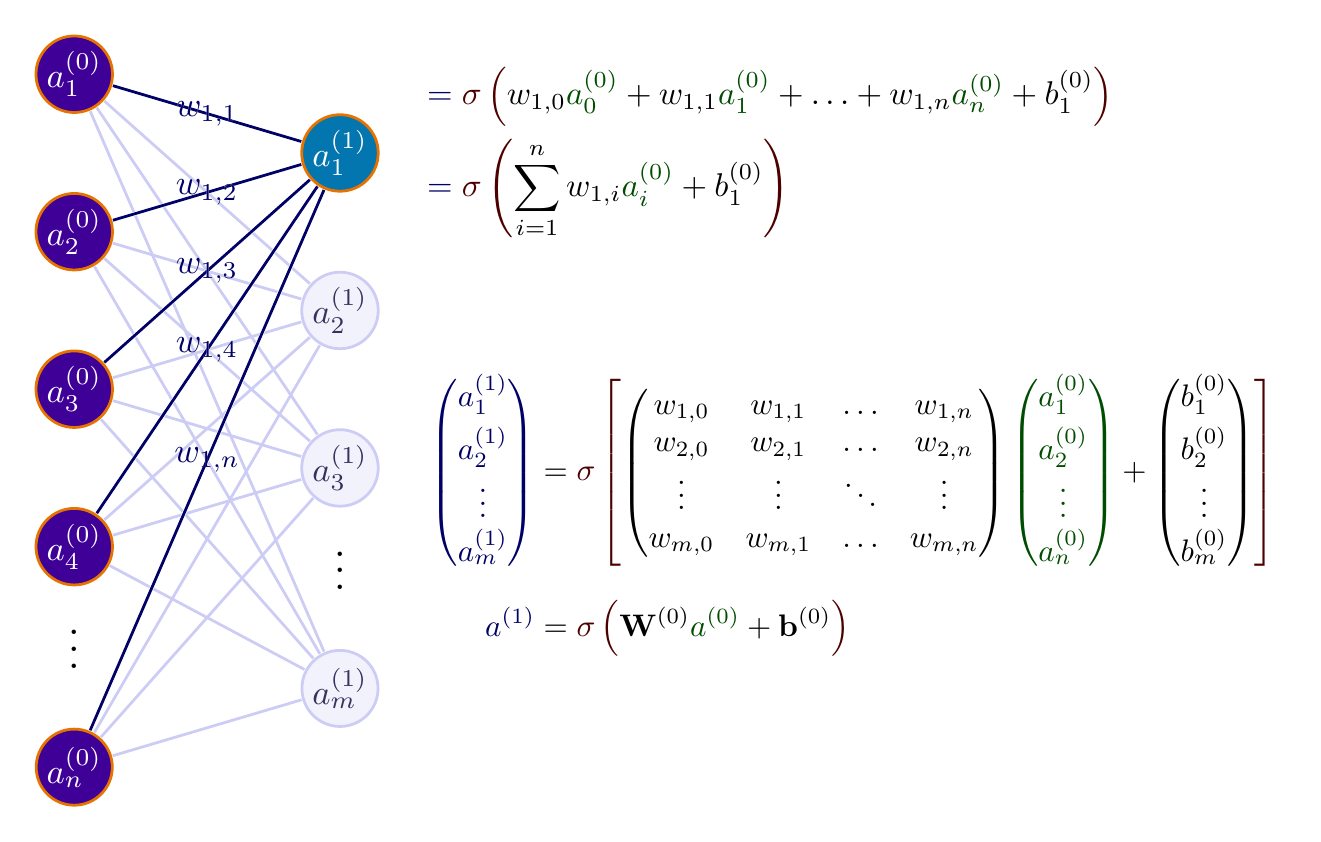}
\caption{\textcolor{VividPurple}{Backpropagation Algorithm, A Weight and Bias Update in Neural Network}}
\label{fig-A9}
\end{figure}

The purpose is to first squeeze the higher dimensional information through a bottleneck through the encoder part involving multiple layers with decreasing nodes as we propagate in the deeper hidden layer. That will encode input data as an encoding vector of discrete probability distribution data values. Each latent vector dimension represents some learned attribute about the data-this probabilistic latent information in a useful low dimensional representation. The bottleneck imposed by network architecture enforces a compressed knowledge representation of the original input at the output to represents meaningful attributes of the original input data. These attributes are correlations between the input feature vectors from data that the network discovers during training. It will learn and leveraged any correlations that exist between input features maps. An optimal layer-designed CNN encoder resists the infinitesimal perturbations in the input to the output feature extraction function task. Using a linear function instead of nonlinear activation functions at each constraining hidden layer, this scheme is similar to the Principal component analysis (PCA) dimensionality reduction technique to get two principal Eigenvalues. \cite{hotelling1933analysis} Next, we will reconstruct the input as lossless compression in the decoder part from low dimensional generalizable latent space information representation. Hence PCA linear search as a lower-dimensional hyperplane in the high-dimensional dataset whereas autoencoders perform a nonlinear search as a lower-dimensional manifold in the high-dimensional dataset. \cite{10.1145/3072959.3073601} An autoencoder should be sensitive enough to the inputs to build a reasonably accurate reconstruction. On the other hand, it must be insensitive enough to the inputs so that model does not solely memorize to overfit the training data. These conditions will enforce only the variations in the input data used to reconstruct the output. The variational autoencoders will provide the continuous latent probability distributions of encoded input data instead of encoding input data in discrete variables values. We can sample these continuous, normally distributed latent probability distributions to generate new probabilities. We will sample the vectors defining the mean, variance, and covariance matrix to build the multivariate Gaussian model from the latent state vector distribution. We use the reparameterization technique for the sampling. While random sampling is not helpful in the backpropagation stage, as shown in the \cref{fig-A9}, we need gradient calculation of each parameter in the network for the final output loss function.

Hence, we reconstruct molecules similar to the input molecules dataset to discover new exciting material. We will build a variation autoencoder that uses the small SMILES strings and encodes them into a latent representation which we sample to generate an entirely new SMILES string and hence a reasonable new realistic molecule. \cite{https://doi.org/10.1002/chem.201604556,PhysRevX.7.021024} Molecules and polymers are discrete chemical objects, and chemical space is approximately estimated to be about $10^{60}$ discrete molecules from the periodic table combinatorial elements. Data-driven approaches are promising in such a vast chemical space, allowing us to interpolate, optimize, explore, and potentially compress it to low dimensional space. It is a potent tool to solve the problem of inverse material design and discovery, which has been one of the goals of the modern chemical industry. Rapid screening of potential drug-like molecules in the pandemic era high requested. Furthermore, these molecules' properties are scanned and investigated at projected low dimensional chemical space by jointly training with different target properties in the variational autoencoder network from this lower-dimensional latent space representation. A probabilistic model constructs where most molecules are at the center of the distribution, and slight variance from mean distribution generated various novel molecules with similar target properties. \cite{doi:10.1021/acscentsci.7b00572} The water solubility of organic molecules is crucial for the degradation and lifetime reliability of organic semiconductor devices application. We will predict the aqueous solubility of organic molecular materials from their chemical composition through deep learning models. We will specify chemical composition by SMILES strings as inputs and load and preprocess data in the split of training and validation sets. The training set is used to train the network and the validation set to check the validity of learning achieved by the training set of data. After that, on a well-trained network, we pass the test dataset to predict the properties from the network. The test dataset is the dataset which network has not prior seen in the training and validation phase. The encoder portion of the variational autoencoders (VAE) network builds through a convolutional neural network (CNN). \cite{schutt2017schnet} The CNN learns low-dimensional numerical representations of a high-dimensional SMILES string dataset and builds confidence.  In variational autoencoder (VAE), as shown in the \cref{fig-1}, encoded latent variables learn as probability distributions rather than discrete values of an autoencoder. VAE is used to generate new SMILES strings. Convolution layers will use translational and rotational invariance in modeling molecules. Hyperparameters and kernel size uses to govern the architecture of the CNN and optimize it. Hyperparameters search for deep neural networks performed by DeepHyper, which automatically searches the deep neural network search space. \cite{8638041} The encoded information generates new molecules in the decoder portion of VAE by a Recurrent Neural Network (RNN),\cite{DBLP:journals/corr/HeGDG17} which takes latent space encoded SMILES and maps back to the input by sampling from a Gaussian distribution and generate new SMILES. We will use a one-hot encoding vector corresponding to each atom on the categorical cross-entropy distribution data to convert back to new SMILES characters of the vector of 0's and 1's ASCII code of length 31. The length of the vector is equivalent to all possible SMILES characters in the data set. Furthermore, our dataset defines a maximum of 40 SMILES characters. Different molecule structures have varying SMILES string lengths. We limit the maximum length to 40 with dummy string characters to make it a suitable size. Therefore each molecule represents by a set of 40 vectors, and each has a length of 31. henceforth a dataset size of 40x31 matrix. We will use the \texttt{Pandas},\cite{reback2020pandas}, \texttt{Seaborn}  \cite{Waskom2021} and \texttt{Numpy},\cite{harris2020array} libraries to process input data and visualization, and \texttt{Keras}, \cite{chollet2015keras} and \texttt{Tensorflow}, \cite{tensorflow2015-whitepaper} and \texttt{Python} libraries, \cite{mckinney-proc-scipy-2010} for the neural network modeling. 
We have used the quantum chemistry \texttt{QM9} molecule database repository from Stanford University. The data science-centric discovery of novel organic molecules requires unbiased and rigorous exploration of chemical compound space. However, due to the sizeable combinatorial possibility of atomic species, substantial uncharted territories persist in exploring organic molecules with optimal properties. In this quest, the \texttt{QM9} dataset published containing B3LYP/6-31G(2df,p) level quantum chemistry calculated geometric, electronic, energetic, and thermodynamic properties for 134000 stable small organic molecules composed of atomic species of C, H, O, N, F out of the GDB-17 chemical universe of 166 billion organic molecules. \texttt{QM9} dataset used for benchmarking hybrid quantum chemistry, deep  learning-based data-driven discovery for systematic identification of new molecular. \cite{doi:10.1021/acs.jctc.9b00181,schutt_unifying_2019,brockherde_bypassing_2017,doi:10.1063/1.4928757}. We will visualize an input data heatmap by using the Seaborn library to show each character's position in the SMILES string molecules. A 40x31 sparse matrix represents each molecule. The bright spots in the heatmap indicate the position at which 1's is found in the matrix. Beyond that, the rows all have a bright spot at index 1, which stands for the extra characters padded onto our string to make all SMILES strings the same length. The input layer of CNN is a training image containing a 40x31 matrix that passes the \texttt{Keras} library. The four convolution layers attempt to learn these SMILES images' unique features relevant to predicting the properties by the input matrix's element-by-element multiplications operation with a kernel filter matrix. Next, for the numerical prediction, we flatten the output of the convolution layer and pass it to ``dense'' regular layers, and further to the last layer is for one-hot prediction. Following the Keras terminology, each fully-connected layer is called a ``dense'' layer. The learning curve is a plot of loss function for the validation and train Mean Squared Error (MSE) and Mean Absolute Error (MAE) as a function of epoch to estimate over-fitting or under-fitting of the model. Epoch defines as passing all training examples through the neural network for once. We compare ground truth data with CNN model predictions through a parity plot by compiling the model. We can also visualize the model metadata through the \texttt{keras2ascii} tool. We also define the VAE loss function and check accuracy metrics values after each epoch. A recurrent neural network (RNN) is used to decode SMILES strings from latent space values by each RNN cell learning from the previous RNN cells, therefore, learning from a time-dependent series of data. We used gated recurrent unit (GRU),\cite{Kyunghyun_Learning,chung2014empirical} as cells in RNN, which solve the vanishing gradient problem of a standard long short term memory (LSTM) RNN cell. \cite{10.1162/neco.1997.9.8.1735} It can take into account some temporal informational history of the network. From a Gaussian distribution, we will randomly sample feed into the decoder. The output of the decoder converted back into SMILES characters. Finally, we save VAE models for future uses using \texttt{save()} and  \texttt{load model()} functions from the \texttt{Keras} library. The loss function uses the network to update parameters to reduce the mean square error between the output property and the actual property. The Optimization is achieved through a stochastic gradient descent optimizer. If both the training error and the validation error decrease, then the neural network trained well. However, if the validation error increases while the training error goes down, this is the case of over-fitting, which can be rectified by regularizing some weights parameters tuning or penalizing the activation in the neural network hidden layer. In the L1 regularization, a scaling parameter uses to tuned activations. In the Kullback–Leibler (KL) divergence scheme, a sparsity parameter introduces, denoting average activation over entire neuron samples. In the KL divergence scheme, a sparsity parameter introduces, denoting average activation over entire neuron samples. At the same time, KL divergence measures the difference between two probability distributions, which can be similar by minimizing the KL divergence in the VAE. For RNN, the loss function defines as reconstruction function error, i.e., decoder efficiency, how close the final output compare to the input to match the probability distributions as closely as possible. VAE generates these novel molecules from scratch with an intuitive understanding of the seen dataset.  An optimal layer-designed RNN decoder resists small finite-sized perturbations in the input to the output reconstruction function task.

\subsection*{Generative Framework}
\begin{figure}[H]
\centering
\includegraphics[scale=0.9]{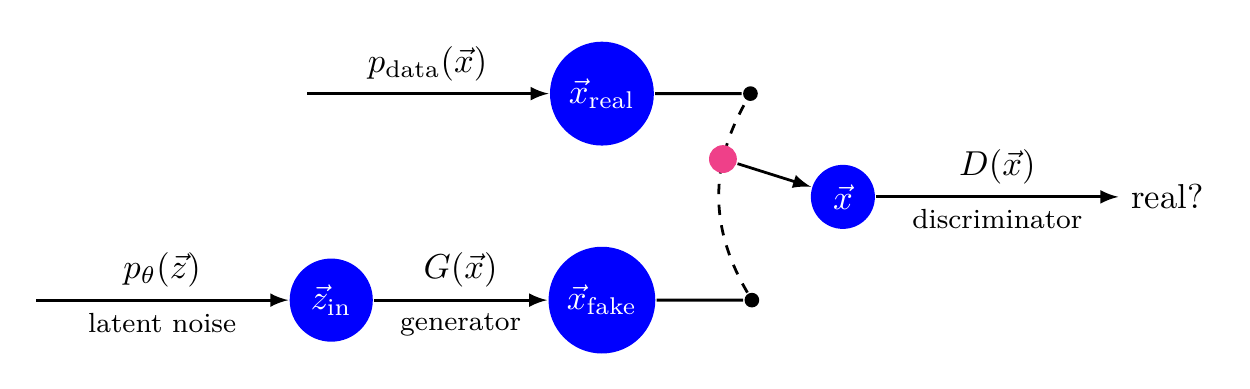}
\caption{\textcolor{VividPurple}{A Generative Algorithm}}
\label{fig-A20}
\end{figure}

A variational autoencoder (VAE) is a generative, as shown in the \cref{fig-A20}, an unsupervised deep learning algorithm that uses unlabeled data to fit a model and generate new data points from sampling trained probability distribution $\hat{{P}}(x)$ over data. In variational autoencoder approaches for one-hot encoded vectored data $ x $ using the definition of a marginal and conditional probability, for random variable $ z $ with the known normal distribution of mean and variance, The $\hat{{P}}(x)$ is,

\begin{equation}\label{eq-VAE-1}
\hat{{P}}(x) = \int\,\hat{{P}}\left(x | z \right) {P}(z)\,dz
\end{equation}

Direct training of $\hat{{P}}\left(x | z \right)$ is computationally tricky. However, creating symmetric distribution $\hat{{P}}\left(z | x \right)$ and training simultaneously is computationally easy to deduce $\hat{{P}}(x)$ from $\hat{{P}}\left(x | z \right)$ with assistance of $\hat{{P}}\left(z | x \right)$ is key in the success of variational autoencoder. Therefore variational autoencoder is a set of two trained conditional probability distributions that operate on the data $ x $ and latent variables $ z $ as latent space $ z $ is encoded compression of $ x $. The first conditional probability is decoder $ p_\theta(x | z) $ for trainable parameters $ \theta $ to be fitted as proceeds from the latent variable $ z $ to $ x $. The second conditional probability is encoder $ q_\phi(z | x) $ and we always know $ p(z) $ as we defined it to a standard normal distribution to communicate through encoder and decoder section. To observe the $ q_\phi(z | x) $ assistance to train the $ x_i $ value with training parameters $ \theta $, we are interested to generate the new $ x_i $, and the variational autoencoder loss function is defined by using the definition of expectation as log likelihood of $ x_i $ as $ \log\left[\hat{{P}}(x_i)\right] $,

\begin{equation}\label{eq-VAE-2}
\log\left[\hat{{P}}(x_i)\right]= \log\left[\int\,p_\theta(x_i | z) {P}(z)\,dz\right] = \log {E}_z\left[p_\theta(x_i | z)\right]
\end{equation}

\begin{figure}[H]
\centering
\includegraphics[scale=0.9]{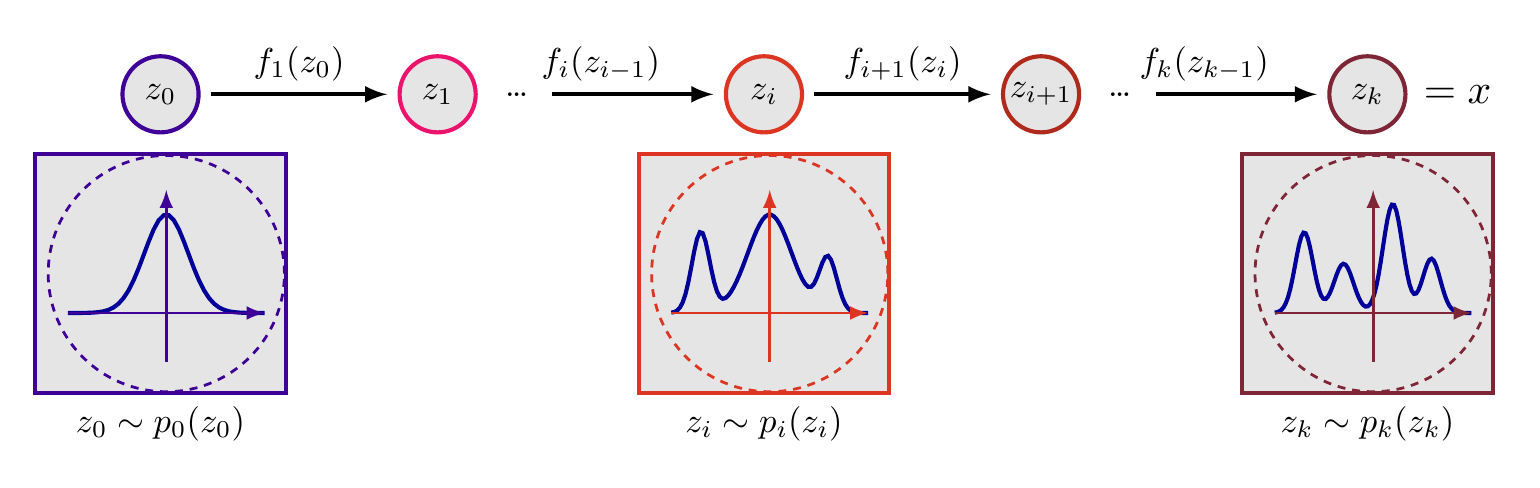}
\caption{\textcolor{VividPurple}{A Normalizing Flow Algorithm}}
\label{fig-A16}
\end{figure}

It is not easy to estimate $ p_\theta(x | z) $ is a neural network as integrating over the latent variable input $ z $ of a neural network is not straightforward. We can integrate over the latent variables, and it is another generative modeling technique known as normalizing flow as shown in the \cref{fig-A16}.  

However here, we approximate the integral by sampling some $ z's $ from  $ p(z) $ as, 

\begin{equation}\label{eq-VAE-3}
\log\textrm{E}_z\left[ p_\theta(x_i | z)\right]\approx \log \left[\frac{1}{N}\sum_j^N  p_\theta(x_i | z_j)\right],\, z_j \sim P(z_j)
\end{equation}

However, grabbing $ z's $ from $ p(z) $ for approximating the integral is inefficient because $ z's $ are likely lead to the observed $ x_i $, and integral is dominated by the $ p_\theta(x_{i} | z_{j}) $ terms. Here we used the $ q(z | x) $ as it can provide efficient guesses for $ z_{j} $. Therefore, $ \log \textrm{E}_z\left[p_\theta(x_i | z)\right] $ is approximated by sampling from $ q(z | x_{i}) $. However sampling from $ q(z | x_{i}) $ is not identical to sampling from $ p(z) $ by adding their ratio to the expression,

\begin{equation}\label{eq-VAE-4}
\log\textrm{E}_z\left[ p_\theta(x_i | z)\right]\approx \log \left[\frac{1}{N}\sum^N_j  p_\theta(x_i | z_j) \frac{P(z_j)}{q_\phi(z_j | x_i)}\right],\, z_j \sim q_\phi(z_j | x_i)
\end{equation}

Where the numerical approximation of expectation term is achieved by the ratio of $ P(z) / q_\phi(z | x) $. The exact expression with respect to $ z \sim q_\phi(z | x_i) $ after inserting the sampling ratio is,

\begin{equation}\label{eq-VAE-5}
\log\textrm{E}_z\left[ p_\theta(x_i | z)\right] = \log\textrm{E}_{z \sim q_\phi(z | x_i)}\left[ p_\theta(x_i | z) \frac{P(z)}{q_\phi(z | x_i)}\right]
\end{equation}

By using the Jensen's inequality for the concave log function and swapping the order of expectation and the log function as, 

\begin{equation}\label{eq-VAE-6}
\log \textrm{E}\left[\ldots\right]\geq \textrm{E}\left[\log \ldots\right]
\end{equation}

The advantage of this transformation is loss function is not an exact estimate of the log-likelihood but becomes a lower bound. By using this transform and using the properties of the log function and separating it into two terms as,

\begin{equation}\label{eq-VAE-7}
\textrm{E}_{z \sim q_\phi(z | x_i)}\left[ \log\left(p_\theta(x_i | z) \frac{P(z)}{q_\phi(z | x_i)}\right)\right] = \textrm{E}_{z \sim q_\phi(z | x_i)}\left[ \log p_\theta(x_i | z)\right] + \textrm{E}_{z \sim q_\phi(z | x_i)}\left[ \log \left(\frac{P(z)}{q_\phi(z | x_i)}\right)\right]
\end{equation}

In the right-hand side of \cref{eq-VAE-7}, w have now to integrate over $q_\phi(z | x ) $ and a standard normal distribution $ P(z) $ and does not contain $ p_\theta(x | z) $ which is computationally expansive to integrate over a neural network input. Another advantage of the VAE is that in the lower dimension latent space, the variable is continuous, and therefore, some type of optimization is possible to add to the loop. Next, we transform the $q_\phi(z | x ) $ to a normal distribution to ensure the integral computes efficiently and use an identity that relates to the Kullback–Leibler divergence to the right-hand side term of \cref{eq-VAE-7} as,  

\begin{equation}\label{eq-VAE-8}
\textrm{E}_{p(x)}\left[ \ln\left(\frac{q(x)}{p(x)}\right)\right] = -\textrm{KL}\left[p(x)|| q(x)\right]
\end{equation}

The Kullback–Leibler divergence is a binary function of two probabilities. By using \cref{eq-VAE-8} in the \cref{eq-VAE-7}, the Log-likelihood approximation evidence lower bound (ELBO) is,

\begin{equation}\label{eq-VAE-9}
\log\left[\hat{\textrm{P}}(x_i)\right] \geq \textrm{E}_{z \sim q_\phi(z | x_i)}\left[ \log p_\theta(x_i | z)\right] -\textrm{KL}\left[q_\phi(z | x_i)|| P(z)\right]
\end{equation}

In the \cref{eq-VAE-9}, the first term on the right-hand side is reconstruction loss, and it assesses, after transforming from $ x \rightarrow z \rightarrow x $, how close we reach to expectation. The second term is Kullback–Leibler divergence measures how close the encoder is to the defined $ P(z) $ standard normal distribution. The integral term is the right-hand side computed analytically, and no sampling is required. The Kullback–Leibler divergence term appeared as correction term accounting that we use encoder $ q_\phi(z | x_i) $ that generates $ z's $ from training data point $ x_i $ and does not use $ P(z) $ directly. We add a minus sign as we want to minimize this loss function in training as,

\begin{equation}\label{eq-VAE-10}
\mathcal{L}(x_i, \phi, \theta) =  -\textrm{E}_{z \sim q_\phi(z | x_i)}\left[ \log p_\theta(x_i | z)\right] +\textrm{KL}\left[q_\phi(z | x_i)|| P(z)\right]
\end{equation}

We approximate the expectation in the reconstruction loss by sampling $ z's $ from the decoder $ q_\phi(z | x) $ using a single sample. We have used the ELBO equation  \cref{eq-VAE-10} for training the dataset. We approximate the expectation in the reconstruction loss by sampling $ z's $ from the decoder $ q_\phi(z | x) $ using a single sample. We have used the ELBO equation  \cref{eq-VAE-10} for training the dataset, and both $ P(z) $ and $ q_\theta(z | x) $  are normal distribution and $ P(z) $ is standard normal distribution with $ (\sigma) $ of unity and  $ (\mu) $ of zero. The Kullback–Leibler divergence between two normal distributions is, 

\begin{equation}\label{eq-VAE-11}
KL(q, p) = \log \frac{\sigma_p}{\sigma_q} + \frac{\sigma_q^2 + (\mu_q - \mu_p)^2}{2 \sigma_p^2} - \frac{1}{2}
\end{equation}

And by using the $ P(z) $ standard normal distribution properties, the \cref{eq-VAE-11} simplify further as,

\begin{equation}\label{eq-VAE-12}
\textrm{KL}\left[(q_\theta(z | x_i))|| P(z)\right] = -\log \sigma_i + \frac{\sigma_i^2}{2} + \frac{\mu_i^2}{2} - \frac{1}{2}
\end{equation}

Where $ \mu_i $, $ \sigma_i $ are the output from  $ q_\phi(z | x_i) $. 

In the \cref{eq-VAE-12} instead of making the latent space perfect normal distribution at this cost, more reconstruction is achieved by modifying the ELBO equation by adding a term $ \beta  $ that adjusts the balance between the reconstruction loss and the KL-divergence as,

\begin{equation}\label{eq-VAE-13}
l = -\textrm{E}_{z \sim q_\phi(z | x_i)}\left[\log p_{\theta}(x_i | z)\right] + \beta\cdot\textrm{KL}\left[(q_\phi(z | x))|| P(z)\right]
\end{equation}

\begin{figure}[H]
\centering
\includegraphics[scale=0.8]{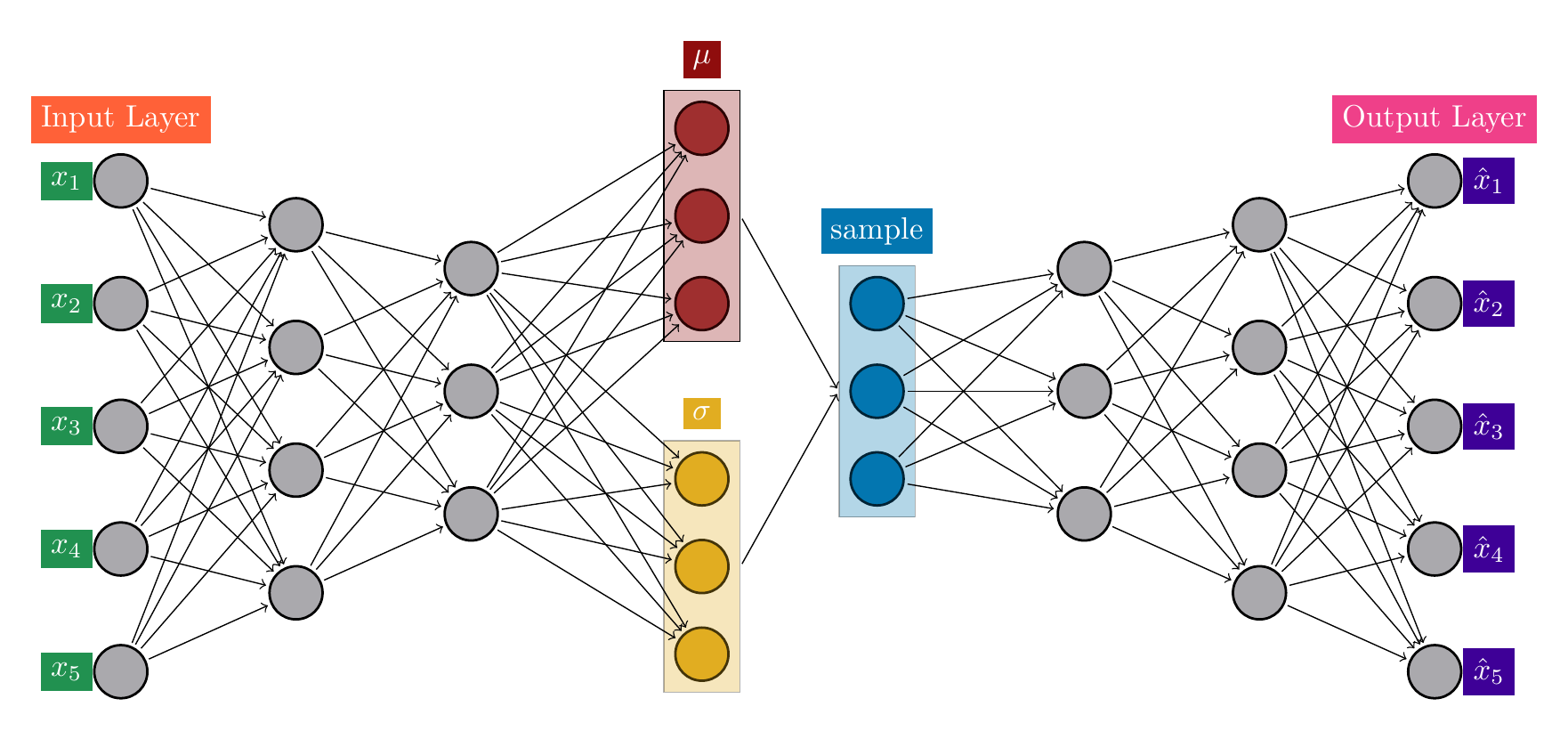}
\caption{\textcolor{VividPurple}{Variational Autoencoder Model for Molecular Design, and Joint Property Prediction}}
\label{fig-3}
\end{figure}

Where $ \beta > 1  $ emphasizes the encoder distribution matching chosen latent standard normal distribution and $ \beta < 1  $ emphasizes reconstruction accuracy. However, the $ \beta  $ can be adjusted toward the opposite direction to strongly match the prior Gaussian distribution to improve the encoder. All latent dimensions of the encoder become genuinely independent, and input features arriving at latent dimensions are orthogonal projection and disentangled at the cost of reconstruction accuracy. A stronger disentangling is desirable if the latent space is more important than generating new samples, as shown in \cref{fig-3}.

\begin{figure}[H]
\centering
\includegraphics[scale=0.6]{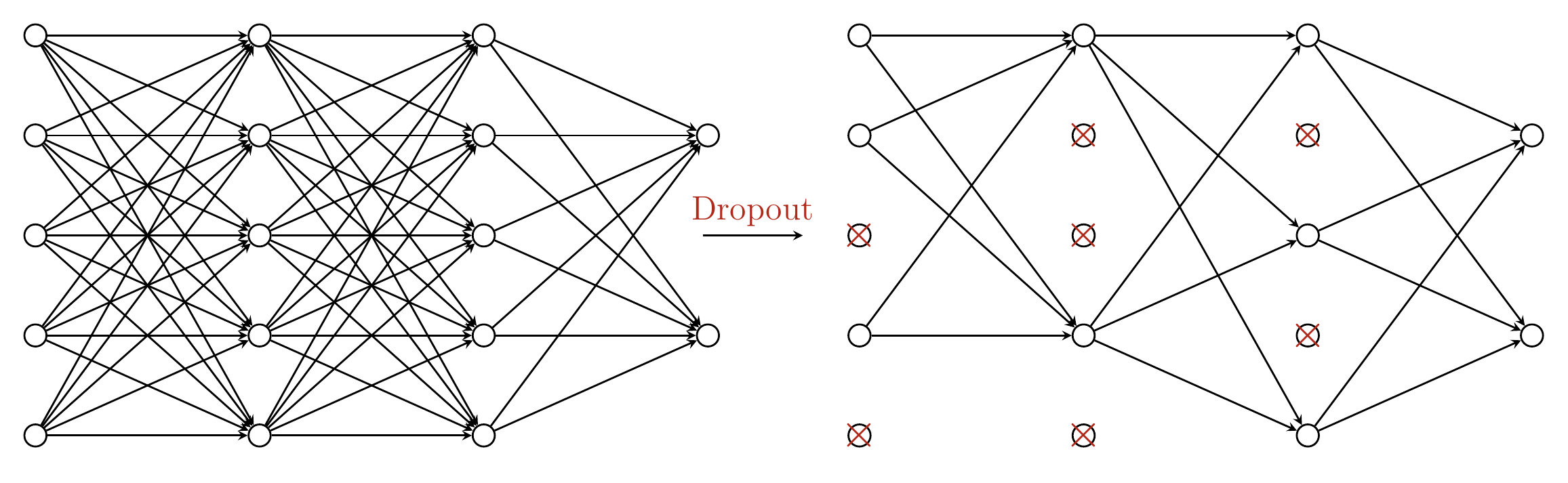}
\caption{\textcolor{VividPurple}{A Dropout Operation}}
\label{fig-A11}
\end{figure}

In the dataset, each molecule is represented by a SMILES and we converted it into a fixed-dimensional vector using one-hot encoding. We feed encoded fixed-dimensional vectors into the encoder of a variational autoencoder. The encoder had five hidden layers with three convolutional layers with 38 rectified linear (ReLu) units and was regularized using dropout, as shown in \cref{fig-A11}. \cite{nair2010rectified,srivastava2014dropout} 

The number of hidden units, the learning rate, and momentum decay were determined using iterations of Bayesian optimization. The network was trained to minimize the RMSE/MSE and MAE of the predicted quantities using the Autograd automatic differentiation package. \cite{maclaurin2015gradient}
\section*{Supervised Machine Learning \& Statistical Regressor}

Machine learning is extensively employed to discover patterns in immense datasets. In this regard, various machine learning algorithms developed, i.e., supervised learning, unsupervised learning, active sequential learning, and reinforcement learning. The supervised learning algorithm tries to deduce a representative function from the dataset that it sees in the training phase. In the unsupervised learning scheme, we find the structure in the data without the labels in the data points to learn. Various algorithms are proposed in the supervised learning regime, including model-based learning, linear model, kernel ridge model, nearest neighbors model, support vector machines, decision trees models such as random forests, Gaussian processes model, and neural network-based modeling. In the decision trees-based random forests modeling, the structure of the model contains a root node, which contains all data and is the starting point of the algorithm. Subsequently, prediction is made at a leaf node, the final branch node after a series of splits subset of data represented at the branches. Based on a feature in the dataset, decision nodes contain a splitting criterion to perform the splitting at a single division. There are various challenges in existing materials science workflows applying the machine learning algorithm to accelerate materials design and discovery research. A basic materials design workflow identifies materials properties, trains models for the properties, predicts properties for new chemical compositions and synthesizes and verifies the predictions. The first challenge related to material science data is that it is relatively expensive and time-consuming to generate extensive data, and model errors increase with a smaller dataset size. The second challenge is biased target data points, as data is often heavily grouped around specific values with a lot of research interest in material science. Although the dataset's total number of data points is significant, the lack of ``negative" data points still may give a biased prediction. The third challenge, especially in materials science, is that the dataset is compositionally grouped around periodic table elements. Moreover, compositions are often heavily grouped around specific elements; this skewed dataset in the machine learning process may bias the model prediction for those periodic table elements. As actual properties of a material depend upon underlying quantum physics orbital chemistry, which may shift dramatically from element to element. However, with the machine learning algorithm trained on compositionally grouped skew datasets, we will fail to predict actual properties as the algorithm has not seen meaningful properties from such a data pattern. There are also challenges in making prediction models highly accurate around a few more accurate training data points and less fluid around other data points in the multi-fidelity modeling. The loss or cost function defines the error metrics and is usually set up uniquely throughout the deep learning neural network. Data cleaning and feature engineering are significant steps in deep-learning modeling; they involve data generation, augmentation, and feature generation. The unreliable models will give all predictions the same values every time as the model has not learned anything from the data points in the training process. In the deep-learning feature generation and feature engineering step, we have converted the real-world scientific data into machine learnable formatted vector data such as structural properties values from input structures, elemental chemical properties values from input compositions, the intensities in SEM/AFM images, and experimental data values from each material. It is critical in the in-building of machine learning models as the model's prediction heavily depends on the seen dataset. We can perform the data imputation to resolve the missing values using the statistical mean, median, or mode procedure. We perform the feature generation using the elemental properties, structural features, or one-hot encoding algorithms in the feature engineering stage. Also, we normalize the feature values to enable us to feed into the machine learning neural network through StandardScaler feature normalization algorithms available in sci-kit-learn that scale the data with a mean of zero and a standard deviation of one. 
Furthermore, we perform the feature selection process to identify the most pertinent features from the full feature matrix for the model to be trained on based on the learning curves and target properties. We have used the ensemble model feature selector that fits the data using a random forest model and selects features based on the resulting random forest feature importance ranking. It will rank the most highly relevant feature from the full feature matrix. Furthermore, a computationally expansive sequential forward selector was used from the ``mlxtend package" with the random forest model. \cite{raschkas_2018_mlxtend} In the sequential forward selector, first, we select the feature through the ensemble model feature selector, identify the top feature, perform the learning curve with only these features, rank them, and select the highest-ranked feature. In the next cycle, we remove that feature from the ensemble model feature selector, perform the learning curve again, rank the feature, and select the highest feature. In this recursive cycle, we perform the number of top-performing features that can best explain the entire dataset in the best possible way. A more methodical feature selection approach is to conduct feature evaluation during every train-test splitting cycle, as it will prevent overfitting. Next, the formulating learning curves assess error versus the number of features and the amount of training data required for selecting the optimal number of features. We have observed that additional training and validation data does not improve model performance in performance matrics after a certain datapoint number. Similarly, in the feature learning curve, we have observed that the model's behavior in training and validation is flat out. More features do not help anything new in the learning phase in the model training. The next crucial step in machine learning modeling is model assessment, optimization, and predictions. The model is trained through available features search to split each feature on a decision tree-based forest model. Next, we search for the split best value for all combinations of features and values, choose the tree based on performance, and repeat the algorithm for the next level of nodes. We fixed the tree route by progressively moving deep into the forest and splitting the data on the right features by assigning values to each node. The learning process of a tree-based random forest decision diagram is training data decide nodes to split data. We only feed the features to use by the algorithm and tune hyperparameters which essentially influence the learning process. While making the predictions based on the trained model, we also have to quantify how fair and confident the prediction is to quantify the learning performance. In this regard, we take the ensemble of the model and use statistical techniques to calculate the variance in the prediction for model assessment. In the deep learning training procedure, the dataset divides into training data, testing data, and validation data. Training data is explicitly used to train the model; therefore, A well-trained model is expected to fit and predict the training data points very well because it has already been seen in the training cycle many times. The validation datapoint was temporarily withheld from the training data during a cross-validation procedure and used to optimize a model before the final model fitting. Finally, to evaluate the learned model performance, a portion of the dataset was withheld in testing data points until a final model was fit and used to predict known test values to compare with predictions to assess model error and uncertainty. Splits are performed in the dataset to robustly train the model on a partially available dataset to estimate the model performance at the correct known data points. The cross-validation defines K-fold, where `K' means how many splits are performed in the dataset. A series of models trained on each fold, with the training data being the rest of all folds, and testing data is from itself fold and assess model error by averaging across all the splits. The cross-validation data split is used for model optimization, and a single train-test split uses for the final model evaluation. Different cross-validation splitting strategies employed no-split, repeated K-fold, repeated K-fold-learn, and left-one group-out. In the no-split strategies, no actual cross-validation happens and is primarily used for model comparison with all the data used in the training and testing stage. In the repeated K-fold cross-validation strategy, the dataset splits into two iterations of random left-one group-out 20\%, 5-fold cross-validation for the selected regression model evaluating the model performance matrix score. The repeated K-fold-learn is similar to the above technique but used to generate a learning curve. Many datasets have subsets of data that belong to distinct groups, and one is often interested in how a model may perform when predicting new data on a brand new group. Therefore left-one group-out cross-validation splitting strategy where one group of data is removed from training and validation and used in testing to observe the model performance. The left-one group-out will give a higher error than the random left-one group-out cross-validation. Also, for the left-one group-out cross-validation, we plotted error metric value versus the group. In the 2-fold cross-validation task, we can hold 50\% of the data to train the model, observe the limiting performance of a model, and how the error matrics change. We have accomplished the model performance test with up to 90\% of the data left out using an increment of different left-one group-out percent cross-validation tests. We then fit a Kernel ridge model to each group to observe RMSE changes with left-one group-out. As the expected model becomes worse with the decreasing amount of data available to train the network, we can constitute a good model with a minimum amount of data for specific performance. Next, we have performed nested cross-validation for the model optimization. The nested cross-validation is a robust scheme as data splitting performs at each nesting level across all the datasets. In one cycle of 5-fold cross-validation nested cross-validation, we will split data into 5-fold cross-validation, which means there are 5 level 1 splits and 5 level 2 splits for each of the level 1 split; hence a total of 25 splits performed in terms of cross-validation. In the nested cross-validation, we got RMSE error matrics higher than the random 5-fold cross-validation RMSE. It is an overly optimistic predictor of model performance when predicting the values of unseen data from a unique group. The most robust model predictor combines nested K-fold cross-validation with a left-one group-out cross-validation scheme. In this scheme, left-one group-out performed at all the training and validation split levels for one cycle; it is equivalent to 125 splits performed in terms of cross-validation. Hence it functions as a good approximation for how the model may perform on new, unseen data at all levels of nested split and is the most robust strategy to evolute the model performance. This joint scheme predicts better RMSE error matrics than the pure left-one group-out cross-validation scheme, as left-one group-out performs at every nested level. Also, it is comparable but more robust to nested 5-fold cross-validation random cross-validation RMSE where no left-one group-out is performed. Later in the directed-message passing neural network approach, it is equivalent to the scaffold cross-validation scheme where molecules with different scaffolds are grouped in different test-train splits to train for the robust model. Some scaffold groups are left out in the training set and only used in the test cycle. The error in the model is primarily quantified on the test data by statistical metrics such as root mean squared error, normalized root mean squared error, coefficient of determination, and mean absolute error. The root mean squared error defines as, \cite{levi2019evaluating}

\begin{equation}\label{eq-MAST-1}
RMSE=\sqrt{{\frac{\sum_{i=1}^{N}\left(y_i-{\hat{y}}_i\right)}{N}}^2}
\end{equation}

The normalized root mean squared error defines as root mean squared error divide by standard deviation, 

\begin{equation}\label{eq-MAST-2}
\frac{RMSE}{\sigma}=\frac{\sqrt{{\frac{\sum_{i=1}^{N}\left(y_i-{\hat{y}}_i\right)}{N}}^2}}{\sigma_y}
\end{equation}

The coefficient of determination defines as, 

\begin{equation}\label{eq-MAST-3}
R^2=\frac{{\frac{\sum_{i=1}^{N}\left(y_i-{\hat{y}}_i\right)}{N}}^2}{{\frac{\sum_{i=1}^{N}\left(y_i-\bar{y}\right)}{N}}^2}
\end{equation}

The mean absolute error defines as,

\begin{equation}\label{eq-MAST-4}
MAE=\frac{\sum_{i=1}^{N}\left|y_i-{\hat{y}}_i\right|}{N}	
\end{equation}

Where $ N $ data point's total number, $ {\hat{y}}_i $ is predicted value, and $ y_i $ is known value for the "i-th" data point, and $ \sigma_y $ is the standard deviation. During the fitting process of model optimization, parameters of the model are established. The decision tree-based modeling leaf nodes and branch values are model parameters decided by the learning process. On the other hand, model hyperparameters aspects of model optimization are controlled by humans to fine-tune the model performance. It also affects the model learning process and hence effet the model parameters fitting in the decision tree-based modeling. The hyperparameters are the forest's maximum depth and the maximum leaf nodes in the trees. The hyperparameter tuning is used to fine-tune the initially built model and estimate its performance to improve it. The hyperparameter optimization is performed through grid search, randomized search, and Bayesian search strategy. We optimize the alpha parameter related to regularization strength that learns inverse transform for the model optimization in the Kernel ridge regressor. The optimization was performed at each train-test split in the random 5-fold cross-validation, and the best model was selected to predict the left-one group-out data. The hyperparameter optimization grid search explored from the lower bound of -5 to the upper bound of 5, with a grid density of 100, in the logarithmic scale. An optimized Kernel ridge regressor gave better error matrics than the non-optimized Kernel ridge model. Instead of specifying a range of values in the grid search, we have also performed a randomized search over two variables, alpha, and gamma, for 100 iterations on a uniform probability distribution for a given split. We have also performed the grid search on the multi-layer perceptron regressor class to identify the best neural network architectures in terms of the number of layers. The hyperparameter optimization with the combination of nested cross-validation and left-one group-out cross-validation will give a more realistic best-case fit model compared to random 5-fold cross-validation. We have estimated the true and predicted errors statistical distributions for the model and recalibrated it with the uncertainty quantification (UQ). The predicted errors are estimated using Bayesian methods by the Gaussian process regressor class or ensemble methods, such as random forest, gradient boosting regressor, or ensemble the models. The Bayesian methods by the Gaussian process regressor class give a direct estimate of the error bar for each predicted data point. In the random forest decision tree regressor, the error bar calculates the standard deviation of each weak learner prediction on each data point at the decision tree. The average is the predicted value, and the standard deviation on that value is the error bar by using a 5-fold cross-validation. The r-statistic plot is a reduced error value defined as the true model residuals error divided by the predicted model error, i.e., error bars. This plot should be a normal distribution, with a mean of zero and a standard deviation of one for the model error estimates that are reasonable estimates of a true error. \cite{Ling_2017,LU2019109075} In the normalized error distribution plot, similar to the r-statistic distribution plot, the distribution of normalized residuals and normal distribution are plotted. The r-statistic normalized model errors distribution value represents in purple color, a normal distribution in blue color, and the distribution of normalized residuals in green color. The normalized residuals calculate as residuals divided by the dataset standard deviation. The r-statistic distribution is skinnier than the normal distribution, having a standard deviation of less than one.\cite{morgan2020opportunities,  Palmer2022May} Similar to the normalized error distribution data, the cumulative normalized error distribution is plotted where all curves will converge to one as the reduced error $x/\sigma$ increases. The r-statistic normalized model errors distribution value represents in purple color, a normal distribution in blue color, and the distribution of normalized residuals in green color. The normalized residuals calculate as residuals divided by the dataset standard deviation. The r-statistic normalized model errors distribution in purple and normalized residuals distribution in a green offshoot faster toward the left from the normal distribution in blue indicates that the denominator of $x/\sigma$ is too large in the model error bars in purple. It indicates that the trained model is overestimating the true error. From the r-statistic distribution data, we have plotted RMS residual versus prediction variance error to estimate true errors residuals correlation with the respective predicted errors. The x-axis represents the predicted errors, and on the y-axis, true error residuals are grouped into a finite number of bins, and each data point is the root mean squared residual of that particular bin. The top histogram plot shows the distribution of data points present in each bin. The dataset standard deviation normalizes both axes; hence they are unitless. The soft gray dashed line represents the $y=x$ identity line, and if the model predicted errors are precisely the same as true errors, all data points ideally fall on this $y=x$ identity line. The dark black dashed line represents a weighted linear fit of the r-statistic distribution data. The slope is less than one and intercepts at non-zero on the y-axis, indicating the model overestimates the true error in r-statistic distribution data. Therefore previously r-statistic distribution had a standard deviation of less than one. Next, we recalibrate these uncertainty estimates to align with the true values more closely. We have plotted r-statistic distribution plots and RMS residual versus prediction variance error plots for both uncalibrated and calibrated data and superposed them \cite{hirschfeld2020uncertainty,musil2019fast} In the r-statistic distribution plot, we observed that the mean is closer to zero, and the standard deviation is closer to one for recalibration data than the uncalibrated data. In the RMS residual versus prediction variance error plot, we have observed that the slope of the fitted line is closer to unity in the recalibration data compared to uncalibrated data. We have also observed that the intercept moves closer to zero, and the coefficient of determination R-squared value improves. We have performed the recalibration process with five numbers of a repeat, and hence a total of 625 nested cross-validation splits are performed in the data points. The recalibration process improves the robustness of linear correlation and improves the model error estimates. We have compared the uncertainty quantification behavior of Bayesian methods by the Gaussian process and previous ensemble-based random forest models. In the Gaussian process regressor, the r-statistic distribution plot has a standard deviation greater than one for uncalibrated data, unlike to ensemble random forest regressor class, indicating the Gaussian process regressor underestimates the true error. Further calibration process pushes the standard deviation close to one. We have not observed any correlation between the true and predicted errors in the RMS residual versus prediction variance error plot for both uncalibrated and calibrated data points. A more computationally expensive ensemble of the Gaussian process regressor model may improve the robustness of linearity and correlated uncertainty estimates after calibrated data points. In the final model predictions stage of machine learning in materials science, we try to identify materials properties of interest and predict their values, a target range for these property values, and their maximum or minimum value. With the limiting constraint of materials search space, we try to identify materials of interest with a specific target material or compositional space properties. However, this is the last stage in material informatics. We designed the whole training and learning stack from the initial stage, considering this goal from data cleaning to augmentation to the modeling process. We have trained and developed the following model through the Scikit-learn standard machine learning models class, such as Kernel ridge regressor (KRR), Random forest regressor (RFR), Gaussian process (Kernel) regressor (GPR), Customised neural network through Keras-Tensorflow class. In the model evolution and automatic hyperparameter optimization stage, we have employed various cross-validation strategies with random left-one group-out, left out a specific group, left out the cluster, and nested cross-validation. We have evaluated the model error bars for the uncertainty quantification. We have also performed error analysis on predictions through average oversplits and the best-worst strategy and performed predictions on left-one group-out data. We have plotted the result and performed analysis summaries using the parity plots as average oversplits, the best-worst split strategy, and the best-worst of all data points. The uncertainty quantification was performed using error distributions, errors by group, histograms, and residuals versus model errors. We have employed the Materials Simulation Toolkit for Machine Learning, a python package backend on sci-kit-learn, and the Keras-TensorFlow modules developed at the University of Kentucky to broaden and accelerate materials property prediction through supervised machine learning. \cite{jacobs2020the}  The tool includes many features and automates the data process flow, importing materials from the online repositories, cleaning, splitting, comparing, evaluating the input dataset, simultaneous execution of preprocessing selection, generating features through feature engineering analysis, feature selection construction, and performance analysis of different models by evaluation metrics. It also automates the baseline random five-fold cross-validation statistical test using Linear regression, Kernel Ridge Regression with radial-basis function kernel, Gaussian process regressor with restarts optimizer of 10, Random forest regressor with 150 estimators, and Multi-layer perceptron regressor neural network with hiden layer size of 20 for the model performance. In the k-fold cross-validation step, the network will first select the features, split the normalized data in train and test, and validate the optimized hyperparameters for that network model. Select different features, train-test splits dataset accumulate new statistics in the next run. After multiple runs, the network will choose the best model based on the lowest error metric generated by the given set of tests and datasets. We can optimize a network by changing the model type, varying the features, varying the test-train data split, and left-one group-out cross-validation of the dataset's group of features. To test the model performance, left-one group-out cross-validation are more computationally demanding and rigorous than random cross-validation because the left-one group-out test data tends to be outside the domain of the training segment groups data; on the other hand, randomly segmented cross-validation data tends to have similar distributions of training and test data. Therefore, left-one group-out test data to have more demanding network production statistics and better predict the real-life scenario. Kernel ridge model from sci-kit-learn has no hyperparameters optimization to optimize model parameters to make predictions on left-one group-out data efficiently from the training set. Nested cross-validation with hyperparameters optimization is built into the model to improve it. Hyperparameter optimizations can perform through the grid-search method, randomized search, and Bayesian-based search method. We grided the space for the alpha parameters in Kernal Ridge regression and optimized the model for the k-fold cross-validation. In every split, we predict the best model from test data, use that model on the left-one group-out data, and improve predictions. Tree-based methods like random forest regressor and gradient boosting are good feature selection choices. Many features will sort ranked weight naturally with more essential features than others and find the essential features. It also conducts a preliminary statistical assessment of model error analysis and uncertainty quantification (UQ) on the learning algorithm. To find the best model performance with an unbiased estimate in the iterative loop from selecting features to training models, optimizing hyper-parameters, and assessing model performance through bootstrapped uncertainty quantification to check the reliability of the models.  It works as a bridge between materials data repositories and deep learning algorithms. Also, it envisioned assisting and facilitating the result reproduction and easier dissemination of the model by streamlining the other existing data-sharing platform infrastructure, such as cyber-infrastructure for sustained scientific innovation (CSSI). \cite{Cyberinfrastructure2021}
\section*{Sequential Learning \& Optimal Design of Experiments}

In the engineering and natural science field, we perform various experiments or numerical simulations to validate a hypothesis, characterize the molecule's properties, or design material with carefully planned goals to target specific material properties. Optimizing the experimental procedure to optimize the time and resources is critical to the project life cycle's success. In the multi-objective optimization problem, brute force exploration of design space to find optimal parameters is hugely inefficient; hence traditionally, the intuition and experience of researcher and scientist is the guiding principle in the design of experimentation. In the data science regime, we can optimize the discovery process using a sequential learning approach by employing neural networks to design the experiments. Further optimization of predicted material properties can be achieved by using additional neural network topology through supervised learning or reinforcement learning to develop predictive models or classify the suggested material through sequential active learning. The first step in active learning comes from an approximation of the error in the prediction from the model. In the design of experiment flow, we can aggregate information from different sources and make a model based on the target properties goal, which gives the quantified uncertainties. We can learn this model in the active learning algorithm based on target properties by defining the information acquisition function and using the uncertainties quantification. We can perform an experiment where uncertainties are the highest. Hence, after the experimentation, we learned the most about the process after performing the subsequent experimentation at that data point, and current knowledge about that data point is the least. We can also train the information acquisition function where mean prediction is the highest and maximum expected improvement occurs at the datapoint's present cycle. We can also train the information acquisition function for the mean value and certainty where maximum likelihood or probability of improvement is expected and reveal that data point in the next cycle will be maximally beneficial. We have color-coded the revealed data points to observe the performance of sequential learning. We can also tune the information acquisition function to the region where the most information is known, and hence uncertainty is least to meet our target properties criteria. We can add these new experimentation results to our existing active learning dataset. Hence, our information acquisition function becomes more robust with each iterative passing cycle and predicts better suggestions. We will hide most data points from the information acquisition function and reveal only a few tens of percentage data points, ensuring the highest value of target properties is not fed into the information acquisition function. Furthermore, we train the neural network on the information acquisition function to minimize uncertainty. After training, we will explore and validate how many minimum cycles it can reveal the maximum target properties compared to a brute force random search. The Lolo library implementation yields the sample-wise uncertainties from simple base learners of the random forest-based decision trees model. The library imbues robust sample-wise uncertainties by combining the infinitesimal-jackknife variance estimates and jackknife-after-bootstrap paired with a Monte-Carlo sampling correction. \cite{JMLR:v15:wager14a, Ling2017Sep} To compute an estimate of expected value, the mean of the predictions over all the decision trees represents the expected value for the particular model as, 

\begin{equation}\label{eq-DOE-1}
 E[M(x)] = \frac{1}{N} \sum\limits_{J}^{n_T} T_J(x) 
\end{equation}

Where, $n_T $ is the number of trees, at point $x$ on tree index $J$ prediction is $ T_J(x) $, and at point $x$ expected value of model prediction is $E[M(x)] $. Next, at point $x$, we will compute the variance value due to training point $i$ by combining the infinitesimal-jackknife and jackknife-after-bootstrap estimates paired with a Monte-Carlo sampling correction,

\begin{equation}\label{eq-DOE-2}
\sigma^2_i(x) = Cov_j[n_{i,J}, T_J(x)]^2 + [\overline{T}_{-i}(x) - \overline{T}(x)]^2 - \frac{e\nu}{N_T} 
\end{equation}

Where on the tree index $ j $, covariance is $Cov_j$, $n_{i,J} $ number of times point $i$ used to train tree index $J$,  Euler's number is $e$, variance over all trees is $\nu$, $\overline{T}_{-i}(x) $ is average prediction over the trees that were fit without using point $i$, $\overline{T}(x) $ is average prediction over all of the tree. We will estimate the uncertainty by adding the contributions of each of the training points to the variance of the test point together with an explicit bias model and a noise threshold as,

\begin{equation}\label{eq-DOE-3}
 \sigma[M(x)] = \sqrt{ \sum\limits_{i=1}^S \mathrm{max}[\sigma^2_i(x), \omega] + \widetilde{\sigma}^{ 2}(x)} 
\end{equation}

Where at point $x$, by training point $i$, the variance is $\sigma^2_i(x) $, $\omega  $ is noise threshold, $ \widetilde{\sigma}^{ 2}(x) $ is explicit bias model, and $S$ is number of training points. Further the noise threshold ($\omega$) is,

\begin{equation}\label{eq-DOE-4}
\omega = \underset{i}{\mathrm{argmin}}\, \sigma^2 [M(x_i)]
\end{equation}

We have defined five different types of information acquisition functions such as maximum uncertainty (MU), maximum expected improvement (MEI), maximum likelihood or probability of improvement (MLI), random prediction, and upper confidence bound (UCB) based on criteria on which data-point is more relevant for the model to query next in the design of the experimentation strategy. The maximum target property value model's prediction over all possible experimentation set $x_i$ is used to define the maximum expected improvement (MEI) information acquisition function as,

\begin{equation}\label{eq-DOE-5}
 x^* = \mathrm{argmax} \enspace E[M(x_i)]
\end{equation}

The maximum likelihood or probability of improvement (MLI) information acquisition function defined with up to our current-cycle best-case value of target property is $x_{best} $ query a region of experimentation sets $x_i$ where the high likelihood of improvement of target property value with sufficient uncertainty as,

\begin{equation}\label{eq-DOE-5}
 x^* = \mathrm{argmax} \enspace \frac{E[M(x_i)] - E[M(x_{\mathrm{best}})]}{\sigma[M(x_i)]}
\end{equation}

The maximum uncertainty (MU) information acquisition function is defined as the strategy that subsequently queries the target property value with the highest uncertainty as,

\begin{equation}\label{eq-DOE-5}
 x^* = \mathrm{argmax} \enspace \sigma[M(x_i)]
\end{equation}

Upper confidence bound (UCB) information acquisition functions is defined as the strategy queries the sample with the maximum value of its mean prediction plus its uncertainty. The upper confidence bound (UCB) information acquisition function is defined as the strategy that queries the maximum target property value of its mean prediction and summation to its uncertainty value as,

\begin{equation}\label{eq-DOE-5}
 x^* = \mathrm{argmax} \enspace [E[M(x_i)] + \sigma[M(x_i)]]
\end{equation}
\section*{Material Discovery through Graph Neural Network}

\begin{figure}[H]
\centering
\includegraphics[scale=0.6]{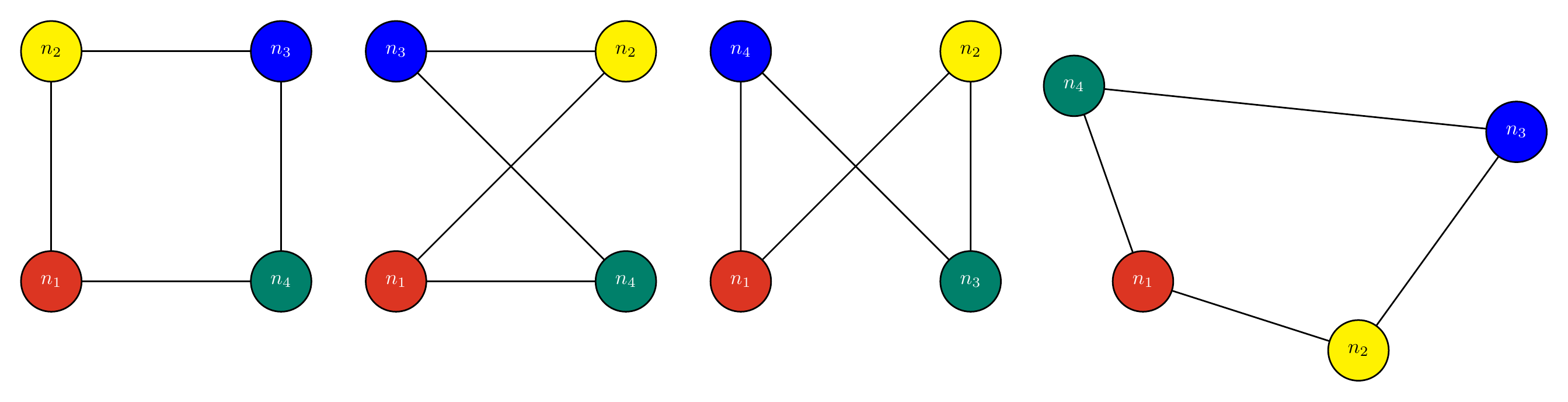}
\caption{\textcolor{VividPurple}{A Graph Neural Network Topology}}
\label{fig-A19}
\end{figure}

\begin{figure}[H]
\centering
\includegraphics[scale=0.4]{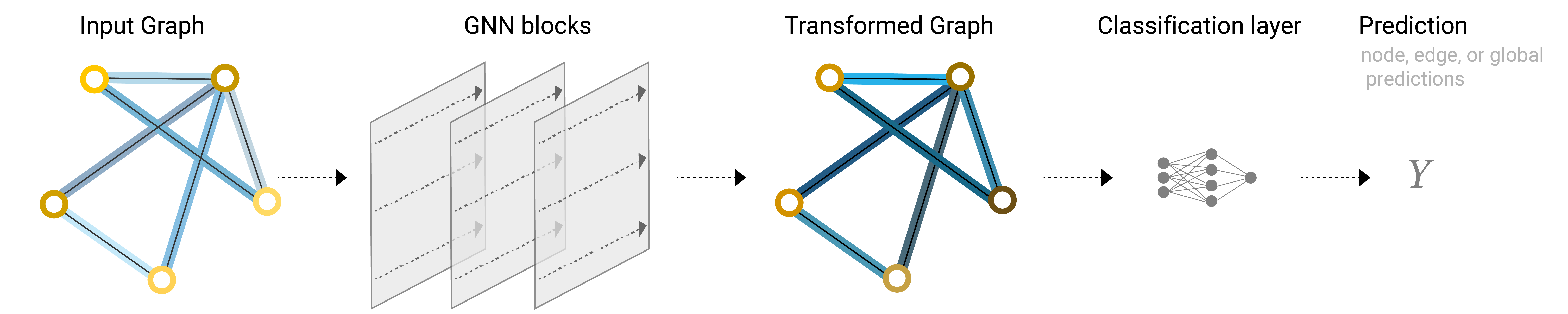}
\caption{\textcolor{VividPurple}{A Graph Neural Network}}
\label{magnet-fig-2}
\end{figure}

We have framed a deep learning strategy to predict quantitatively accurate and desirable material properties by constructing a relationship between the molecular structure and its property through a material graph-based neural network approach as shown in \cref{fig-A19}, and \cref{magnet-fig-2}. The central goal of natural science is deriving the relationship between material structure and its behavior properties. The underlying mathematical and physical development of quantum mechanics and field theory of nature has accomplished it in the whole of chemistry, and a large part of physics, as Paul Dirac himself stated in 1929. However, even after around a hundred years of numerical and computational resource development, he remarked that the exact application of these laws is computationally cumbersome. The solution of these mathematical equations has their intricacy even on state of the art, the most powerful supercomputer cluster available to humanity. For example, a new molecule's material diffusivity property or dynamic stability approximate calculations through an ab-initio density functional theory solution take weeks. To mitigate this computational bottleneck, with the recent development of machine learning techniques in the information science domain. The natural science research paradigm is shifting from the computational solution of physical science laws to data-driven science and materials informatics, where decades-long experimental and conventional calculations produced substantial data to construct surrogate machine learning models. These deep learning trained models proposed and predicted novel molecule structures and properties directly from the existing dataset with infinitesimal computational time and resources. Deep learning is the study of computer algorithms that improve and evolve automatically through experience by the fundamental statistical law of information science. In this regard, recently, deep fake networks and deep belief networks have been demonstrated, which construct on-demand a synthetic human face from existing human face attributes images available on the internet-no such human born on this planet or seen by these networks. The materials graph networks implemented on the topology of graph network proposed by \texttt{DeepMind} and \texttt{Google} whose architecture is motivated by the ultimate goal is a combinatorial generalization. Combinatorial generalization aims to train a network from the limited number of examples, learn the relation rules between them, apply them to the unlimited number of instances or observations, and achieve this objective structural representation of a graph as information inference is the best suitable topology. \cite{Battaglia_Interaction_2016,DBLP:journals/corr/abs-1806-01261}

The graph network's structural representation of information constructs through relational reasoning of the attributes of ``objects'' or ``elements''. These ``elements-object'' are defined as the ``entity'' of the network. The property between these ``entities'' is defined as ``relation''. The ``Rule'' of graph network is 
functions that map ``entities'' and ``relations'' to other ``entities'' and ``relations''. For the material graph network, the ``entities'' are mapped as the atom of the materials, and the ``relation'' is the chemical bond of the materials species. The ``rule'' is mapped as a convolution operation of the materials network. The inductive bias of the network is a set of assumptions that the learner uses to predict outputs of given inputs that it has not encountered. The inductive bias arrives from the localized convolution operation in the network as the graph network the structure is arbitrary; hence the relative inductive bias is arbitrary and depends upon the final structure of the graph network as shown in \cref{magnet-fig-2}.\cite{doi:10.1021/acs.chemmater.9b01294} The performance of these networks is surprisingly very high, and an ordinary human cannot classify which image is an imaginary human face created by the algorithm and which one belongs to a natural person. We have used this revolutionizing information science approach in the natural science field to construct new materials-the network trained with the existing materials dataset, which we accumulated over the decades. The eventual goal is to create custom-tailored new physical species and materials of desirable properties, which can assist in faster the quantum mechanical ab-initio calculations and accelerating the design of experiments as shown in the \cref{magnet-fig-2} below. The inputs to the material networks are, e.g., molecular structure and atomic composition, and output are properties of interest, e.g., bandgap and formation energy. Recently for the estimation of material properties through the neural network various, e.g., crystal graph convolutional neural networks(CGCNN), \cite{PhysRevLett.120.145301} SchNet,\cite{doi:10.1063/1.5019779} and SchNetPack \cite{doi:10.1021/acs.jctc.8b00908} are demonstrated. In material informatics, deep learning networks for materials use primarily two-way. The first approach is compositional model fingerprinting, where we describe complex materials in terms of compositional level information, e.g., chemical structure formula.\cite{pilania2019data} These chemical elements' properties aggregate the statistics to predict the material properties. The compositional level machine learning gives a good performance, but they can not distinguish between the polymorphs with the same chemical composition formula but different crystal structures. \cite{meredig_combinatorial_2014} The second approach uses structural model fingerprinting, where we use machine learning interatomic potentials that describe the local atomistic environment to the overall material properties. This scheme can accurately construct the local potential energy surface to describe a particular material system. However, unfortunately, this approach has a bottleneck as limited to a small number of elements. We have to train one network customized for individual chemical spaces. As the chemical species expand in the periodic table elements and crystal structures, the model network also grows extensively, limiting its universal general-purpose usability. The graph neural- networks approach from the computer science domain is borrowed in materials space to mitigate these challenges. In this approach, materials molecular structure fingerprint as a graph network where the atoms map as graph nodes and various chemical bonds are edges. Furthermore, we propagate information from graph nodes through edges to subsequently connected neighboring nodes during the training phase. This localized convolutional operation will accurately learn about the local atomics environment of molecular structure as per training weights and biases. Hence also referred to as a graph convolutional neural network. \cite{bruna2013spectral, defferrard2016convolutional} After the succeeding hidden convolutional layers, we encode abstracted information inference at the output graph layer, containing the entire molecular structure. We fingerprint the output graph layer as an encoded feature vector for the target properties variable of the materials in the feedforward or any other neural network. The advantage of this material graph network approach is that the descriptor is general-purpose, relatively accurate compared to the pure composition model, transferable, and applicable to all the elements and all kinds of available materials datasets.

\begin{figure}[H]
\centering
\includegraphics[scale=0.6]{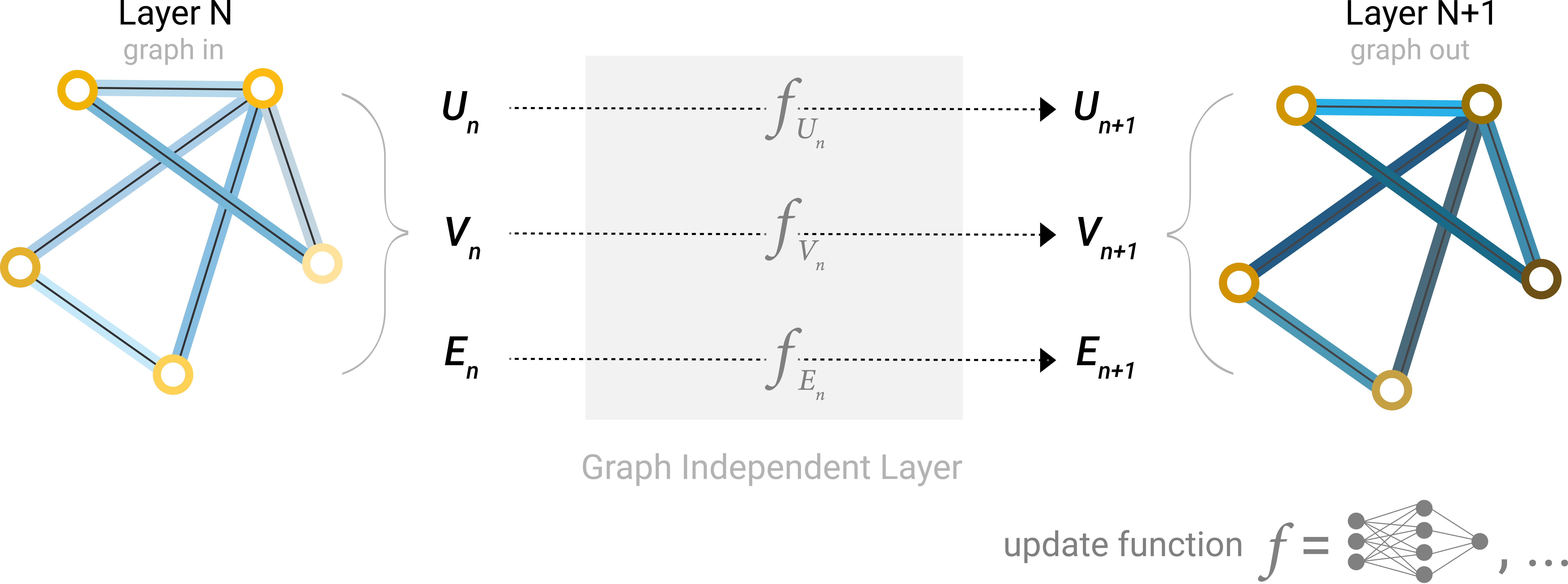}
\caption{\textcolor{VividPurple}{A Graph Neural Network Building-Block Update Function}}
\label{magnet-fig-3}
\end{figure}

\begin{figure}[H]
\centering
\includegraphics[scale=0.4]{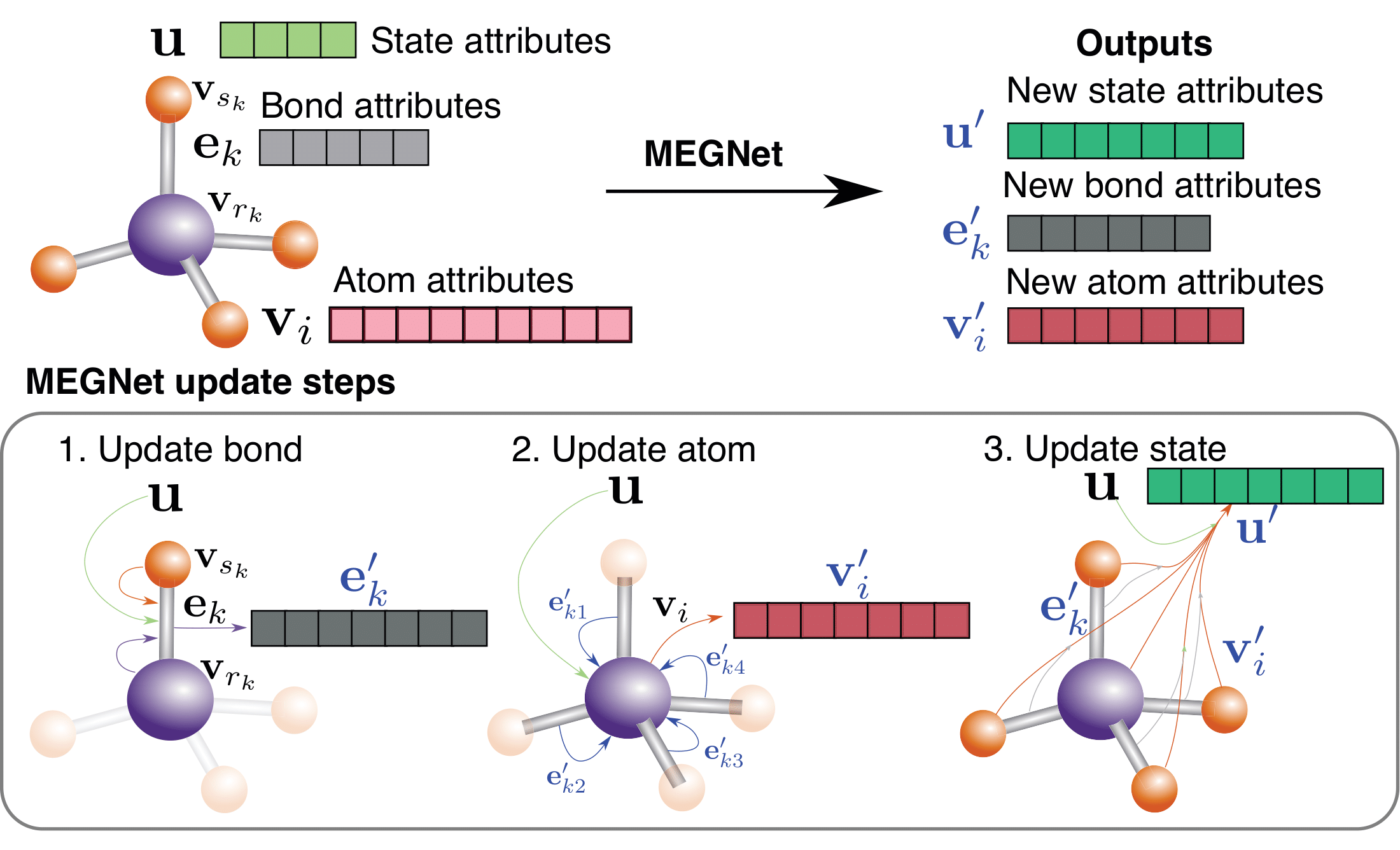}
\caption{\textcolor{VividPurple}{Material Graph Neural Network (MEGNet) Building-Block  Information Flow Model} \cite{doi:10.1021/acs.chemmater.9b01294}}
\label{magnet-fig-4}
\end{figure}

In addition to the nodes and the edges in the material graph network, we have also incorporated the attributes of the global state, where information in global states is an independent network structure and depends upon external conditions, e.g., temperature, pressure variable on the molecules. We use gaussian expanded distance for bonds in atomic representation to evaluate these bond states. In the graph convolution stage, we first update the bond state's information by using neighboring atoms, previous bond states, and the global stage parameters; afterward, we update the atom state as the graph nodes and ultimately update the global state's parameters in the material graph as shown in \cref{magnet-fig-3} and \cref{magnet-fig-4}. We use the artificial neural networks activation function to approximate the functions to train and approximate any continuous function through the universal approximation theorem. \cite{cybenko1989approximation}

In materials science, we encounter many datasets with varying accuracy and precision, e.g., density functional datasets with varying degrees of functional PBE, HSE, exchange-correlation potential, and experimental data. To utilize the available dataset optimally and mitigate highly accurate data scares regimes, we have to eventually mix up the highly accurate dataset with the abundantly available, less precise cheap dataset. It will motivate to design a multi-fidelity graph network where a wider accuracy variation of datasets uses to train the network for improved prediction efficiency. The Fidelity information is not related to the structure of the graph networks, and hence it can be added as categorical variables in the global states of the graph networks. By this approach, we can efficiently incorporate the dataset from a different source in the single graph network for training purposes to accurately predict the property of the materials. We use the molecular material state attributes as fidelity information inputs by incorporating a fidelity-to-state embedding subnetwork in the original graph network.

The dataset generated for the same properties of the materials by the different physical calculation principle methods classifies as different fidelity states. Therefore, the graph network evolves its structure by embedding the different fidelity states using different state vectors. The advantage of a multi-fidelity network scheme is that a trained model on multiple datasets can predict the material's properties for one fidelity and the combined fidelity. Hence the calculation data and experimental dataset can be seamlessly used for the network's training.

\subsection*{Materials Graph Networks Formulation}

\begin{figure}[H]
\centering
\includegraphics[scale=0.4]{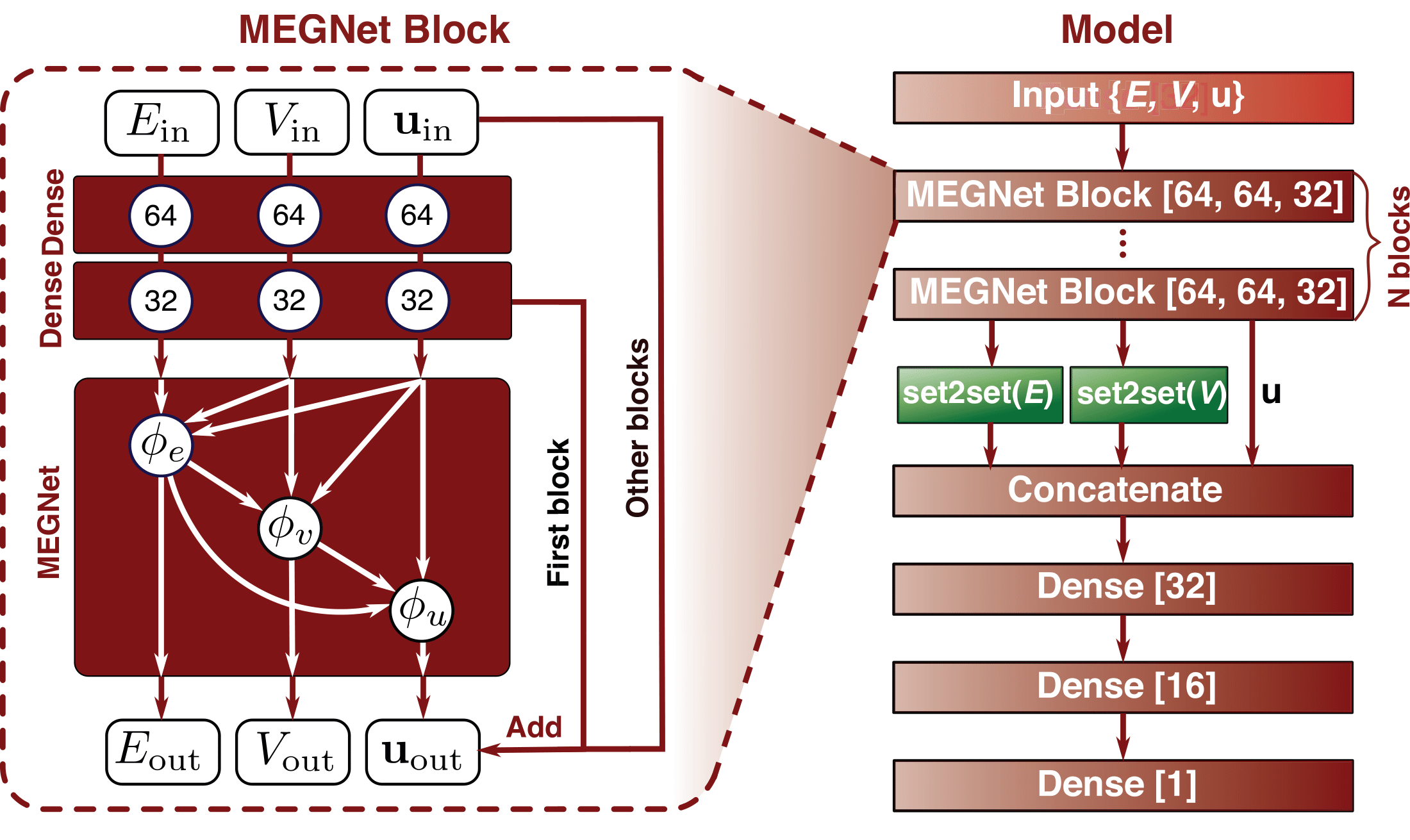}
\caption{\textcolor{VividPurple}{Material Graph Network Schematic Flow} \cite{doi:10.1021/acs.chemmater.9b01294}}
\label{magnet-fig-5}
\end{figure}

The graph neural network approach uses the combinatorial generalization and relational reasoning of graph theory with the function approximation of the neural network. \cite{duvenaud2015convolutional,battaglia2016interaction, chang2016compositional, Chen2022Nov} The input to the graph convolutional neural network architecture has an atomic node or vertex $V$, with a bond edge of $E$, and global state $\bf{u}$ attributes. The atoms in the graph layer of $i$ have the atomic attribute vector $V$, which is a set of $\bf{v_i}$ in a $N^v$ size atomic system. Similarly, the bond edge attribute vector $E=\{(\bf{e}_{k}, r_{k},s_{k})\}_{k=1:\mathcal{N}(j)}$ for bond $k$ in the layer $i$, sending message with index $s_{k}$ and receiving message with index $r_{k}$, and total number of bonds is $\mathcal{N}(j)$. Subsequently, The molecule/crystal-level physical state attributes stored in the global state vector $\bf{u}$ updates the state information for the atoms, bonds, and new global state. A graph neural network constitute a series of update operations that map an input graph $G=(E,V,\bf{u})$ to an output graph $G^{(i)}=(E^{(i)}, V^{(i)}, \bf{u}^{(i)})$. In the graph neural network, from the previous set of bond edge attributes $E=\{(\bf{e}_{k}, r_{k},s_{k})\}_{k=1:\mathcal{N}(j)}$ a new bond attributes first updated using attributes from itself as information flows from atomic attributes $V=\{\bf{v}_{j}\}_{j=1:N^{\mathrm{v}}}$, and the global state attributes $\bf{u}$ as shown in \cref{magnet-fig-5}. The second step is the update of the atomic attributes, and in the final step, global state attributes are updated to represent a new graph. 

\begin{equation}\label{eq-Atomset-1}
\mathbf{e}_{k}^{(i)}=\phi_{e}\left[\mathbf{e}_{k}^{(i-1)} \oplus  \mathbf{v}_{s_{k}}^{(i-1)} \oplus  \mathbf{v}_{r_{k}}^{(i-1)} \oplus  \mathbf{u}^{(i-1)}\right]
\end{equation}

Where $\oplus $ is concatenation operator, and bond edge update function is $\phi_e$. After bond updates, all the atomic attribute $\bf{v}_{j}^{(i)}$ updated using the self past cycle attribute, binding bonds attributes, and global state vector $\mathbf{u}^{(i-1)}$ as,

\begin{equation}\label{eq-Atomset-2}
\mathbf{v}_{j}^{(i)}=\phi_{v}\left[\mathbf{v}_{j}^{(i-1)} \oplus  \mathbf{v}_{k \in \mathcal{N}(j)}^{(i-1)} \oplus  \mathbf{e}_{l, r_{l}=j}^{(i)} \oplus  \mathbf{u}^{(i-1)}\right]
\end{equation}

Where the number of bonds connected to atom node $j$ is $k \in \mathcal{N}(j)$, and atomic update function is $\phi_v$. Long-range interactions can be incorporated if the more convolution operation performs on localized atom-bond connectivity without the aggregation step. However, it will increase the graph structure vector. In the \cref{eq-Atomset-2} by taking the average of bonds connected to atom $j$, a local pooling operation is performed and acts as a graph structure vector aggregator. In the last step of graph update, using information from itself and all the atomic nodes and bonds edges, global state attributes $\bf{u}$ update as,

\begin{equation}\label{eq-Atomset-3}
\mathbf{u}^{(i)}=\phi_{u}\left[\frac{1}{N_{b}} \sum_{k} \mathbf{e}_{k}^{(i)} \oplus  \frac{1}{N_{a}} \sum_{j} \mathbf{v}_{j}^{(i)} \oplus  \mathbf{u}^{(i-1)}\right]
\end{equation}

Where global state update function is $\phi_u$ and exchange information at a larger scale, it incorporates the global structure of a graph and the input physical state and attributes of the materials. The model performance determines through the update functions $\phi_e$, $\phi_v$ and $\phi_u$. The weights and biases for atomic nodes, bond edges, and global state updates function are different. 

\subsection*{Atomset Formulation}
\begin{figure}[H]
\centering
\includegraphics[scale=0.3]{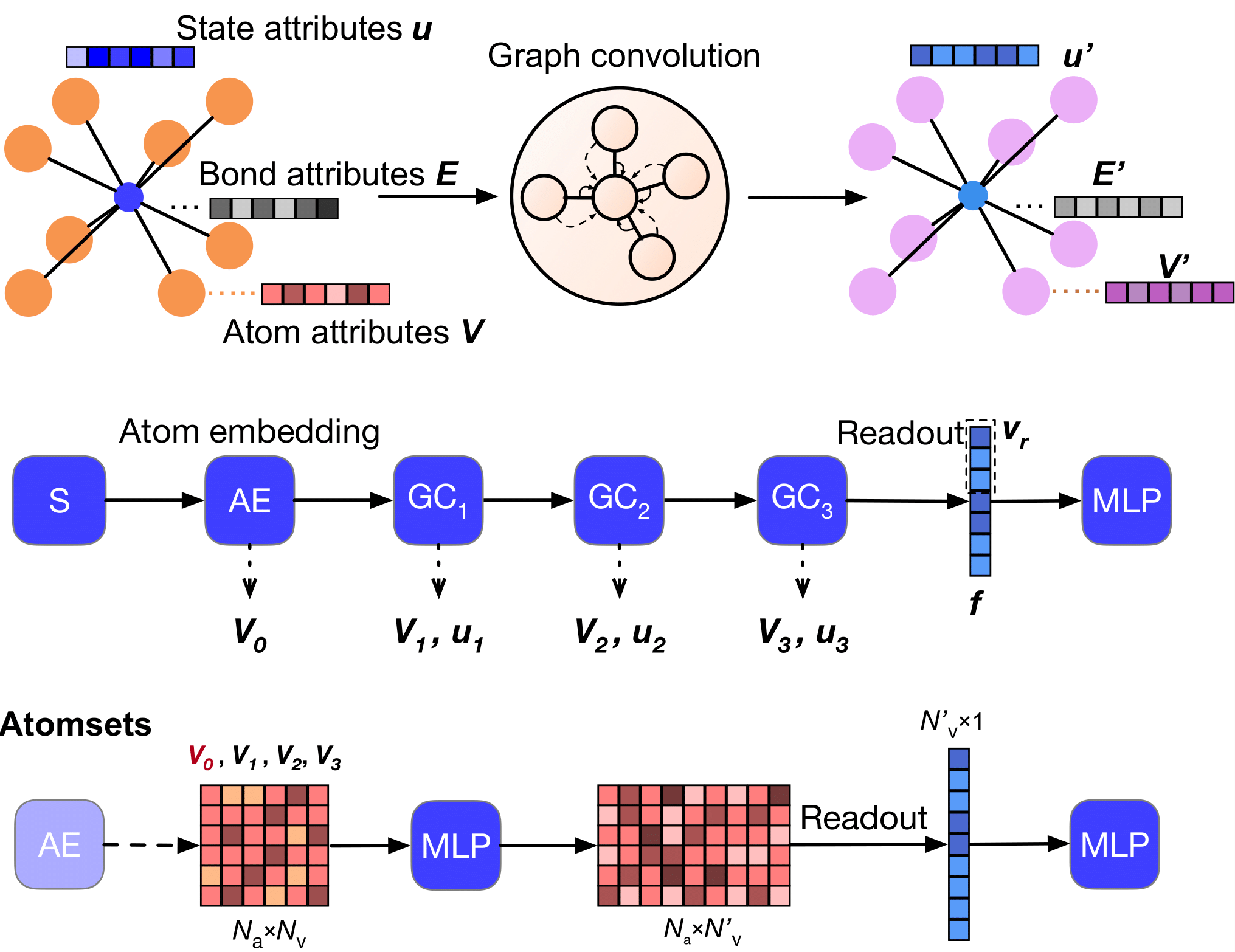}
\caption{\textcolor{VividPurple}{Atomset Formulation Building-Block and Graph Convolution Neural Network (GCNN) Architecture} }
\label{Atomset-fig-1}
\end{figure}

The input to the graph convolutional neural network architecture has atom (V), bond (E), and state (u) attributes. The output is a vectorized structure graph that evolves with the training dataset and is passed to the embedding layer to serialize the vectors to further feed to convolution or multi-layer perceptron or recurrent neural network layers for property prediction as shown in \cref{Atomset-fig-1}. For the graph convolutional layer of $i$, the atom, bond, and state features in the graph convolution are updated as, 

\begin{equation}\label{eq-Atomset-1}
\mathbf{e}_{k}^{(i)}=\phi_{e}\left[\mathbf{e}_{k}^{(i-1)}, \mathbf{v}_{S_{k}}^{(i-1)}, \mathbf{v}_{r_{k}}^{(i-1)}, \mathbf{u}^{(i-1)}\right]
\end{equation}

Where in the layer $i$, bond attributes $\mathbf{e}_{k}^{(i)}$ of the bond $k$, and atom attributes $\mathbf{v}_{j}^{(i)}$ of the atom $j$. Atom connected through bond $k$, sending message indices is $s_{k}$ and receiving message indices is $r_{k}$ to update the state information for the atom, bond, and state features through the
bond update functions $\phi_e$ which is approximated using a multi-layer perceptrons network to learn and train.

\begin{equation}\label{eq-Atomset-2}
\mathbf{v}_{j}^{(i)}=\phi_{v}\left[\mathbf{v}_{j}^{(i-1)}, \mathbf{v}_{k \in \mathcal{N}(j)}^{(i-1)}, \mathbf{e}_{l, r_{l}=j}^{(i)}, \mathbf{u}^{(i-1)}\right]
\end{equation}

Where atom $j$ neighbor atom index is $\mathcal{N}(j)$ and atom $j$ connected through bonds $\mathbf{e}_{l, r_{l}=j}^{(i)}$ receive message from $r_{1}$ atom index $j$.

\begin{equation}\label{eq-Atomset-3}
\mathbf{u}^{(i)}=\phi_{u}\left[\frac{1}{N_{b}} \sum_{k} \mathbf{e}_{k}^{(i)}, \frac{1}{N_{a}} \sum_{j} \mathbf{v}_{j}^{(i)}, \mathbf{u}^{(i-1)}\right]
\end{equation}

As in the graph network, the vectorized structure graph size increases the feature matrices size. Varying numbers of permutationally invariant atoms are used to construct a statistical graph readout function to make the graph approach computationally efficient. The readout function by applying atomic-node/vertex and edge-bond attribute set embedded vector sets into one vector. Afterward, to generate the final output, the readout atomic-node/vertex vectors, edge-bond vectors, and state vectors are concatenated to further pass to multi-layer perceptrons. One approach uses a linear mean readout function along the atom number dimension that averages the feature vectors.

\begin{equation}\label{eq-Atomset-4}
\overline{\mathbf{x}}=\frac{\sum_{i} W_{i} \mathbf{x}_{i}}{\sum_{i} W_{i}}
\end{equation}

\begin{figure}[H]
\centering
\includegraphics[scale=0.9]{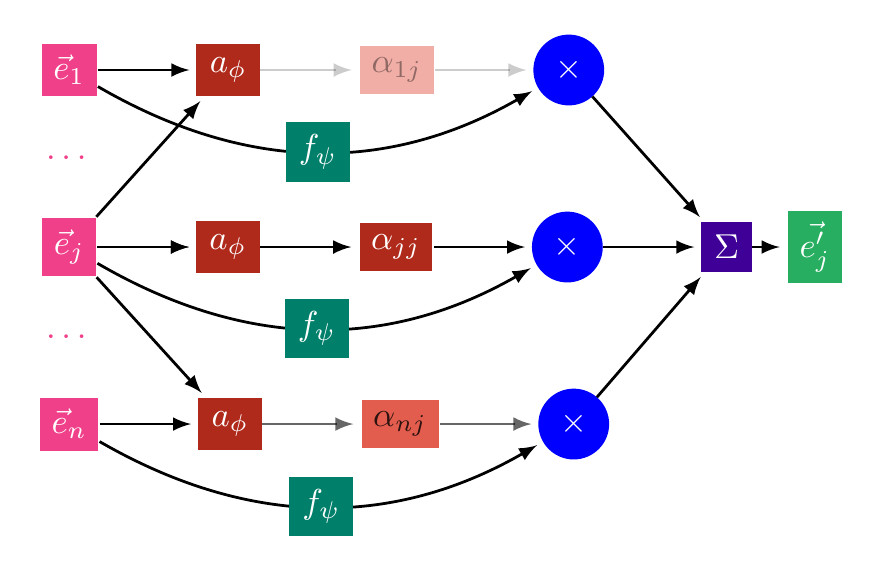}
\caption{\textcolor{VividPurple}{An Attention Mechanism}}
\label{fig-A14}
\end{figure}

Where for atom $i$, $w_{i}$ is the weight, and $\mathbf{x}_{i}$ is the feature row vector. Another approach is the weight-modified attention-based order-invariant sequence-to-sequence (seq2seq) readout function as shown in \cref{fig-A14}. \cite{Sutskever2014Sep} 

The vectorized structure graph memory feature vectors define as, \cite{https://doi.org/10.48550/arxiv.1511.06391}

\begin{equation}\label{eq-Atomset-5}
\mathbf{m}_{i}=\mathbf{x}_{i} \mathbf{W}+\mathbf{b}
\end{equation}

Where learning weight $\mathbf{W}$, biase $\mathbf{b}$, and initialize at $\mathbf{q}_{0}{ }^{*}=\mathbf{0}$. We update the long short-term memory (LSTM) at time step $t$ as per the attention mechanisms,

\begin{equation}\label{eq-Atomset-6}
\begin{aligned}
&\mathbf{q}_{\mathbf{t}}=\operatorname{LSTM}\left(\mathbf{q}_{\mathbf{t}-\mathbf{1}}^{*}\right) \\
&e_{i, t}=\mathbf{m}_{\mathbf{i}} \cdot \mathbf{q}_{\mathbf{t}} \\
&a_{i, t}=\frac{w_{i} \exp \left(e_{i, t}\right)}{\sum_{j} w_{j} \exp \left(e_{j, t}\right)} \\
&\mathbf{r}_{\mathbf{t}}=\sum_{i} a_{i, t} \mathbf{m}_{\mathbf{i}} \\
&\mathbf{q}_{\mathbf{t}^{*}}=\mathbf{q}_{\mathbf{t}} \oplus \mathbf{r}_{\mathbf{t}}
\end{aligned}
\end{equation}

Compared to the linear mean graph readout function, which is faster, the weighted sequence-to-sequence (seq2seq) graph readout function in the three algorithmic steps trains model weights and is slower. More flexible and complex relationships between the input and output can be encoded, and the choice of graph readout is a trade-off between model accuracy and computational efficiency. We have used the \texttt{QM9} molecule dataset containing more than 130,000 molecules and more than thirteen different properties calculated on each molecule. We first construct the graph network model from the \texttt{QM9} dataset by reading the atomic structure position as graph element objects. To find the initial neighbor's graph edges, a crystal graph uses the atomic structure position with a user-defined cutoff distance. Gaussian basis centers distance and width used for the expansion of the graph network. Afterward, from the properties list of the \texttt{QM9} dataset, we identify the target property we have to train the model, which will evolve the graph network topology. We perform our training validation and test using the target property's 80 percent, 10 percent, and 10 percent split. For the multi-fidelity graph network, we have to define different fidelity from the training dataset as the different extrinsic states. For the crystalline material, the material graph network can also be extended by adding the physics law by the symmetry properties of the unit cell or the number of atoms in a Wigner Seitz cell to construct the structure vector of the graph. Incorporating it will be more physical for the intrinsic properties prediction by the material graph network. 
\section*{Message-Passing Neural Network (MPNN) Topology}

A directed message passing neural network (D-MPNN) is a deep learning algorithm in the recent past successfully used in predicting various single and multi-fidelity molecular optical, solvation, reaction transfer learning, toxicity properties, infrared spectra, and estimating uncertainty to identify new antibiotics in the chemical, biological and pharmaceutical area. \cite{stokes2020a, DBLP:journals/corr/abs-2002-03244, heid2022machine, chung2022group,  Withnall2020Dec} D-MPNN algorithm implemented in the open-source Chemprop python package at MIT, D-MPNN algorithm iteratively aggregates local chemical features to predict properties.\cite{chemprop} In the Deep learning algorithms, molecular SMILES strings representing bit-vectors are fed to perform subsequent neural network tensorial matrix operations to train the underline network topology. The SMILES strings are mapped into bit-vectors or integer-vectors using the fingerprint algorithm. The standard strategy in the fixed width deterministic fingerprinting algorithm is to break the SMILES string into substructure blocks and look for similarities to map into either bit one or, bit zero representation at a particular index of the fixed width bit-vector string. The other strategy is to count the substructure blocks in the SMILES string and make a bit- or integer-vector based on the periodicity of substructure count. Another strategy is to use the descriptor constructed based upon relevant local properties of the underline molecules to map the SMILES string. \cite{pub.1084684694} Nevertheless, out of many descriptors set, humans chose the relevant ones based on underline applicable physics or chemistry. However, from the vast availability of descriptors set, which descriptors are best suited for the properties predication goals are completely heuristic-in advance, challenging to suggest which one gives better performance in terms of training resource and network topology. In contrast, in graph-based directed message passing neural network topology, the SMILES molecules are mapped on the graph vertex and edges, learn the variable size bit-vector used to represent the molecule, and later train on the network to learn the molecule properties by mapping from bit-vector to properties. The advantage of the D-MPNN algorithm is that it allows for learning both bit-vector fingerprints and properties, and the bit-vectors are end-to-end, automatically generated, continuous featurization of atom bonds and edges, and fully differentiable leading to better prediction capabilities. \cite{doi:10.1021/acs.jcim.9b00237} The bit-vector remains constant in the fixed-size fingerprint approach and Morgan fingerprinting approach throughout the training cycle. However, in the message-passing algorithm approach, the bit-vector is flexible and can learn features from nearby nodes and edges in the graph representing nearby atoms and bonds in the molecules. The message passing algorithm decomposes the whole network topology into a message which passes on the feed-forward graph network nodes representing each atom and graph edges representing each bond in the molecules. \cite{zhou2020graph} However, in the message-passing algorithm approach, the bit-vector is flexible and can learn features from nearby nodes and edges in the graph representing nearby atoms and bonds in the molecules. The message passing algorithm decomposes the whole network topology into a message which passes on the feed-forward graph network nodes representing each atom and graph edges representing each bond in the molecules. Convolution operations are performed on each atom to learn the local feature or environment of the atoms in a molecule, learning information about its nearest neighbors. The subsequent convolution performs to learn about the second, third, and next-nearest neighbors. After inquiring about enough statistics from the neighbor atoms and updating the convolution operations on the bit-vector are aggregated to pass the atom and bond level representations to a single learn bit-vector that represents the entire molecule. Hence each molecule is represented by a separate message-passing neural network bit-vector, and in the multiple molecule input approach, those bit-vectors are concatenated. These molecule-level bit-vectors are passed further to the feed-forward perceptron neural network to predict the molecular properties through regression, classification, or multi-class regression tasks or passed to the variational autoencoder network to generate the new SMILES strings. Also, we can extract the message passing algorithm-generated fingerprint bit-vectors and use it for other dimensionality reduction algorithms or with an ensemble of Morgan fingerprints to perform interpretability on the model based on substructures. There are hyperparameters in the message passing neural network approach related to the number of convolutions performed at the node, depth, the sizes of layers in the networks, and the total number of parameters in training. Also, data splitting performs randomly or on a scaffold or cross-validation basis with various loss functions to estimate. The advantage of the graph-based neural network approach is that the fingerprint bit-vector learns the local geometry in some higher dimensional space; however, the successive convolution operation makes the algorithm slow. A message-passing algorithm can overcome this. However, for a good representation of the predicted model, it needs more data than other neural network approaches. It also needs to learn the bit-vectors local environment from the dataset and infer the dataset properties. A more rigorous data splitting strategy such as scaffold balance is used where SMILES molecules are splitting based on their underlying scaffold. It imitates the families that the molecule belongs to, and hence test data have a high probability with a different scaffold from the training data set. We have found scaffold-balanced splitting gives the mean absolute error, and root mean square errors are higher for the test dataset; however, it gives a more reliable prediction model with the model's generalization. In the ensemble approach, different models run with the same underline network topology and identical hyperparameters with a different subset of training dataset but using different random initializations. Moreover, hence arrive at the different local minima after the training. Taking the average ensemble of these different members to predict the test dataset is more robust than an individual model, as many counter-intuitive effects are cancel-out. The other advantage of using an ensemble is that it can get the variance of predictions of that ensemble in addition to the mean. Furthermore, it can estimate the epistemic uncertainty in the prediction for the uncertainty quantification purpose in the trained model itself or based on the training dataset shape and size coverage. The variance of predictions is large when the prediction points are far away from the party line, emphasizing the model's uncertainty. We have also projected the high dimensional fingerprinting bit-vectors from the message-passing neural network into the lower dimension to understand which fingerprinting bit-vectors are critical for the model prediction and compare them with the RDKit based Morgan Fingerprints. We perform dimensionality reduction through principal component analysis (PCA) to reduce these three hundred thousand dimensional bit-vectors down to two dimensions for visualization purposes and color each data point according to its target property. In the principal component analysis, we have encountered the general tendency to move the top right corner towards the bottom left corner on the central axis in the PCA analysis. We have observed a monotomous shift in target property, which gives insight into how the model learns the high dimensional manifold properties and how it separates the space or maps onto a lower dimension to make compelling predictions. For the multi-fidelity dataset, we train the model on a cheap computation dataset for an extensive time and freeze the message-passing neural network hyperparameters. We only trained on the feed-forward perceptron network for the expensive experimental dataset. Therefore, only the feed-forward network parameter slightly shifted, ensuring the bulk trained message-passing neural network stays at its local minima's central limit. This approach can effectively train the data-hungry message-passing neural network for a relatively small experimental dataset without requiring an exotic fingerprinting algorithm based upon underlined physical or chemical properties.

\subsection*{Message Passing Neural Networks}

\begin{figure}[H]
\centering
\includegraphics[scale=0.7]{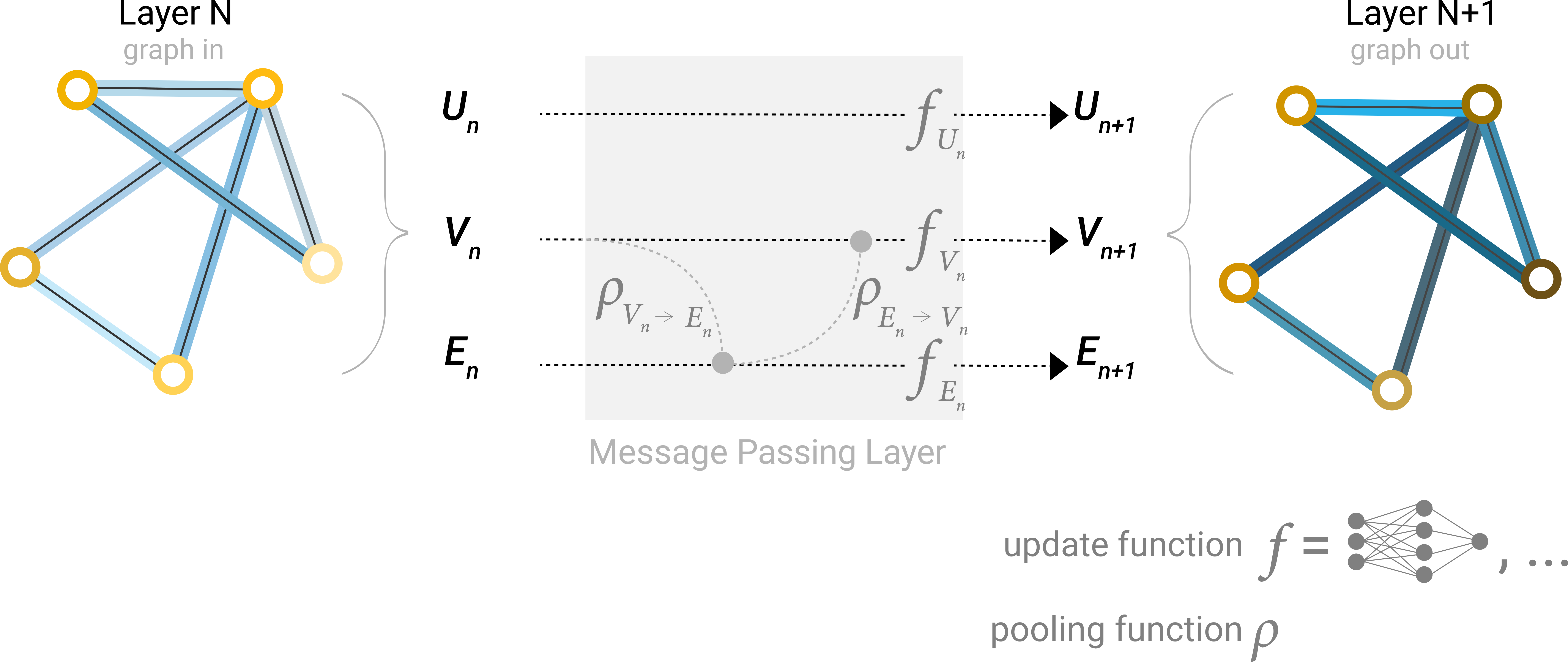}
\caption{\textcolor{VividPurple}{Message-Passing Neural Network Topology}}
\label{fig-DMPNN-01}
\end{figure}

The Graph convolution neural networks (GCNN) were embraced to estimate molecular properties prediction in the MatErials Graph Network (MEGNet) by exploiting the convolution operation. Further, the message-passing neural network (MPNN) was proposed to predict target property in a dataset using the Graph neural networks (GNN) as a backbone and adopted for the message-passing paradigm as shown in \cref{fig-DMPNN-01}. \cite{li2015gated} This neural network topology further extended to the Directed message-passing neural network (D-MPNN) by updating representations of directed bonds rather than atoms. Furthermore, it also improves by combining the computed molecule-level features algorithm with the learned molecular representation of the message-passing neural network. 

D-MPNN is originally called \texttt{structure2vec} to differentiate it from the generic MPNN architecture. MPNN architecture is an undirected graph $G$ that operates on the atomic nodes, or vertex features  $x_v$, and bond edge features $e_{vw}$. The graph $G$ evolves in two phases; at first, message-passing phase information is transmitted across all the molecules to build a neural representation. Next, the second readout-phase final representation of the molecule predicts target properties. \cite{dai2016discriminative} In total $T$ steps of message-passing phase, on every $t$ step update each vertex $v$ hidden variable states $h_v^t$ through vertex update function $U_t$, and associated messages $m_v^t$ using message update function $M_t$ as follow,

\begin{equation}\label{MPNN-eq1}
\begin{aligned}
m_v^{t+1} &= \sum_{w \in N(v)} M_t(h_v^t, h_w^t, e_{vw}) \\
h_v^{t+1} &= U_t(h_v^t, m_v^{t+1})
\end{aligned}
\end{equation}

Where on an undirected graph, $G$ set of neighbors of $v$ is $N(v)$, and $h_v^0$ is some function of the initial atom features $x_v$. The property prediction performed in the readout phase is based on the final hidden states through a readout function $R$ as, 

\begin{equation}\label{MPNN-eq2}
\hat{y} = R(\{h_v^T | v \in G\})
\end{equation}

For the single property prediction task, output $\hat{y}$ is scalar, and in MPNN multitask configuration, $\hat{y}$ is a vector to predict multiple properties. Molecular graphs trained to predict each molecule and loss function computed based on ground truth values and predicted outputs. The end-to-end entire model train in the message-passing and readout phases as the gradient of the loss function backpropagating. 

\subsection*{Directed Message Passing Neural Networks}
\begin{figure}[H]
\centering
\includegraphics[scale=0.35]{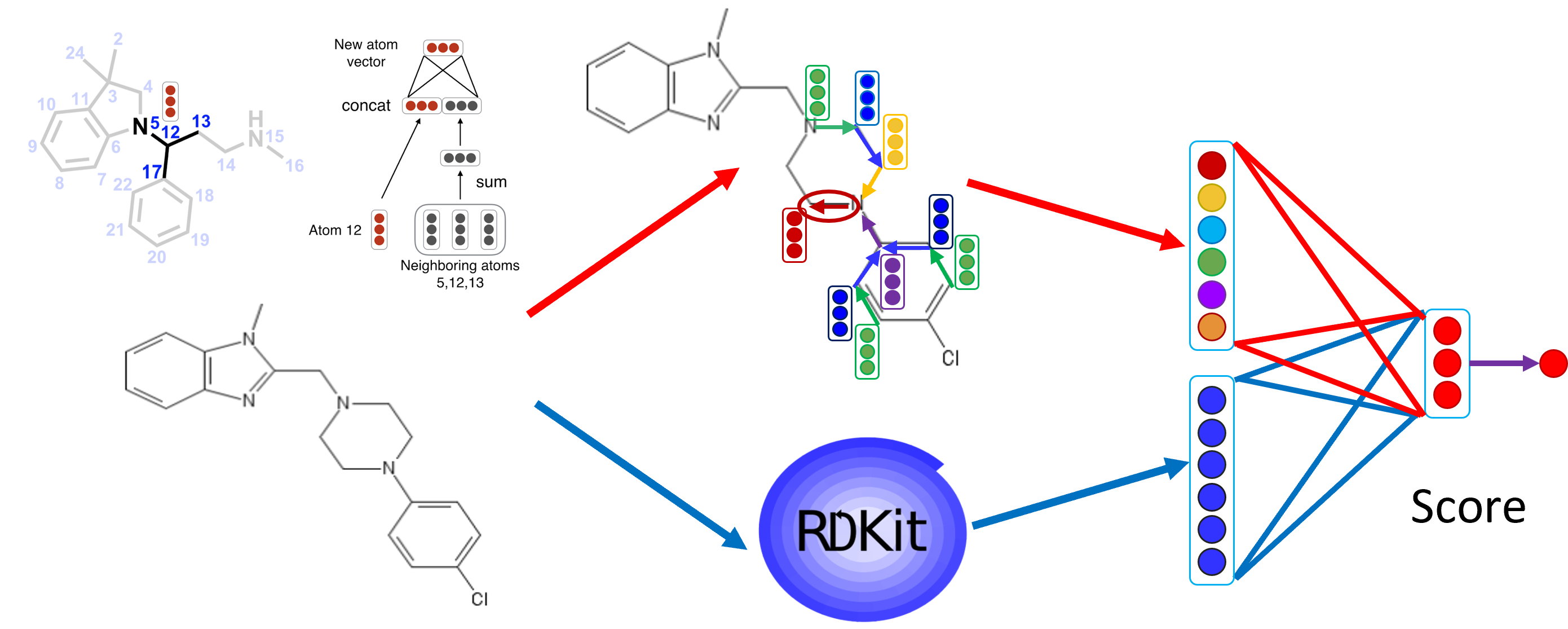}
\caption{\textcolor{VividPurple}{A Directed Message Passing Neural Network (D-MPNN) for Material Design}}
\label{fig-DMPNN-02}
\end{figure}

In the generic MPNN topology, messages sent during the message-passing phase update the atomic node or vertices; in comparison, the Directed MPNN (D-MPNN) message passes to directed bond edges. In the generic MPNN topology, messages sent during the message-passing phase update the atomic node or vertices; in comparison, the Directed MPNN (D-MPNN) message passes to directed bond edges. Directed MPNN designed to overcome the totters excursions that introduce noise into the graph topology as shown in \cref{fig-DMPNN-02}. Directed MPNN avoid messages to passed along a path which form $v_1v_2 \cdots v_n$, for some $i$ and  $v_i=v_{i+2}$. \cite{mahe2004extensions} In the D-MPNN algorithm, as shown in the \cref{fig-DMPNN-01} in the following iteration message will be parsed $1\rightarrow2$ to only atomic nodes/vertex $3$, and $4$ in comparison to the MPNN where the message also propagates through node $1$ and create an unnecessary loop in trajectory. 

This approach is inspired by the probabilistic graphical models' belief propagation algorithm in the atom-based message-passing paradigm for constructing fingerprints. The underlying assumption is that using directed bonds will decrease symmetry and improve network topology performance. \cite{koller2009probabilistic,coughlan2009tutorial}. The bond-edge hidden states $h_{vw}^0$ are initialized to parse the directed message-passing for a learning matrix $W_i \in \mathbb{R}^{h \times h_i}$ as,

\begin{equation}\label{DMPNN-eq1}
h_{vw}^0 = \tau(W_i\ {\tt cat}(x_v, e_{vw}))
\end{equation}

Where ReLU activation function $\tau$, \cite{nair2010relu} and $x_v$ atom features concatenation for atom $v$, and bond features concatenation $e_{vw}$ for bond $vw$, is ${\tt cat}(x_v, e_{vw}) \in \mathbb{R}^{h_i}$. The MPNN algorithm work on atomic node-based hidden states $h_v^t$ and messages $m_v^t$, whereas D-MPNN operates on hidden states $h_{vw}^t$ and messages $m_{vw}^t$ and direction information included as $h_{vw}^t$ and $m_{vw}^t$ are in the distinct direction $h_{wv}^t$ and $m_{wv}^t$. The Directed message-passing equations update the next cycle hidden states $h_{vw}^{t+1}$, and message $m_{vw}^{t+1}$ for $t \in \{1, \dots, T\}$ as,

\begin{equation}\label{DMPNN-eq2}
\begin{aligned}
m_{vw}^{t+1} &= \sum_{k \in \{N(v) \setminus w\}} h_{kv}^t = \sum_{k \in \{N(v) \setminus w\}} M_t(x_v, x_k, h_{kv}^t) \\
h_{vw}^{t+1} &= \tau(h_{vw}^0 + W_m m_{vw}^{t+1}) = U_t(h_{vw}^t, m_{vw}^{t+1})
\end{aligned}
\end{equation}

Where edge update function $U_t$, message passing functions $M_t$, $M_t(x_v, x_w, h_{vw}^t) = h_{vw}^t$, and $m_{vw}^{t+1}$ is independent of previous iteration reverse direction message $m_{wv}^t$. On every step at the same neural network, edge update function $U_t$ defines as,

\begin{equation}\label{DMPNN-eq3}
U_t(h_{vw}^t, m_{vw}^{t+1}) = U(h_{vw}^t, m_{vw}^{t+1}) = \tau(h_{vw}^0 + W_m m_{vw}^{t+1})
\end{equation}

Where learning a matrix of hidden size $h$ is $W_m \in \mathbb{R}^{h \times h}$, and for the edge, the inclusion of $h_{vw}^0$ provides a skip connection at every step to the original MPNN feature vector. Next, computing the summation of incoming bond features to return to an atomic node/vertex representation of the molecule for a learning matrix $W_a \in \mathbb{R}^{h \times h}$ as,

\begin{equation}\label{DMPNN-eq4}
\begin{aligned}
m_v = \sum_{k \in N(v)} h_{kv}^T = \sum_{w \in N(v)} h_{vw}^T\\
h_v = \tau(W_a {\tt cat}(x_v, m_v)) = \tau(W_a {\tt cat}(x_v, m_v))
\end{aligned}
\end{equation}

The readout phase of MPNN and D-MPNN is the same, and the molecule's feature vector is obtained by summing over the atomic node hidden states in the readout function $R$ as,

\begin{equation}\label{DMPNN-eq5}
h = \sum_{v \in G} h_v.
\end{equation}

The target property predicted $\hat{y} = f(h)$ by feed-forward neural network through $f(\cdot)$.
\section*{Active Learning and Bayesian Optimization Algorithm}
\begin{figure}[H]
\centering
\includegraphics[scale=0.6]{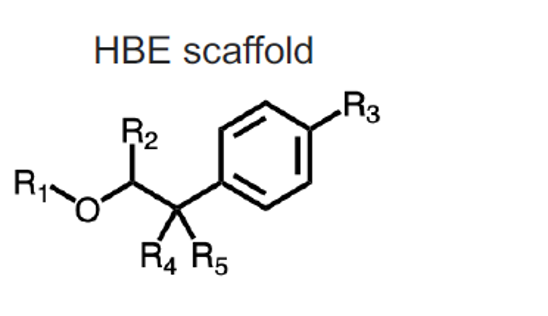}
\caption{\textcolor{VividPurple}{Homo-Benzylic Ethers (HBE) Scaffold}}
\label{bayesopt-fig-2}
\end{figure}

The design and discovery of novel materials with improved physical, chemical, optical, magnetic, and thermal properties is a combinatorial search problem from the elemental space of the periodic table. The natural science community faces a monumental challenge in performing intractable high-accuracy quantum simulations and high-throughput experiments on this colossal search space. Due to the vast statistical nature of this problem, one potential statistical solution explores in the form of employing the active learning (AL) technique. The active learning approach is from the semi-supervised deep learning (DL) class systematically exploring the colossal search space with the least evaluation of evolutionary candidates. The collective use of active learning with the addition of the ``a-prior'' distribution from the Bayesian optimization (BO) approach can expedite the ideal molecular search for the demanded desired properties such as pharmaceutical molecular search and screening, clean energy, electrode-electrolyte interfaces reactions, and its storage application in multivalent battery for sustainable development in the era of climate change. We use the molecular dataset's  SMILES. First, we generate the desired features by enoding and performing the dimensionality reduction. After the preprocessing, we drive the Bayesian optimization cycles, quantify the uncertainty through the Gaussian process regression model, and evaluate an Acquisition function. This  Acquisition function will query the search space and select the novel possible candidates, which we use to tune the model performance. In rechargeable batteries such as lithium-ion, charge-carrying components reside inside the electrodes, unlike the redox flow batteries. The advantage is charge-carrying components, the redox-active molecules or redox fluid dissolved within the electrolyte and stored in a separate external tank. It will separate the energy and power capacity, which can be enhanced by increasing the storage size without redesigning the power converting unit. This silent feature alone makes redox flow batteries a promising stationary energy storage system for integrating intermittent systems between renewable energy sources and the electrical grid. The non-aqueous Redox flow battery (RFB) has advantages for integrating intermittent renewable energy sources into the electrical grid. It separates the energy and power capacity and potentially yields high energy density at a lower cost. The non-aqueous redox flow batteries have active organic materials, which provide higher energy density. The central challenge associated with the operation of redox flow batteries is the electrolyte-electrode interface during the various charge-discharge cycles. The charge-carrying components reactivate soluble and potentially passivate electrodes, reducing the flow batteries' lifetime. One sustainable solution to this problem is to engineer recyclable organic molecules as charge-carrying components, such as Homobenzylic ethers (HBE). Homobenzylic ethers (HBE) are a desirable molecular scaffold in the redox flow battery for the mesolytic cleavage motif as shown in \cref{bayesopt-fig-2}. HBE should have suitable redox potential windows to operate, high solubility, ease of synthesis, and electrochemical reversibility. We first find Homobenzylic ethers molecules with a suitable redox potential window to realize this.

Table S4 List of functional groups and their corresponding scaffold sites used for populating the HBE database of 112,000 entries. Substituent groups are Ph: phenyl, Et: ethyl, OMe: methoxy, iPr: isopropyl, OPr: propoxy, EtOMe: methoxyethyl, and Pr: propyl.

\begin{table}[H]
\resizebox{\textwidth}{!}{%
\begin{tabular}{|l|l|}
\hline Scaffold Site & Functional group \\
\hline R1 & $-\mathrm{Me},-\mathrm{Et},-\mathrm{Pr},-\mathrm{Ph},-\mathrm{CN},-\mathrm{Eth},-\mathrm{COMe},-\mathrm{C}(\mathrm{Me}) \mathrm{Me},-\mathrm{CCOMe}$ \\
\hline R2 & $-\mathrm{N}(\mathrm{Me})_{3}^{+},-\mathrm{COMe},-\mathrm{Et},-\mathrm{OCMe}, \mathrm{N}(\mathrm{Me})_{2},-\mathrm{NO}_{2},-\mathrm{C}(=\mathrm{O}),-\mathrm{Pr},-\mathrm{-C(Me)Me}, -\mathrm{-CCOMe}, -\mathrm{-C(Me)OMe}$ \\
\hline R3 & $-\mathrm{N}(\mathrm{Me})_{3}^{+},-\mathrm{Me},-\mathrm{COMe},-\mathrm{Br},-\mathrm{C}(=\mathrm{O}),-\mathrm{OEth},-\mathrm{Pr},-\mathrm{-C(Me)Me},-\mathrm{C}(\mathrm{COMe}),- \mathrm{C}(\mathrm{Me})(\mathrm{OMe})$ \\
\hline R4-R5 & $-\mathrm{N}(\mathrm{Me})_{3}^{+},-\mathrm{OMe},-\mathrm{COMe},-\mathrm{Br},-\mathrm{N}(\mathrm{Me})\mathrm{Me},-\mathrm{Et},-\mathrm{OEt},-\mathrm{NO}_{2},-\mathrm{C}(=\mathrm{O}),-\mathrm{Pr},-\mathrm{C}(\mathrm{Me})\mathrm{Me},-\mathrm{C}(\mathrm{COMe}),-\mathrm{C}(\mathrm{Me})(\mathrm{OMe})$ \\
\hline
\end{tabular}}
\end{table}

There is ample space for molecular candidates for desirable Homobenzylic ethers in the scaffold. These are different substituents in the scaffold from R1 to R5 to apply, and we can replace them with a different functional group in the SMILES string. These possibilities will explode the molecule's space to a $ 10^5$ combination, making it expensive and time-consuming to synthesize and characterize the Oxidation potential. We can speed up the process by performing Density Functional Theory (DFT) calculations starting from the SMILES string and mapping it into a 3-D geometry feature through a chem-informatics RDKit package.  \cite{greg_landrum_2022_6483170} We can generate a variety of chemicals, and physical features from the SMILES string through the RDKit package, such as molecular weight, number of carbon atoms, and aromatic rings in the molecule. We can generate 125 features, so a vector of 125 components can represent each molecule represented by the SMILES string.
Moreover, principal component analysis is used to identify the critical feature selection. Furthermore, geometry optimization thorough DFT Gaussian quantum chemistry package and calculate Gibbs free energy and oxidation potential energy. However, for the search space of $ 10^5$, it is still a var time consuming even for high-throughput DFT on the cloud-scale cyber-infrastructure. In the conventional supervised learning scheme, we can take a small random sample from the dataset, perform DFT calculation, use it for a regression model based on a particular feature, and then predict the unknown test dataset. However, potential problems arrive with the extrapolation of the few training datasets. Another active learning approach to enhance prediction accuracy is to train machine learning models using a few DFT-computed data and estimate the oxidation potential energy. We need agent-based active learning models that make accurate predictions after training to achieve high prediction accuracy. However, we also need machine learning models that guide training data selection. \cite{mitchell_statistical_1969} The Gaussian process regression (GPR) surrogate model and agent-based acquisition function implement this active learning scheme. The Gaussian process regression surrogate model predicts the properties of the output based upon the input feature similarities or differences, and hence it also predicts the uncertainty in the prediction. The similarities and differences between data points calculate using kernel functions, which measure the feature distances between these data points. These distances quantify as a square exponential or radial basis function used in a Gaussian process regression,

\begin{equation}\label{bayesopt-eq-1}
\mathrm{K}\left(x_{1}, x_{2}\right)=\exp \left[-\frac{1}{2} \frac{\left(x_{1}-x_{2}\right)^{2}}{l^{2}}\right]
\end{equation}

Gaussian process regression will deliver the normal distribution of similarities and differences with its mean and standard deviation, which quantifies the uncertainty of Gaussian process regression prediction. It underlines the uncertainty of Gaussian process regression exploited to identify the next point to query where the uncertainty is maximum. Hence, we have the least confident about the prediction at that point, and in this way, we have more knowledge about the underline distribution through the successive query. It is the Bayesian optimization incorporated active learning scheme. The underlying knowledge can also be explored through various acquisition functions, including exploration and exploitation strategies. The acquisition function formulates an optimal strategy toward an objective by guiding the subsequent evaluation through Bayesian optimization. The acquisition function uses the prediction produced by the Gaussian process regression surrogate model to guide the selection of the following candidate to query from the data. The acquisition function can quantify the uncertainty in the prediction by Upper Confidence Bound (UCB), Probability of improvement (PI), and Expected improvement (EI) matrix. The Upper Confidence Bound (UCB) acquisition function defines as,

\begin{equation}\label{bayesopt-eq-2}
\begin{aligned}
UCB(x) &=\mu(x)+\xi * \sigma(x) \\
x_{\text{next}} &=\operatorname{argmax}(UCB(x))
\end{aligned}
\end{equation}

Where $\mu(x)$ is mean $\sigma(x)$ and standard deviation of Gaussian process predication. The Probability of improvement (PI) acquisition function defines as,

\begin{equation}\label{bayesopt-eq-3}
\begin{aligned}
P I(x)&=P\left(y_{x}^{\prime} \geq y_{\text{current~Best}}+\xi\right) =\Phi(Z) \\
\mathrm{Z}&=\frac{\mu(x)-y_{\text{current~Best}}-\xi}{\sigma(x)} \\
x_{\text{next}}&=\operatorname{argmax}(P I(x))
\end{aligned}
\end{equation}

Where $\Phi(Z)$  is the cumulative distribution function. The Expected improvement (EI)  acquisition function defines as,

\begin{equation}
\begin{aligned}\label{bayesopt-eq-4}
E I(x)&= \left\{\begin{array}{l}
\left(\mu(x)-y_{\text{current~Best}}-\xi\right) \Phi(Z)+\sigma(x) \phi(x), \sigma(x)>0 \\
0, \quad \sigma(x)=0
\end{array} \right. \\
x_{\text {next }}&=\operatorname{argmax}(E I(x))
\end{aligned}
\end{equation}

Where $\phi(x)$ is probability density function.

These acquisition functions define fitness score as it chooses the current best value and guides the next candidate's selection. The next candidate has the highest probability of improvement over the current best value. Furthermore, this improvement would also be maximized. In this way, starting with a small number of DFT calculated oxidation potentials, we can query the molecule space and arrive at a molecule for which the network expects maximum oxidation potential. We can perform an actual DFT calculation to verify the network claim on that particular molecule. After performing the DFT calculation, we have the new oxidation potential data. We added the starting small number of molecular lists and re-ran the active learning network for improvement. We have also checked the performance of our Bayesian optimization using the known dataset for the oxidation potential and tuned our model based on the existing DFT oxidation potential dataset. From the principal components analysis calculation, some of those descriptors that have dominating principal components, e.g., total polar surface area number of benzene rings, play a significant role and affect the oxidation potential of a molecule.

\subsection*{Gaussian Process Regression (GPR)}

An infinite-dimensional multivariate Gaussian distribution is a Gaussian process with any dataset label collection being jointly Gaussian distributed. Gaussian process regression (GPR) is Bayesian-based non-parametric regression and is not limited by a functional form and is helpful for small datasets by providing uncertainty measurements on the predictions. Gaussian process regression estimates the probability distribution over all admissible functions that may fit the dataset instead of estimating a specific function parameters probability distribution. Unlike supervised machine learning algorithms, the Bayesian technique infers probability distribution over all possible values and does not learn exact values for all the parameters in a linear fitting function as,

\begin{equation}\label{Debug-eq-1}
y=w x+\epsilon
\end{equation}

We incorporate prior knowledge about the space of functions by selecting the mean and covariance functions in the Gaussian process prior. By using the Bayes's rule, for a specified prior distribution on the function space $ p(w) $  of a parameter $ w $, and based on evidence observing the relocating probabilities as,

\begin{equation}\label{Debug-eq-3}
\text { posterior }=\frac{\text { likelihood } \times \text { prior }}{\text { marginal likelihood }}
\end{equation}

\begin{equation}\label{Debug-eq-2}
p(w \mid y, X)=\frac{p(y \mid X, w) p(w)}{p(y \mid X)}
\end{equation}

Where the posterior distribution update using the training data is $ p(w \mid y, X) $ containing the information for both a-prior distribution and training dataset. A Gaussian process prior with mean function m(x), and covariance function $ k\left(x, x^{\prime}\right) $ as, 

\begin{equation}\label{Debug-eq-5}
f(x) \sim G P\left(m(x), k\left(x, x^{\prime}\right)\right)
\end{equation}

To estimate predictive posterior distribution $ f^{*} $ on the unseen test points observation, $ x^{*} $ predictions are estimated by weighting all possible predictions by their estimated posterior distribution,

\begin{equation}\label{Debug-eq-4}
p\left(f^{*} \mid x^{*}, y, X\right)=\int_{w} p\left(f^{*} \mid x^{*}, w\right) p(w \mid y, X) d w
\end{equation}

We considered the Gaussian prior and likelihood for the tractable integration and solved for the predictive distribution $ f^{*} $ and the uncertainty quantification estimated through point prediction mean and variance value. Independently, identically distributed (i.i.d) Gaussian noise $\epsilon$ $\sim N\left(0, \sigma^{2}\right)$ also incorporated to the labels by summing the label distribution and noise distribution. The Gaussian process regression was implemented on several libraries sci-kit-learn Gaussian process package, Gpytorch, GPy. \cite{Gardner2018Sep} By definition of Gaussian process prior, training and test data points are joint multivariate Gaussian distributed and from the distribution,

\begin{equation}\label{Debug-eq-6}
\left[\begin{array}{c}
y \\
f_{*}
\end{array}\right] \sim \mathscr{N}\left(\left[\begin{array}{c}
\mu \\
\mu_{*}
\end{array}\right],\left[\begin{array}{cc}
K(X, X)+\sigma_{n}^{2} I & K\left(X, X_{*}\right) \\
K\left(X_{*}, X\right) & K\left(X_{*}, X_{*}\right)
\end{array}\right]\right)
\end{equation}

Where covariance kernel matrix $K$ elements evaluated at observations for the corresponding covariance function. In this way, the training subset is used for model selection. During model selection, the mean function and covariance kernel function are tuned. The mean function is the mean of the training dataset and is typically constant or zero value. The kernel function can take any form, such as constant kernel, linear kernel, Radial basis function kernel, square exponential kernel, and Matern kernel. However, it must follow the semi-positive definite and symmetric properties of a kernel. The radial basis function (RBF) kernel is the composition of a constant kernel with a radial basis function. The advantage of the radial basis function is that it encodes the smoothness of the function, Hence used for inputs and outputs space similarity corresponds as, 

\begin{equation}\label{Debug-eq-7}
k\left(x, x^{\prime}\right)=\sigma_{f}^{2} \exp \left(-\frac{1}{2 \ell^{2}}\left\|x-x^{\prime}\right\|^{2}\right)
\end{equation}

Where two hyperparameters the lengthscale $ \ell $, and signal variance $ \sigma^{2} $. The hyperparameters' initial value and bounds are chosen from the sci-kit-learn library. Labels independently, identically distributed (i.i.d) noise variance $ \alpha$, and normalized distance $y$ tuned for a constant mean function; it is zero for false and true for training data mean value. For covariance kernel function, another approach of hyperparameter tuning is to maximize the logarithmic marginal likelihood of the training data. Multiple optimizers with different initializations restart approaches are used if the logarithmic marginal likelihood is not convex. Also, a gradient-based optimizer improves efficiency. Conditioned data and test observation estimate the predictive posterior distribution, and because of the Gaussian process prior, The predictive posterior is tractable with normal distribution described by the mean $ \bar{f}^{*} $ and variances from the diagonal of the covariance matrix $ \Sigma^{*} $ by inverting the K-matrix as, 

\begin{equation}\label{Debug-eq-8}
\begin{aligned}
f^{*} \mid X, y, X^{*} &\sim \mathscr{N}\left(\bar{f}^{*}, \Sigma^{*}\right) \\
\bar{f}^{*}&=\mu^{*}+K\left(X^{*}, X\right)\left[K(X, X)+\sigma_{n}^{2} I\right]^{-1}(y-\mu) \\
\Sigma^{*}&=K\left(X^{*}, X^{*}\right)-K\left(X^{*}, X\right)\left[K(X, X)+\sigma_{n}^{2} I\right]^{-1} K\left(X, X^{*}\right)
\end{aligned}
\end{equation}

The K-matrix inversion scales cubically with the number of training points. The Gaussian process regression prediction function gives the inference on the dataset. Next, regression model performance on test observations calculates through the predictions mean squared error (MSE).

Gaussian process regression is a distribution over functions that collect random variables with a joint Gaussian distribution. In the non-parametric supervised learning regime, Gaussian process regression quantified the uncertainty for a random variable $X \sim \mathcal{N}(\bm{\mu},\bm{\Sigma})$ with a mean of $\bm{\mu}$, in the Gaussian distribution define as,

\begin{equation}\label{bayesopt-eq-4-1}
P(\bm{x} ; \bm{\mu}, \bm{\Sigma})=\frac{1}{(2 \pi)^{\frac{d}{2}}|\bm{\Sigma}|} exp^{-\frac{1}{2}\left((\bm{x}-\bm{\mu})^{\top} \bm{\Sigma}^{-1}(\bm{x}-\bm{\mu})\right)}
\end{equation}

Where $d$ data dimension, and $\bm{\Sigma}$ is covariance matrix. By constraining the probability distribution at training points, the Gaussian process regression, instead of learning parameters $\bm{W}$, the learning function inference over the test points by constructing the covariance matrix and effective for small datasets. From the data distribution of training points set $\overline{\bm{X}}$, training labels set $\bm{y}$, and test points set $\overline{\bm{X}_*}$, the prediction functions ${y_*}$ for a noise-free case is,

\begin{equation}\label{bayesopt-eq-4-2}
\begin{aligned}
& {y}_{*} \sim \mathcal{N}( \bm{K}\left(\overline{\bm{X}_{*}}, \overline{\bm{X}}\right) \bm{K}(\overline{\bm{X}}, \overline{\bm{X}})^{-1} \bm{y}, \\
& \left.\bm{K}\left(\overline{\bm{X}_{*}}, \overline{\bm{X}_{*}}\right)-\bm{K}\left(\overline{\bm{X}_{*}}, \overline{\bm{X}}\right) \bm{K}(\overline{\bm{X}}, \overline{\bm{X}})^{-1} \bm{K}\left(\overline{\bm{X}}, \overline{\bm{X}_{*}}\right)\right)\,,
\end{aligned}
\end{equation}

Where kernel function $k$ such as squared exponential or periodic kernels defines the covariance between two points in dataset as $\bm{K}_{i,j} = k(\bm{x}_i,\bm{x}_j)$. The selection of $k$ depends upon the dataset statistics and an important design decision in Gaussian process regression. However, for the larger datasets, Gaussian process regression is a computationally expensive algorithm as the inference step involves matrix inversion operation and which scales as $\mathcal{O}(N^3)$. Inference approximation techniques to increase the scalability and applicability of the Gaussian process regression are actively explored. For the training-set labeled property $\mathrm{y}$, feature matrix $\bm{X}$, and test-set feature matrix $\bm{X}_{*}$, the predicted mean $\boldsymbol{\mu} *$, and variance $\boldsymbol{\Sigma}_{*}$ is,

\begin{equation}\label{bayesopt-eq-5}
\begin{aligned}
\boldsymbol{\mu}_{*}&=\bm{K}_{*}^{\mathrm{T}} \bm{K}_{\mathrm{y}}^{-1} \bm{y} \\
\boldsymbol{\Sigma}_{*}&=\bm{K}_{* *}-\bm{K}_{*}^{\mathrm{T}} \bm{K}_{*}^{-1} \bm{K}_{*}
\end{aligned}
\end{equation}

Where $\mathrm{K}=k(\mathrm{X}, \mathrm{X}), \mathrm{K}_{*}=$ $k\left(\bm{X}, \bm{X}_{*}\right)$, and $\bm{K}_{* *}=k\left(\bm{X}_{*}, \bm{X}_{*}\right)$. The covariance function $k\left(\bm{x}, \bm{x}^{\prime}\right)$ for the feature vectors $x$ and $x^{\prime}$ based on Matérn kernel $(\nu$ $=1.5)$ is,

\begin{equation}\label{bayesopt-eq-6}
k\left(\bm{x}, \bm{x}^{\prime}\right)=\left[1+\frac{\sqrt{3}\left\|\bm{x}-\bm{x}^{\prime}\right\|}{\sigma_{1}}\right] \exp \left(-\frac{\sqrt{3}\left\|\bm{x}-\bm{x}^{\prime}\right\|}{\sigma_{1}}\right)+\sigma_{\mathrm{n}}^{2}
\end{equation}

Where data length scale $\sigma_{1}$, and data expected noise level $\sigma_{\mathrm{n}}$ hyperparameters are determined during model training maximum likelihood estimation. Next, Gaussian process regression model's $f(x)$, predicted mean $\mu(\mathrm{x})$, and standard deviation $\sigma(\mathrm{x})$ used to computes the expected improvement (EI) acquisition function as, \cite{bassman2018active}

\begin{equation}\label{bayesopt-eq-7}
\begin{aligned}
&\mathrm{EI}(\bm{x})= \begin{cases}\left(\mu(\bm{x})-f\left(\bm{x}^{+}\right)\right) \Phi(Z)+\sigma(\bm{x}) \phi(Z) & \sigma(\bm{x})>0 \\
0 & \sigma(\bm{x})=0\end{cases} \\
&Z=\frac{\mu(\bm{x})-f\left(\bm{x}^{+}\right)-\epsilon}{\sigma(\bm{x})}
\end{aligned}
\end{equation}

Where $\mathrm{x}^{+}$ is molecules feature vector, the best estimate of a target property of molecules is $f\left(\mathrm{x}^{+}\right)$, the probability density function (PDF) $\phi(Z)$, and cumulative density function  (CDF) is $\Phi(Z)$. In the \cref{bayesopt-eq-7}, $\epsilon$ parameter determines the exploration during the optimization process. The trade-off between exploration and exploitation achieved an optimal yield at a constant value of $0.01$ in our calculation. We have used the Sci-kit-learn machine learning package inbuilt Gaussian process regression model. 
\section*{Evolutionary Algorithm-based Neural Network}

\begin{figure}[H]
\centering
\includegraphics[scale=0.7]{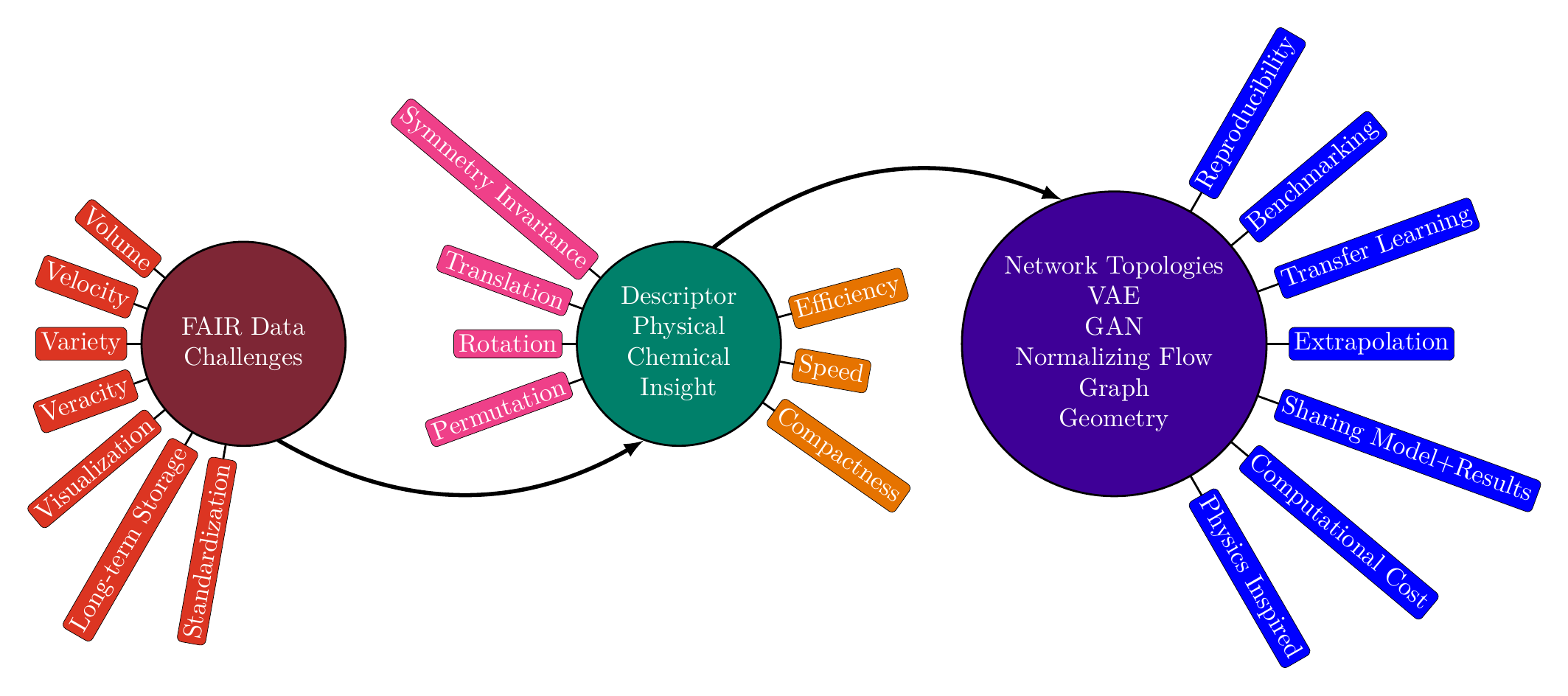}
\caption{\textcolor{VividPurple}{A FAIR Data Challenges, Descriptor, and Network Topologies}}
\label{fig-A18}
\end{figure}

We have implemented the genetic algorithm-based neural network topology for predicting the material properties in the data-scarce regime of material science. To enhance the prediction accuracy of the neural network in the deficient data regime of physical and material science. We have custom-defined and optimized the generic activation functions and weights as hyperparameter optimization. This neural network (NN) by utilizing a genetic algorithm (GA), which enriches the data statistics by employing its population selection, crossover, and objective mutation function as an evolutionary algorithm (EA).  \cite{turing1950i,fraser1957simulation} The infusion of the genetic algorithm into the neural network enhances the accurate prediction and balances the interpretability implemented through \texttt{DEAP} computation package. \cite{DEAP_JMLR2012} 
The advantage of incorporating genetic algorithms is that they do not require much data and simulate the behaviors of entities in a population. By augmenting the genetic algorithms in the neural network, we can generate more data and optimize the hyperparameter search space by random mutation. The Genetic algorithms will generate the many species of neural network and, out of this population of the network, generate the best network based on the fitness selection from the genetic tournament. Fitness factors will enhance the probability of a particular network reproducing more to optimize the hyperparameters for a neural network's best cost or loss function. Genetic algorithms' crossover and mutation functions will create many offspring networks with modified network weights and activation functions. In this hybrid approach, the network will learn better about its hyper-parameters space during the training phase. This hybrid approach can be used to learn the interpretable physical loss of material science. The coupled Parsimonious NN and GA network topology predict the material property more accurately as the network learns non-linear activation function better by interpreting the natural science empirical relations. \cite{desai2020parsimonious} The coupled Parsimonious NN and GA network implement the improved descriptor in the data-scarce regime. The neural network used to express the reactive force field in the molecular system, \cite{yoo2021neural} molecular dynamics simulations through the neural network active learning to predict the melting temperature of alloys. Furthermore, neural network-based interatomic models have recently been implemented on emerging exascale architectures. \cite{DESAI2022108156} The machine learning models have been used in various applications to perform image classification and segmentation, detect objects, text translation, play games, self-driving mobility, and unsupervised reinforcement learning robotics. In these selected application domains, network performance reaches and surpasses human capability. However, much of this superhuman performance arises from pretraining the network on the big datasets, e.g., millions of images in Imagenet, billions of text structures in GPT3, and thousands of hours of a game played by an agent in the reinforcement learning application. These custom-trained network models have been encouraging and significantly advancing the data science field in the past few years. However, all the state-of-the-art deep learning network algorithms failed when trained on an inadequate dataset. Nevertheless, such a massive collection of training data is not available in the natural science and engineering domain due to expansive experiments or the computational cumbersome and time-consuming numerical simulations from the physical laws of nature. Therefore, we are in a bottleneck situation where deep learning algorithms work well in certain areas. However, we do not have quality data and are in the data scare regime for the effective employment of these networks. There is a remarkable initiative to develop, manage, share, and stewardship the massive scientific data vault, e.g., findable, accessible, interoperable, and reusable (FAIR) principle framework as shown in \cref{fig-A18}. \cite{wilkinson2016the,hunt2021sim2ls} 

In this data deficient small data regime, to provide the confidence, physical interpretability, and discovering empirical relations from data to the predictive results of the information-centric network techniques, we have to evolve the existing network topologies. There are various methods in the hyperparameter search space domain to optimize the existing neural networks up to a certain extent. The inclusion of genetic algorithms in neural networks to balance interpretability and accuracy from a given dataset was recently proposed as a parsimonious neural network topology. Conventional neural network models make predictions of the output based on the inputs quantity of interest. They have multiple hidden layers in a deep network to achieve this, and each layer has copious neurons. However, all the neurons in each generic layer have identical non-linear activation functions and weight and bias values gained from the small training dataset. To enhance the predictability and confidence of the network by custom-defined activation function of each neuron in the same layer, we have a more extensive combination of activation functions to explore the search space deeply to discover the empirical relation from the statistical mechanics. Consequently, the genetic algorithm will determine each neuron's activation functions and weight values in each layer defined by a combination of network training data and fitness score from the coupled genetic algorithm, which favors the robustness of reproducibility. Therefore this combined fitness objective function for which the network predicts the output result is more interpretable and with a greater degree of confidence. The combined fitness objective function $ \mathcal{O.F.} $ for genetic optimization defines as,

\begin{equation}\label{parsi-eq1}
\mathcal{O.F.}=10{log}_{10}{\bigg[\mathcal{L}_{MSE}\bigg]}+p\bigg[\sum_{i=1}^{N_{N}} w_{i}^{2}+\sum_{j=1}^{N_{w}} \mathcal{F}_{GA}\left(w_{j}\right)\bigg]
\end{equation}

Where $ \mathcal{L}_{MSE} $ the conventional mean squared error on the test set of the neural network indicates the model performance and avoids overfitting in the training and testing backpropagation stage, the logarithmic of $ \mathcal{L}_{MSE} $ suppresses the wide range of errors in comparison to $ \mathcal{F}_{GA} $ the genetic algorithm fitness function. The second term in the equation maps $ i^{th}$ activation functions ``\texttt{linear}'', ``\texttt{ReLU}'', ``\texttt{tanh}'', and ``\texttt{ELU}'' to the corresponding complexity score of 0, 1, 2, and 3. Hence linear activation is favored over non-linear activation function, and $ p $ is a parsimony factor that gives a fitness score to individual networks balancing the interpretability and accuracy. The $ \mathcal{F}_{GA} $ defined as,

\begin{equation}\label{parsi-parsi-eq2}
\mathcal{F}_{GA}\left(w_{j}\right) =
\begin{cases}
0 & \text{if, $ w_{j}=0$}\\
1 & \text{if, $ w_{j}= \frac{1}{2},1,2,\frac{\Delta{t}}{2},\Delta{t},2\Delta{t}$}\\
2 & \text{if, $ w_{j}=$ trainable}
\end{cases}    
\end{equation}

\begin{figure}[H]
\centering
\includegraphics[scale=0.8]{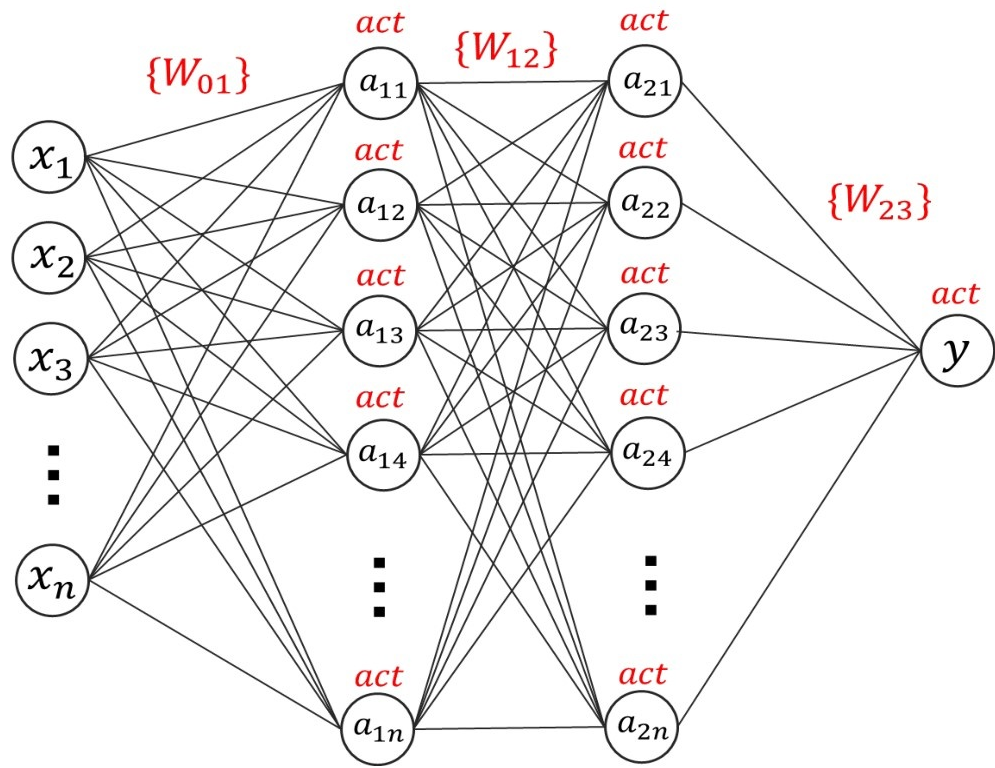}
\caption{\textcolor{VividPurple}{A Genetic Algorithm-based Neural Network}}
\label{parsi-fig-1}
\end{figure}

Where genetic algorithm fitness favors fixed, simple weights over arbitrary values generated optimizable training weights. For weight value of 0 has assigned a complexity score of 0, other fixed weight values are assigned the complexity score of 1, and the trainable weight is assigned a complexity score of 2. A genetic algorithm encodes and translates the neural network into a list of an individual set of attributes as genes in a generation as shown in \cref{parsi-fig-1}. The genetic algorithm builds many individual networks, each containing its fitness factor containing mean square error term, parsimony factor, individual neuron weights, and activation function in different hidden layers, representing a sequence of gene attribution-furthermore, this collective population of networks assigned as parent generation. A genetic algorithm evolves this network topology through probabilistic crossovers and mutations to generate a new offspring generation of networks. In the crossover operation of the genetic algorithm, we swap some of the weights and activation functions of the neurons in the networks and generate a new generation of offspring networks. In mutation operation, we replace some of the weights and activation functions of the neurons in the individual network. For all these offspring generations of networks, the neural network again performs backpropagation to determine the mean square loss and corresponding fitness factors, further used to generate the new generation network class. A tournament-based selection rule determines the top fittest individual network among the network players. Each successive evolution of the generation of the network by a genetic algorithm reproduces high reliability. We quantify the complexity and fitness by combining these two terms in the parsimonious neural network. The backpropagation and network training performed on the \texttt{Keras} package and parsimony coefficient are evaluated through \texttt{Python's} \texttt{DEAP} genetic algorithm package. The fittest individual genes based on specific fitness selection criteria filter out the other older individual networks. After N-generation, the evolved set of weights and activation functions is used to determine the empirical relation between input variables and output as an interpretable equation. A parsimonious network favors simple, linear activation over nonlinear activation functions and weights of zeros and ones over arbitrary complex trained weights and hence concisely interprets an empirical relation.

The data science-based materials design and discovery is the inverse problem, so we have the desired property values. We want to design the molecules and polymer that contain those properties. The forward problem is that we have material at hand and want to predict its properties through the statistical regression technique. \cite{segler2018planning, brown2019guacamol,shen2020deep} We have employed the generalized generative genetic algorithm in Python to design novel organic materials and polymers. \cite{nigam2021janus,nigam2019augmenting,kamal2021novel} In the algorithm, we can create various species with unique environments and unique cultures. Combining these environmental and cultural factors influences the evolution of materials and polymers by facilitating migration between various sub-groups. \cite{Kim2021Jan} Molecules and Polymers which are become invalid after recrossing SMILES and SELFIES are supposed to be a little more robust for generative models. \cite{Degen2008Sep,polyG2G,Kuenneth2021Apr,Shetty2021} For the novel polymer design, we use proxy properties to infer the desired properties for the materials, like a high-temperature dielectric, which can withstand high temperatures. The glass transition temperature might be a proxy for temperature resistance. The Dielectric constant is related to energy density and high energy density in the smaller capacitor. Moreover, the bandgap charge exit barrier is related to the ability of the material to withstand high electric fields in the breakdown regime. So we can use the genetic algorithm-based evolutionary process to synthesize these desired material properties. In the case of polymers, we can use monomer and SMILES string to break the polymer chain along chemically interesting substructures. In this regard, we use the open-source cheminformatics and machine-learning software RDKit-based breaking of retrosynthetically interesting chemical substructures (BRICS) decomposition algorithm. Then in the evolution process, we can combine the crossover and mutation phase at their joints or the blocks. We will randomly recombine the chemically interesting fragmented substructure blocks with a different monomer of the polymers and construct a new generation of the polymer. Furthermore, after a sufficient number of the evolution of the polymer generation, we integrate the machine learning property predictors algorithm to predict the properties of these new generations of polymers. To validate the outcome of this evolution, we use the retrosynthesis algorithm at the next stage to evalute realistic systhesiblity of the generated polymers. To validate the outcome of this evolution, we use the retrosynthesis algorithm at the next stage to evaluate the realistic synthesizability of the generated polymers. \cite{Chen2021Sep} To fingerprint the SMILES string, we have used the RDKit-based Morgan fingerprint and RDKit 2D-based physio-chemical-based fingerprint to incorporate the organic molecules and polymer chain's essential physical and chemical traits. To estimate the desired physical properties in the genetic algorithm, we have designed the clamping fitness function for every evolutionary generation to push the genetic algorithm to give the expected outcome. We weighted averages out the fitness for different target properties in the clamping fitness scores. The gray region is the target region for the properties. In the figure, the red region is the average of previous generation parental properties, and the blue region is the current generation child properties average. The evolution of average properties with the standard deviation monotonically increases or decreases as per the defined cost function with the evolution of the next generation. Hence, the genetic algorithm will chase the desired outcome and try to incorporate those traits into the evolutionary generation. These fitness scores can be tailored for the multi properties optimization, and hence the statistical aggregate will contain the evolutionary characteristics, which possibly contain those desired properties. Afterward, we will feed this evolutionary aggregate to the deep learning algorithm to predict the desired properties; consequently, this gives the evolution a positive defined outcome and can be a sample from the aggregate. Therefore, We can combine the genetic algorithm with the deep learning algorithm for the generative method. It is comparable to other deep learning-based Gernatative algorithms such as variational autoencoder, generative adversarial network, graph to graph translation, and normalizing flows for the molecules and polymer design. However, the advantage of a genetic algorithm is that it is faster to explore the vast chemical space like organic molecular species and polymers. Genetic algorithms get lower diversity in the subspace blocks with successive evolution. It can mitigate by using migration between different subgroups containing different fitness scores in the evolution to enhance genetic diversity. The genetic diversity in the polymer and molecules SMILES string in the evolution phase is estimated through the Jaccard similarity coefficient or Tanimoto coefficient of the newly generated SMILES and SELFIES string. Hence after achieving the target goal for the one property in the subgroup, it can migrate to another subgroup, which is at the evolutionary stage for another property goal. It will enhance the genetic diversity and create a vast chemical space explored for the next generation of evolution for that subgroup targeting the subgroup property goal while containing the parental traits from the previously emigrated subgroup. \cite{verhellen2020illuminating}

\subsection*{Jaccard Similarity Coefficient or Tanimoto Coefficient}

The Jaccard similarity coefficient measures the similarity between two sets of objects or two asymmetric binary vectors $i$ and $j$. Four frequency quantities compute from the binary dataset for two customers consisting of $n$ binary attributes for Jaccard similarity measurement. First, we calculate for both $i$ and $j$ vectors, quantity $a=$ number of attributes equal to 1. Second, we calculate for both $i$ and $j$ vectors, quantity $b=$ number of attributes equal 1 for vector $j$ but 0 for vector $i$. Third, we calculate for both $i$ and $j$ vectors, quantity $c=$ number of attributes equal 1 for vector $i$ but 0 for vector $j$. In last, we calculate for both $i$ and $j$ vectors, quantity $d=$ number of attributes equal to 0. \cite{jaccard1912the,tanimoto1958an, rogers1960a} The Jaccard similarity index for these attributes define as,

\begin{equation}\label{Jaccard-eq1}
J(i, j)=\operatorname{sim}(i, j)=\frac{a}{a+b+c}
\end{equation}

In the binary classification task, for true positive $TP$, false-positive $FP$, true negative $TN$, and false-negative $FN$, The confusion matrices evaluated through the Jaccard similarity index as,

\begin{equation}\label{Jaccard-eq2}
\text {Jaccard Index}=\frac{TP}{TP+FP+FN}
\end{equation}
\section*{t-Stochastic Neighborhood Embedding (t-SNE) Algorithm}


The t-Stochastic Neighborhood Embedding (t-SNE) algorithm is proposed to cluster the data in higher dimensions by performing non-parametric mapping to lower dimensions. However, the increased sample sizes did not scale very well; a large perplexity hyperparameter consumes extensive memory. It did not leverage features like Principal component analysis loadings to drive data clustering. To speed up the FItSNE algorithm was proposed with the initial step of k-nearest neighbor with the Barnes-Hut procedure at the cost of extensive computational memory consumption on a computer cluster. Also, in the t-SNE global structure is not preserved. The distance between data clusters does not have any mean, and only the local distance between data points within the cluster has statistical meaning. Due to these reasons, the t-SNE is not employed as the general dimensionality reduction algorithm in 100s of higher dimension manifold components but is only efficient for embedding clusters in two or three dimensions for visualization.\cite{pmlr-v5-maaten09a,vanderMaaten2012Apr}

The first \cref{tSNE-eq-1} in the four equations of the t-Stochastic Neighborhood Embedding (t-SNE) neighbor-graph algorithm defines in the high-dimensional space, distances between two points satisfying the symmetry rule and given by a Gaussian probability, \cite{maaten2008visualizing}

\begin{equation}\label{tSNE-eq-1}
\begin{aligned}
p_{j \mid i}&=\frac{\exp \left(-\left\|x_{i}-x_{j}\right\|^{2} / 2 \sigma_{i}^{2}\right)}{\sum_{k \neq i} \exp \left(-\left\|x_{i}-x_{k}\right\|^{2} / 2 \sigma_{i}^{2}\right)}, \\ 
p_{i j}&=\frac{p_{i \mid j}+p_{j \mid i}}{2 N}
\end{aligned}
\end{equation}

In \cref{tSNE-eq-1} the local structures of the data in the t-SNE are preserved through the bandwidth of the exponential probability $ \sigma $ parameter, which senses the locality of pair data points by the probability of the pairwise Euclidean distances. For large distant points $ X $, with a smaller $ \sigma $, Euclidean distance decays exponentially and becomes zero; on the other hand, it grows faster for small $ X $ nearest neighbors pair data points. In contrast with a large value of $ \sigma $, close points and probability of the pairwise Euclidean distances become comparable, and probability became unity on the limit $ \sigma \rightarrow \infty $ with pair-points becoming equidistant. In the higher dimensions, by Taylor series expansion up to second-order approximation at $ \sigma \rightarrow \infty $, The pairwise Euclidean distances probability becomes a power law as, 

\begin{equation}\label{tSNE-eq-4-1}
\begin{aligned}
P(X) &\approx e^{-\frac{X^{2}}{2 \sigma^{2}}} \\
P(X) &\underset{\sigma \rightarrow \infty}{\longrightarrow} 1-\frac{X^{2}}{2 \sigma^{2}}+\frac{X^{4}}{8 \sigma^{4}}-\ldots
\end{aligned}
\end{equation}

It is equivalent to the multi-dimensional scaling (MDS) cost function, preserving global pair-point distances irrespective pair points are nearby or far apart in a higher-dimensional manifold. At a large value of $\sigma$, long-range interactions between the data points are preserved in the t-SNE algorithm. So, it handles both local and global distances. In contrast, at the $ \sigma \rightarrow 0 $ limiting case, in the higher dimensional manifold, probability distance is exceptionally local and resembles the Dirac delta-function behavior.

\begin{equation}\label{tSNE-eq-4-2}
P(X) \underset{\sigma \rightarrow 0}{\longrightarrow} \delta_{\sigma}(X)
\end{equation}

Therefore perplexity values are critical in the algorithm and recommended finite perplexity values between 5 and 50 for a good compromise between local and global information preservation following the square root law $\approx N^{\wedge}(1 / 2)$, for a sample size of N. The \cref{tSNE-eq-2} define a perplexity constraint in the higher dimension determining for each sample an optimal $ \sigma $, as

\begin{equation}\label{tSNE-eq-2}
\text { Perplexity }=2^{-\sum \limits_{j} p_{j \mid i} \log _{2} p_{j \mid i}}
\end{equation}

The \cref{tSNE-eq-3} defines low-dimensional embedding for student t-distribution, determining the distances between the pairs of points. The crowding problem in embedding at lower dimensions is overcome through student t-distribution heavy tails as,

\begin{equation}\label{tSNE-eq-3}
q_{i j}=\frac{\left(1+\left\|y_{i}-y_{j}\right\|^{2}\right)^{-1}}{\sum_{k \neq l}\left(1+\left\|y_{k}-y_{l}\right\|^{2}\right)^{-1}}
\end{equation}

Finally, in \cref{tSNE-eq-4}, higher dimensional probability on lower-dimensional probability is projected through the Kullback-Leibler divergence loss function. The gradient in the analytical form of Kullback-Leibler divergence is estimated through Gradient Descent optimization as,

\begin{equation}\label{tSNE-eq-4}
\begin{aligned}
D_{KL}\left(P_{i} \| Q_{i}\right)&=\sum_{i} \sum_{j} p_{j \mid i} \log \frac{p_{j \mid i}}{q_{j \mid i}} \\
\frac{\partial D_{KL}}{\partial y_{i}}&=4 \sum_{j}\left(p_{i j}-q_{i j}\right)\left(y_{i}-y_{j}\right)\left(1+\left\|y_{i}-y_{j}\right\|^{2}\right)^{-1}
\end{aligned}
\end{equation}

Assuming in the Kullback-Leibler divergence loss function, higher dimensional space distance is $X$, and lower dimensional distance is $Y$, the locality of t-SNE is

\begin{equation}\label{tSNE-eq-4-3}
\begin{aligned}
P(X) &\approx e^{-X^{2}} \\
Q(Y) &\approx \frac{1}{1+Y^{2}}
\end{aligned}
\end{equation}

From \cref{tSNE-eq-4} definition,

\begin{equation}\label{tSNE-eq-4-4}
D_{KL}(X, Y)=P(X) \log \left(\frac{{P}(X)}{Q(Y)}\right)=P(X) \log P(X)-P(X) \log Q(Y)
\end{equation}

The first term in the right-hand side of \cref{tSNE-eq-4-4}, for both small and large values of $X$ close to zero. In the case of a large value of $X$, as the exponential pre-factor reaches faster zero than the logarithm term reaches $ −\infty $, the first term goes to zero. In the case of smaller $X$, the exponent evolves close to $ 1 $, as $ log(1)=0 $; hence the first term evolves to zero. Therefore only the second term is used to understand locality in the KL-divergence as intuitively,

\begin{equation}\label{tSNE-eq-4-5}
D_{KL}(X, Y) \approx-P(X) \log Q(Y)=e^{-X^{2}} \log \left(1+Y^{2}\right)
\end{equation}

We deduce that the t-SNE algorithm guaranteed that the pair points closer in the higher dimensions remain closer in lower dimensions. However, it does not enforce any such restriction for pair points far apart in higher dimensions to stay far apart in lower dimensions after transformation. Therefore, except that $ \sigma $ goes to the $ \infty $ case, it preserves only the local information in the data structure. The student t-distribution's heavy tails push apart the pair-points in higher dimensions even further in lower dimensions and provide some estimate on global distance information. However, the analytical solution of the Kullback-Leibler divergence cost function offsets this discretion in favor of a computationally efficient solution. Therefore the global distance information is not entirely preserved.

\section*{Uniform Manifold Approximation and Projection for Dimension Reduction (UMAP)}

To overcome the shortcoming of t-SNE, Uniform Manifold Approximation and Projection for dimension reduction (UMAP) a semi-empirical neighbor-graph machine learning algorithm proposed for dimensionality reduction. \cite{sainburg2021parametric,Dalmia2021UMAPConnectivity} UMAP differs from the t-SNE as it uses exponential probability distribution in higher dimensions, unlike the Euclidean distances measure of t-SNE. However, in the algorithm, Euclidean distances can also be incorporated. Also, unlike t-SNE presence of normalization denominator in \cref{tSNE-eq-1}, in UMAP, higher or lower dimensional probabilities are not normalized and reduce computation time for the high-dimensional graph as integration or summation is a computationally costly operation. This procedure is equivalent to Markov chain Monte-carlo (MCMC) approximator for integral denominator calculation by the Bayes rule. However, these probabilities are scaled for the segment [0, 1] from the functional form. The probability measure for distance metric in the higher dimensional manifold by a locally adaptive exponential kernel for each data point is,

\begin{equation}\label{tSNE-eq-5}
p_{i \mid j}=e^{-\frac{d\left(x_{i}, x_{j}\right)-\rho_{i}}{\sigma_{i}}}
\end{equation}

Where $ \rho $ is the distance of each ``i-th'' data point to its first nearest neighbor, the distance metric varies from point to point while ensuring local connectivity in the higher dimensional manifold. These fuzzy simplicial sets lead to nearest neighbor graph construction in UMAP. The $ \rho $ local connectivity parameter bonds together dense regions, sparse regions of fragmented manifold, and isolated points of a locally broken manifold through an adaptive exponential kernel that considers the local data connectivity in the UMAP. Another difference is that while perplexity in t-SNE defines as \cref{tSNE-eq-2}, UMAP instead uses the nearest neighbors number. The nearest neighbor-k without the log2 function as,

\begin{equation}\label{tSNE-eq-6}
k=2^{\sum \limits_{i} p_{i j}}	
\end{equation}

It is a much faster step for UMAP than t-SNE, where a full entropy calculate. Here in the definition of the nearest neighborhood number, the logarithmic part dropped. Algorithmically in t-SNE, it is computed through the expansive Taylor series expansion, and a logarithmic prefactor with a linear polynomial term adds not much information as logarithmic is slower. Also, the higher dimensional probability normalization step of \cref{tSNE-eq-1} omits in UMAP, which needs computation of integration-summation in the t-SNE algorithm. And the higher dimensional probabilities are symmetrized as, 

\begin{equation}\label{tSNE-eq-7}
p_{i j}=p_{i \mid j}+p_{j \mid i}-p_{i \mid j} p_{j \mid i}
\end{equation}

The symmetrization is performed on the weights of the graph as the locally varying neighborhood distance $ \rho $ evolves within the graph, and UMAP glues all such neighborhood-graph together. Also unlike t-SNE student t-distribution, in the UMAP, lower-dimensional distance probabilities are modelled through curves families $1 /\left(1+a^{*} y^{\wedge}(2 b)\right)$ without a normalization procedure as,

\begin{equation}\label{tSNE-eq-8}
q_{i j}=\left(1+a\left(y_{i}-y_{j}\right)^{2 b}\right)^{-1}
\end{equation}

The default UMAP hyperparameters for $a \approx 1.93$, and $b \approx 0.79$ is for minimum lower-dimensional distance of 0.001. However, the hyperparameter $a$ and $b$ can be estimated from non-linear least-square fitting for a piecewise minimum lower-dimensional distance function as,

\begin{equation}\label{tSNE-eq-9}
\left(1+a\left(y_{i}-y_{j}\right)^{2 b}\right)^{-1} \approx \begin{cases}1 & \text{if} y_{i}-y_{j}\leq\min_{-}\text{dist} \\ e^{-\left(y_{i}-y_{j}\right)-\min_{-} \text{dist}} & \text {if}y_{i}-y_{j}>\min_{-}\text{dist}\end{cases}
\end{equation}

Also, unlike the t-SNE Kullback-Leibler divergence loss function, the UMAP uses a binary cross-entropy cost function as,

\begin{equation}\label{tSNE-eq-10}
C E(X, Y)=\sum_{i} \sum_{j}\left[p_{i j}(X) \log \left(\frac{p_{i j}(X)}{q_{i j}(Y)}\right)+\left(1-p_{i j}(X)\right) \log \left(\frac{1-p_{i j}(X)}{1-q_{i j}(Y)}\right)\right]
\end{equation}

On the right-hand side of \cref{tSNE-eq-10}, an additional second term for the binary cross-entropy cost function assists in capturing the global data structure that t-SNE misses. By ignoring the $ p(X) $ constant terms in the cross-entropy and differentiating it for the gradient of the cross-entropy through the stochastic gradient descent (SGD) as,

\begin{equation}\label{tSNE-eq-11}
C E\left(X, d_{i j}\right)=\sum_{j}\left[-P(X) \log Q\left(d_{i j}\right)+(1-P(X)) \log \left(1-Q\left(d_{i j}\right)\right)\right]
\end{equation}

In the second step, UMAP becomes faster than t-SNE or FItSNE as the gradients are calculated from a random subset of Stochastic gradient descent (SGD) samples compared to all in regular gradient descent (GD). It will optimize memory consumption. Furthermore, normalization is not required, similar to the first step of higher-dimensional probabilities for lower-dimensional probabilities, and we do not perform normalization and optimize low-dimensional embeddings. For nearest neighbor search, t-SNE uses tree-based algorithms which scale exponentially with the number of dimensions, and hence more than two or three dimensions embedding is too slow.

Where the terms,
\begin{equation}\label{tSNE-eq-12}
d_{i j}=y_{i}-y_{j} ; \quad
Q\left(d_{i j}\right)=\frac{1}{1+a d_{i j}^{2 b}} ; \quad 1-Q\left(d_{i j}\right)=\frac{a d_{i j}^{2 b}}{1+a d_{i j}^{2 b}} ; \quad \frac{\delta Q}{\delta d_{i j}}=-\frac{2 a b d_{i j}^{2 b-1}}{\left(1+a d_{i j}^{2 b}\right)^{2}}
\end{equation}

\begin{equation}\label{tSNE-eq-13}
\begin{aligned}
\frac{\delta C E}{\delta y_{i}}&=\sum_{j}\left[-\frac{P(X)}{Q\left(d_{i j}\right)} \frac{\delta Q}{\delta d_{i j}}+\frac{1-P(X)}{1-Q\left(d_{i j}\right)} \frac{\delta Q}{\delta d_{i j}}\right] \\
&= \sum_{j}\left[\left(-P(X)\left(1+a d_{i j}^{2 b}\right)+\frac{(1-P(X))\left(1+a d_{i j}^{2 b}\right)}{\left(a d_{i j}^{2 b}\right)}\right) \frac{\delta Q}{\delta d_{i j}}\right]
\end{aligned}
\end{equation}

\begin{equation}\label{tSNE-eq-14}
\frac{\delta C E}{\delta y_{i}}=\sum_{j}\left[\frac{2 a b d_{i j}^{2(b-1)} P(X)}{1+a d_{i j}^{2 b}}-\frac{2 b(1-P(X))}{d_{i j}^{2}\left(1+a d_{i j}^{2 b}\right)}\right]\left(y_{i}-y_{j}\right)
\end{equation}

In the t-SNE, lower-dimensional coordinates are normal random initialized compared to the lower dimensional initial coordinates in UMAP assigned through graph Laplacian by Linderman and Steinerberger's hypothesis. \cite{steinerberger2015a} According to the hypothesis constructing the graph Laplacian in UMAP is equivalent to the t-SNE initial stage, minimizing KL-divergence with early exaggeration. It makes UMAP less susceptible to varying results from each run cycle as t-SNE is notorious for giving different results for every run cycle. However, in both algorithms, the initial condition has an insignificant effect on the final lower-dimensional representation. In the dimensionality reduction algorithm, the matrix factorization and neighborhood graph approach combine through various methodologies such as spectral embedding, spectral clustering, diffusion maps, Laplacian eigenmaps, and graph Laplacian. At the start of the technique, A KNN-graph is constructed, and matrix algebra adjacency and degree matrices regularize it by constructing it into the Laplacian matrix. The Laplacian matrix is factored out in the final stage by solving the eigenvalue decomposition problem as,

\begin{equation}\label{tSNE-eq-15}
L=D^{1 / 2}(D-A) D^{1 / 2}
\end{equation}

We have used the sci-kit-learn library and the spectral-embedding function to display the dataset's initial lower-dimensional coordinates. From the \cref{tSNE-eq-10} UMAP binary cross-entropy cost function is,

\begin{equation}\label{tSNE-eq-16}
C E(X, Y)=P(X) \log \left(\frac{P(X)}{Q(Y)}\right)+(1-P(X)) \log \left(\frac{1-P(X)}{1-Q(Y)}\right)
\end{equation}

\begin{equation}\label{tSNE-eq-17}
C E(X, Y)=e^{-X^{2}} \log \left[e^{-X^{2}}\left(1+Y^{2}\right)\right]+\left(1-e^{-X^{2}}\right) \log \left(\frac{\left(1-e^{-X^{2}}\right)\left(1+Y^{2}\right)}{Y^{2}}\right)
\end{equation}

\begin{equation}\label{tSNE-eq-18}
C E(X, Y)\approx e^{-X^{2}} \log \left(1+Y^{2}\right)+\left(1-e^{-X^{2}}\right) \log \left(\frac{1+Y^{2}}{Y^{2}}\right)
\end{equation}

UMAP better balances the preservation of the local-global structure. From \cref{tSNE-eq-16}, \cref{tSNE-eq-17}, and\cref{tSNE-eq-18} the similar to t-SNE at small limit values of $ X $, due to pre-factor second-term vanishing, the polynomial function is faster than the logarithmic function. Furthermore, similar to t-SNE, coordinates $ Y \rightarrow 0 $ are very small to minimize the penalty.

\begin{equation}\label{tSNE-eq-19}
X \rightarrow 0: C E(X, Y) \approx \log \left(1+Y^{2}\right)
\end{equation}

In contrast to t-SNE at opposite $ X \rightarrow \infty $, large limit values of $ X $, first term vanishes, and the second term pre-factor becomes 1, giving rise to,

\begin{equation}\label{tSNE-eq-20}
X \rightarrow \infty: C E(X, Y) \approx \log \left(\frac{1+Y^{2}}{Y^{2}}\right)
\end{equation}

As the logarithm function denominator contains $ Y $, the penalty increases heavily for a small value of $ Y $. Thus a large value of $ Y $, the logarithmic function ratio becomes $ 1 $, with zero penalties. Consequently, we gain $ Y \rightarrow \infty $ at $ X \rightarrow \infty $, it will preserve the global distances while mapping higher-dimensional space to lower. UMAP efficiently works with any number of dimensionality reductions without requiring the pre-dimensionality reduction plugging step of principal component analysis, autoencoder.
\section*{Material Modelling through Batch Reification and Fusion Optimization Algorithm}

We have expressed the Batch Reification Fusion Optimization Framework in reference to accelerated material design and employed the Bayesian optimization scheme with multi-fidelity model fusion. The concept of integrated computational materials design (ICMD) and engineering (ICME)  framework, such as materials genome initiative (MGI), \cite{council2004accelerating,holdren2011materials} aims to identify and accelerate the material design and discovery from the standard materials development cycle from years and decades to months. We combine and integrate experimentally and computational techniques material as systems and materials by design to achieve this objective proposed by Gregory Olson et al. ​​by identifying the relationship in-between the process, structure, and properties of the material.\cite{olson2000materials,xiong2016cybermaterials} The potential avenue of these initiatives is that we have many models and experiments in material science which we can probe to identify process, structure, property, and performance relationships. Furthermore, develop the high throughput experiments by leveraging the computational framework for more efficient additive manufacturing and understanding. To discover and design new materials, we must invert the process, structure, property, and performance trajectory. From the requirement of specific performance, we have to query back the process and structure of a material-a robust solution achieved by the Bayesian optimization scheme, which is the Black-box optimization technique. Optimizing the accurate model is a computationally cumbersome task. Therefore, we first create a surrogate model of an unknown function to query and acquire information. We use a gaussian process and construct an acquisition function that queries information and identifies the next best possible point to query. The surrogate model evaluated the next best possible point and a posterior recalculated for the successive queries. We update the information acquisition function and construct a more robust model cyclic. We can extract most information that we are allowed to extract from the surrogate model in the Bayesian limits. This iterative process significantly reduces the cost of optimization. Many acquisition functions can be queried, e.g., expected improvement, Thompson sampling, probability of improvement, greedy sampling, upper confidence bound, knowledge gradient,  and GP-Hedge portfolio optimization. Bayesian optimizers are usually not stuck at local optima and are suitable for global optimization. Bayesian optimization techniques have much flexibility to optimize for the different clusters of material properties data. In physical, chemical, and material science, with theory development. A wide variety of models work at broader accuracy levels containing distinct levels of microscopic descriptions of nature. The material properties identify and characterized by the empirical, continuum, finite element, and quantum-field theory-based models. All allowed us to predict a material property in its range of operation under underlying theoretical assumptions and hold some information. Therefore, using a multi-fidelity fusion model, we aim to utilize information from all available reduced-order models and evaluate the most computationally expensive and accurate ground truth model in the minimum optimization step. Using the Bayesian optimization technique, we can also capture the black box model for the regional gap space between the cluster of material data where neither constitutes model predicts the properties effectively and efficiently. In this way, by the multi-fidelity fusion, we can stitch the information model gap between the specific regional space, which arises from various reduced-order cheap models. We have expressed the Batch Reification Fusion Optimization Framework in accelerated material design and employed the sequential and batch Bayesian optimization scheme that combines two multi-fidelity optimization approaches of reification and fusion method optimization. The integrated, highly accurate procedures speed up the optimization rate for the accelerated material high-throughput computational models. Furthermore, the experimental design of materials guidance can leverage the vast array of straightforward and relatively approximate empirical models to material target properties accurate finite element method, Density-functional theory, and phase-field models developed over the decades. The reification-fusion technique is a multi-fidelity fusion approach that aims to vary fidelity model fusion to increase the model accuracy by correcting their errors. Subsequently, varying fidelity fused models correlation between each pair was estimated in the reification procedure. It enables the models with known correlation to be fused, increasing accuracy regarding the ground truth over any individual employed models. Above mentioned optimization procedure decreases the number of ground truth queries and reduces the optimization cost. The batch-based Bayesian optimization strategy argues that determination of true hyperparameters optimization for a ground truth is model is not possible. Therefore, the black box function of the actual distribution model can be approximated by determining the hyperparameters for a surrogate model. Under this premise, we sample and construct surrogate models for each set of hyperparameters from a given distribution. These different surrogate models will estimate different next-best points to query. The k-manifold clustering technique is used to cluster these queries into the number of batch size clusters. \cite{joy2020batch,ghoreishi2019efficient} The Fusion approach has three steps; first, we include the difference between a particular surrogate model and its ground truth information source called model discrepancy, which adjusts the variance of the model. In a second step, we reify the model, which is the process of information reification that describes the estimation of the correlation between the different surrogate models. In the last step, we make a fused model with known correlation as it is significantly easier than fusion under unknown correlation. The model fidelity is the model's accuracy, and the difference between a reduced-order model and the ground truth model is the model's discrepancy. We create a Gaussian process surrogate model, which provides mean and variance information but does not evaluate the accuracy of the surrogate model. To estimate model accuracy, we also evaluate the difference between the mean of the Gaussian process surrogate model and the ground truth function at points where the truth function evaluate. We calculate this difference and fit it to a second Gaussian process. The mean of the second Gaussian process provides the adjustment variance, which we add or fuse to the first Gaussian process surrogate model to estimate its accuracy concerning ground truth. Therefore, the variance at those points where the truth function evaluates infers the model's accuracy by this fusing process. In the Reification process, we assume one of the surrogate models is true. We calculate the errors relative to that model for all other models. We calculate errors for each pairwise reified surrogate model and get the covariance to calculate the correlation coefficients. In the Fusion process, we solve the minimization problem varying over the design space proposed by Winkler \cite{winkler1981combining} and get fused mean and fused variance and average correlation coefficients. We use this fused model as the Black Box for the Bayesian optimization. Next, we will discuss the theoretical background behind these optimization techniques.

\subsection*{Statistical Reification Model Correlation Estimation}

Reification meaning is making the model something more concrete and abstract. In order to estimate the correlation between the models, we first assign one of the models to be the "Truth". Assuming that we have two models that need to fuse as,

\begin{equation}\label{eq-BAREFOOT-1}
\begin{aligned} 
y = f_1(x) = \bar{f}_1(x) + \delta_1(x)\\
y = f_2(x) = \bar{f}_2(x) + \delta_2(x)		
\end{aligned}
\end{equation}

We can assume that model 1 is the truth model. In other words, we reify model 1. And so, the errors of the two models are defined as follows as,

\begin{equation}\label{eq-BAREFOOT-2}
\begin{aligned} 
\tilde{f}_1(x^*) = f_1(x^*) - \bar{f}_1(x^*) = \delta_1(x^*)\\
\tilde{f}_2(x^*) = f_2(x^*) - \bar{f}_2(x^*) = \bar{f}_1(x^*)-\bar{f}_2(x^*) + \delta_1(x^*)		
\end{aligned}
\end{equation}

Before calculating the correlation, we need to calculate the mean square errors. In the present and following sections, the notation will be that $\sigma_1^2$ refers to the total variance of the error for model 1, and we also need to calculate the covariance,

\begin{equation}\label{eq-BAREFOOT-3}
\begin{aligned} 
\mathbb{E} [\tilde{f}_1(x^*)^2] =  \mathbb{E}[\delta_1(x^*)] = \sigma_1^2\\
\mathbb{E} [\tilde{f}_2(x^*)^2] =  \mathbb{E}[(\bar{f}_1(x^*)-\bar{f}_2(x^*))^2] + \mathbb{E}[\delta_1(x^*)]\\
\mathbb{E} [\tilde{f}_2(x^*)^2] = (\bar{f}_1(x^*)-\bar{f}_2(x^*))^2 + \sigma_1^2	
\mathbb{E}[\tilde{f}_1(x^*)\tilde{f}_2(x^*)] = \sigma_1^2	
\end{aligned}
\end{equation}

Furthermore, finally, we can calculate the Pearson correlation coefficient. The subscript refers to the model that has reified as,

\begin{equation}\label{eq-BAREFOOT-4}
\rho_1(x^*) = \frac{\sigma_1^2}{\sigma_1\sigma_2} = \frac{\sigma_1}{\sqrt{(\bar{f}_1(x^*)-\bar{f}_2(x^*))^2 + \sigma_1^2}}		
\end{equation}

To complete the reification process, we repeat the steps above, but this time we reify model 2 and calculate the Pearson correlation coefficient. After we have the correlation coefficient from reifying both models, we calculate the variance adjusted mean correlation coefficient to use as the correlation coefficient in the subsequent calculations,

\begin{equation}\label{eq-BAREFOOT-5}
\bar{\rho}(x^*) = \frac{\sigma_2^2}{\sigma_1^2 + \sigma_2^2}\rho_1(x^*) + \frac{\sigma_1^2}{\sigma_1^2 + \sigma_2^2}\rho_2(x^*)
\end{equation}

This shows how we would reify two models. When there are more reduced-order models, we continue with the same pattern, use every pairwise combination of models, and calculate the average correlation coefficient. The code used to do this is shown below. In this case, we require the mean and covariance matrix for the set of points at which we will be calculating the correlation coefficients. Using matrix functions in \texttt{Numpy}, we create two three-dimensional matrices and their transposes to use element-wise operations between these four matrices to calculate the average correlation coefficients quickly and efficiently.

\subsection*{Model Fusion}
The final step in the process is the model fusion, which summarizes as finding the values of k in the following equation,

\begin{equation}\label{eq-BAREFOOT-6}
\begin{aligned} 
y = k_1(x^*)f_1(x^*) +k_2(x^*)f_2(x^*)
\end{aligned}
\end{equation}
To find these values, we need to solve a minimization problem,
\begin{equation}\label{eq-BAREFOOT-7}
\begin{aligned}
\underset{\mathbf{k}}{\min}~~\mathbf{k}^T\Sigma \mathbf{k}~\text{subject to}~k_1+k_2 = 1 
\end{aligned}
\end{equation}
Where,
\begin{equation}\label{eq-BAREFOOT-8}
\Sigma = 
\begin{bmatrix}
\sigma_1^2 & \bar{\rho}\sigma_1\sigma_2 \\
\bar{\rho}\sigma_2\sigma_1 & \sigma_2^2 \\
\end{bmatrix}
\end{equation}
This approach would not be possible if we did not know the correlation between the models. Furthermore, this makes the reification step crucial in the current approach. The solution to this minimization problem and the fused mean, and variance are as follows,

\begin{equation}\label{eq-BAREFOOT-9}
\begin{aligned} 
\mathbb{E}[y] = \frac{(\sigma^2_2 - \bar{\rho}\sigma_1\sigma_2)\bar{f}_1(x^*) + (\sigma^2_1 - \bar{\rho}\sigma_1\sigma_2)\bar{f}_2(x^*)}{\sigma^2_1+\sigma^2_2-2\bar{\rho}\sigma_1\sigma_2}
\end{aligned}
\end{equation}

\begin{equation}\label{eq-BAREFOOT-10}
\begin{aligned} 
{Var}(y) = \frac{(1-\bar{\rho}^2)\sigma^2_1\sigma^2_2}{\sigma^2_1+\sigma^2_2-2\bar{\rho}\sigma_1\sigma_2}
\end{aligned}
\end{equation}

Again relies on the matrix manipulation techniques in \texttt{Numpy} to ensure quick calculations. The approach for calculating the weights of each mean prediction is when the covariance matrix is known.

\subsection*{Statistical Mechanics' Generalized Lindemann Criterion}

Lindemann criterion is the oldest and most widely used empirical relation to predict the melting curves of solids. Around 110 years back, in 1910, Lindemann proposed the empirical relation to calculate the melting curve in the solid. \cite{Lindemann_1910} The generalized Lindemann melting relation formulates from the statistical mechanical partition function, and the generalized approach can derive a variety of melting relations. \cite{PhysRev.184.233,grimvall1974correlation} According to its statement from the statistical-mechanical point of view, a solid melts when the mean-square amplitude of atomic vibrations from its equilibrium position reaches a certain fraction of lattice spacing. These Thermally driven vibrational amplitude disorders in atoms or molecular systems define by the system-averaged Lindemann index as,

\begin{equation}\label{parsi-eq3}
q= \frac{1}{N} \sum_{i} \Bigg[\frac{1}{N-1} \sum_{j \neq i} \frac{\sqrt{\left\langle r_{i j}^{2}\right\rangle_{T}-\left\langle r_{i j}\right\rangle_{T}^{2}}}{\left\langle r_{i j}\right\rangle_{T}}\Bigg] 
\end{equation}

Where $ r_{ij} $ average distance between $ i^{th} $  and $ j^{th} $ atom, brackets $ \left\langle \right\rangle_{T}$ thermal average at temperature $ T $. The material is considered melting when the average vibrational amplitude disorder exceeds a certain fraction of the interatomic distance. Melting and Debye temperature of solid empirically correlated with atomic mass, the interatomic distance, and interatomic interaction strength. The melting temperatures $ T_{m} $ is,

\begin{equation}\label{parsi-eq4}
T_{m}=\left(\frac{k_{B}}{9\hbar^{2}}\right)f^{2}a^{2}mT_{D}^{2}
\end{equation}

Where $ f $ is fitting parameter, $ a $ is interatomic distance, $ m $ is atomic mass, and $ T_{D} $ is Debye temperature. The one-phase generalized Lindemann melting criteria formulates as the statistical mechanical partition function. Furthermore, it applies to any functional form of intermolecular potential, and Van der Waals intermolecular force is used to calculate thermodynamic properties of even inert gases' melting curves. The Simon–Glatzel empirical equation,\cite{simon1929bemerkungen} Kraut, and Kennedy melting law, \cite{kraut1966new,kraut1966new2} all follow the generalized Lindemann point of view. Simon–Glatzel's empirical equation describes the dependence of pressure on the melting temperature of a solid. These empirical relations are constituted in natural science based on observation and easily fit with the experimental data with suitable parameter values. The Lindemann melting criterion arose from statistical mechanics' arguments and expanded to the two-dimensional classical solids. \cite{PhysRevResearch.2.012040} Similar empirical relations were investigated for thermal expansion, cohesive energy, elastic properties, grain boundary, surface, creep activation energies, heat fusion, vacancy formation energy, viscosity, line dislocation diffusion, and recrystallization temperature. For $N$ particles ensemble system in configuration space, the statistical partition function $Q$ is,

\begin{equation}\label{parsi-eq5}
Q=\frac{1}{N!}\int \cdots \int \exp \left[-U\left(\vec{r}_{1}, \vec{r}_{2}, \cdots \vec{r}_{N}\right) \beta\right] d \vec{r}_{1} d \vec{r}_{2} \cdots d \vec{r}_{N}
\end{equation}

Where $U\left(\vec{r}_{1}, \vec{r}_{2}, \cdots \vec{r}_{N}\right)$ is the energy of entire system, $\vec{r}$ is coordinates of the entire system with volume, $V$, and reduced coordinates space $\vec{\lambda}$ is defined as $\vec{\lambda}=\vec{r}/V^{1/3}$, and in the reduced configuration space, the partition function $Q^{*}$ from \cref{parsi-eq5} is,

\begin{equation}\label{parsi-eq6}
Q=\frac{V^{N}}{N!} \int \cdots \int \exp \Big\{-\big[U(\vec{\lambda}_{1}, \vec{\lambda}_{2}, \cdots, \vec{\lambda}_{N})- U(0)\big] \beta\Big\} d \vec{\lambda}_{1} \cdots d \vec{\lambda}_{N} \exp\Big\{-U(0) \beta\Big\} = V^{N} Q^{*}\exp\Big\{\frac{-U(0) \beta}{N!}\Big\} 
\end{equation}

Where $U(0)$ system energy with all atoms at lattice sites with zero vibration and reduced configuration space partition function $Q^{*}$ is,

\begin{equation}\label{parsi-eq7}
Q^{*}= \iint \exp\Big\{-[U(\vec{\lambda}_{1}, \vec{\lambda}_{N})-U(0)] \beta\Big\} d \vec{\lambda}_{1} \cdots d \vec{\lambda}_{N}
\end{equation}

Lindemann's view of the melting process is based on the transition from the real configurational space to reduce configurational space. The microscopic arrangements in space remain identical as electronic cores at different temperatures in the melting process interpenetrate each other, and modify the atom's effective sizes. However, the atom's effective volumes to the total system volume remain constant as atoms occupy the same fraction of configurational phase space. Therefore, in reducing configurational phase space from \cref{parsi-eq7}, $Q^{*}$ is constant at all melting temperatures $T_{m}$ and all volumes $V_{m}$ along the melting curve.

\begin{equation}\label{parsi-eq8}
Q^{*}\left(T_{m}, V_{m}\right)=\text {constant}
\end{equation}

From this statistical mechanics argument, we now consider the material in a single-particle cell model with $N$ lattice cells. Each cell of volume $V$ has one confined molecule and moves in the potential field of its neighbors' cells. The single-particle cell $Q_{1}$ configuration partition function is 

\begin{equation}\label{parsi-eq9}
Q_{1}=v_{f}^{N} \exp\{-N E(0) / 2 \beta\}
\end{equation}

Where $E(0)$ is the potential at the center of the cell, single-particle free volume $v_{f}$ from integrating over entire cell volume $v$ is,

\begin{equation}\label{parsi-eq10}
v_{f}=\int_{v} \exp{-[E(\vec{r})-E(0)] \beta} d \vec{r}
\end{equation}

Where $E(\vec{r})$ is the effective potential field in which the particle moves inside the cell. From the argument of reduced configuration space defined in \cref{parsi-eq6}, for the reduced coordinates space $\vec{\lambda}$ in the single-particle cell model, reduced configuration space partition function $Q_{1}^{*}$ is,

\begin{equation}\label{parsi-eq11}
Q_{1}^{*}=v^{N} v_{f}^{* N} \exp{-N E(0) / 2 \beta}
\end{equation}

and with similar argument and from \cref{parsi-eq10} reduced $ v_{f}^{*} $ single-particle cell volume is,

\begin{equation}\label{parsi-eq12}
v_{f}^{*}=\int_{0}^{1} \exp{-[E(\vec{\lambda})-E(0)] \beta} d \vec{\lambda}
\end{equation}

\begin{equation}\label{parsi-eq13}
v_{f}=v v_{f}^{*}
\end{equation}

Consequently, in a single-particle cell model, we will observe the same scaled picture at all points along the melting curve from \cref{parsi-eq8} melting postulate, 

\begin{equation}\label{parsi-eq14}
v_{f}^{*}\left(T_{m}, v_{m}\right)=\text {constant}
\end{equation}

Now in the single-particle cell from \cref{parsi-eq9}, the free energy is defined as,

\begin{equation}\label{parsi-eq15}
A=-k T \ln Q_{1}
\end{equation}

And, combining from \cref{parsi-eq10} and \cref{parsi-eq12} in the \cref{parsi-eq15}, the free energy of the solid in the single-particle cell-model is,

\begin{equation}\label{parsi-eq16}
A=\frac{1}{2} N E(0)-N k T \ln v-N k T \ln v_{f}^{*}
\end{equation}

The \cref{parsi-eq16} is generic and obtained from the reduced configuration phase-space partition function and full intermolecular potential in the single-particle cell model. Furthermore, all the thermodynamic properties along the melting curve were obtained by conserving $T_{m}$ and $v_{m}$. 
\section*{Neural Network Hallucinations \& Remedies}

A mathematically illogical cost or loss expression in the deep learning paradigm also flies through the network if the mathematical expression is valid. The deep learning network capacity is large enough to compensate for illogical expression. Therefore, the model's performance may look reasonable enough to infer from the error-laden deep learning model; however, the cost or loss expression is not as intended design. The chances of falling into such an illogical trap and mitigating and debugging these silent failures are challenging tasks in the deep learning field. Polymer informatics is the intersection of polymer science and machine learning, and both fields are incredibly flexible with their compositions. The polymer universe has a massively staggering number of combinatorial possibilities. It can be classified as a linear organic polymer, homocyclic, heterocyclic aromatic polymer, mixed polymer, co-polymer, network polymer, Group-IV organo-polymer, e.g., organo-Si/Ge/Sn, and metal-organo polymer, and each class has its sub-classifications. Optimal polymer material for any given application often requires property values conversely related to each other. For example, energy storage applications need high energy density capacitors requiring a polymer with a large bandgap with insulating property and a large dielectric constant which effectively shields the electric field. However, they are conversely related to most materials. These contrasting requirements will make the polymer space delving highly significant in searching for an experimentally synthesizable, cost-effective, and sustainable process material with desired properties. On the other hand, Machine learning is a computational neural network set to learn if it tends to improve performance on some tasks, given more and more experience in terms of data. Neural networks are also incredibly flexible machine learning models. These flexibilities arise with the trade-off time cost needed to debug a neural network on silent failures and become challenging and inefficient when the model predicts the wrong outcome. A neural network architecture's capacity in the abstract defines the number of different functions a given network architecture can approximate. Furthermore, by increasing the number of hidden layers, network capacity can be increased for a particular architecture. However, it will again increase the debugging risk. Therefore, we have to optimize the network prediction outcome in terms of capacity for efficient neural network employment. It can be achieved by first employing the network with a minimum capacity of one, i.e., one hidden layer, and training the network on the data to overfit the model, so we first over learn and the network will memorize the data instead of the practical learning. Afterward, we will increase the capacity, i.e., the number of hidden layers, and add the generalization of the network layers by the Regularization technique to learn the optimal hidden layer model from the data to find the optimal capacity of a neural network for the given task. Therefore, searching the polymer space efficiently with a machine learning model requires additional scrutiny of the predicted results before entrusting the model outcome. These virtual neural experiments help guide physical experiments to verify the predicted properties. \cite{Tran2020Nov,Chandrasekaran2020Jun,Lightstone2020Jun,Chen2020May,Zhu2020Jul,Venkatram2019Oct,Jha2019Jan,Mannodi-Kanakkithodi2021,Mannodi-Kanakkithodi2016May,Huan2015Jul,BibEntry2021Jun} Furthermore, to increase the complexity of polymer science problems, there are complex phenomenons such as polaron and bi-polaron transport. Physical understanding is yet incomplete, and indeed, the physics-based simulations method such as density functional theory and Ab-initio molecular dynamics do not adequately cover the entirety of governing principles. Moreover, beyond computation correction, the resources to estimate properties through these methods are incredibly expansive. Furthermore, these simulation environments do not have the building prediction statistics in their physical formulation. Due to these considerations, a data science-based statistical exploratory scheme has the enormous advantage of exploring and predicting the polymer space. Multi-layer perceptrons and graph neural networks class of network can be debugged through the net-debugger toolbox. There are five tests we should perform the debug a neural network sanity check. First, test the tensor shape and size in between model output and training label. In the neural network, broadcasting is performed, a programming constructs in which tensors are stretched into higher dimensions and later project the latent information in lower dimensions space. The second test is the input-independent baseline test. We train the models on two different data sets, where one is a real dataset and another is a transformed dataset with an output that is not dependent on the input to make the loss maximum for this model. We compared these two models' results, and the test will pass if, after the number of the epochs, the dependent loss is much smaller than the independent loss, where losses are expected to be maximum. In the third overfit small batch test, we use a small batch of data that can completely overfit. It will check that the model network architecture has enough capacity, i.e., the hidden layers. The neural network is composed of several hidden layers, and its mathematical implication is a matrix multiplication operation followed by a nonlinear activation function operation on the input information. We need non-negligible gradients in the backpropagation step to get the model to improve and learn efficiently. Therefore, the nonlinear activation function has a massive impact on the gradients in our model. The fourth test we perform is a chart dependencies test for the correct implementation of matrix multiplication. We observe and record from the training dataset the input-output pair. Furthermore, each pair is known as an instance, and each instance has its associated features. Each instance is independent, and information should not be passed in between separate instances in the neural network. The final test we performed is overfitting the entire training data. It returns the capacity of the most miniature architecture that can overfit all of the training data and work as starting architecture. We grow this network with the regularization step, hyperparameters optimization, and splitting of the dataset into training, validation, and test set.  
\bibliographystyle{IEEEtran}
\bibliography{Materials_Informatics_An_Algorithmic_Design_Rule}

\end{document}